\documentclass[11pt,english]{article}
\usepackage[utf8]{inputenc}
\usepackage{amsmath,amssymb,amsthm}
\usepackage{color}
\usepackage{soul}
\usepackage{setspace}
\usepackage{natbib}

\usepackage{hyperref,comment}
\usepackage{xcolor}
\newcommand{\pdfcolor}{blue!80!black}
\hypersetup{
    colorlinks,
    linkcolor={\pdfcolor},
    citecolor={\pdfcolor},
    urlcolor={\pdfcolor}
}
\usepackage[title]{appendix}

\usepackage{quoting}
\quotingsetup{vskip=1pt}

\usepackage{titlesec}
\titlespacing{\section}{0pt}{6pt plus 4pt minus 2pt}{6pt plus 2pt minus 2pt}
\titlespacing{\subsection}{0pt}{6pt plus 4pt minus 2pt}{6pt plus 2pt minus 2pt}
\titlespacing{\subsubsection}{0pt}{6pt plus 4pt minus 2pt}{6pt plus 2pt minus 2pt}

\usepackage{enumitem}
\setlist[itemize]{leftmargin=1cm, noitemsep, topsep=1pt}
\setlist[enumerate]{leftmargin=1cm, noitemsep, topsep=1pt}

\title{
\Large Fair Allocation of Vaccines, Ventilators and Antiviral Treatments: Leaving No Ethical Value Behind in Health Care Rationing\thanks{
This version supersedes: ``Leaving No Ethical Value Behind: Triage Protocol Design for Pandemic Rationing,''
NBER Working Paper 26951, April 2020.  We are grateful for input from several experts in bioethics, emergency healthcare, 
and webinar participants at Johns Hopkins University, London Business School, Stony Brook International Conference on Game Theory, 
Australasian Microeconomic Theory Seminars,
the HELP! (HEaLth and Pandemics) Econ Working Group Seminar Series, 
Penntheon Webinar Series, 
Stockholm School of Economics Corona Economic Research Network Webinar Series,  UK Virtual Seminars in Economic Theory, NBER Workshop
on Market Design, Harvard-MIT Theory Workshop, Virtual Market Design Seminar Series, University of Tokyo Theory Seminar, CUHK-HKU-HKUST Theory Seminar, North Carolina State University Theory Seminar, University of Birmingham Seminar, Brazilian School of Economics Seminar,
University of Southern California Theory Seminar, 54th Memorial Gilbert Lecture at University of Rochester, and the University of Alabama Culverhouse  Seminar. Nikhil Agarwal, Alex Rees-Jones, Robert Truog, Iv\'{a}n Werning, and Douglas White provided helpful comments. Feedback from an anonymous Associate Editor,  
three anonymous referees, David Delacr\'{e}taz, Fuhito Kojima, Govind Persad, and Alex Teytelboym was particularly valuable.}}
\author{Parag A. Pathak \quad \quad Tayfun S\"{o}nmez\\ \quad \quad M. Utku \"{U}nver \quad \quad M. Bumin Yenmez\thanks{Pathak: Department of Economics, Massachusetts Institute of Technology, and
NBER,  email:~\textsf{ppathak@mit.edu}, S\"{o}nmez:
Department of Economics, Boston College, email:~\href{mailto:sonmezt@bc.edu}{\tt sonmezt@bc.edu},
\"{U}nver: Department of Economics, Boston College, and Professorial Research Fellow, Deakin University email:~\href{mailto:unver@bc.edu}{\tt unver@bc.edu},
Yenmez: Department of Economics, Boston College, email:~\href{mailto:bumin.yenmez@bc.edu}{\tt bumin.yenmez@bc.edu}.}}
\date{\footnotesize \textit{First Draft}: April 2020\\ \textit{Current Draft}: May 2023\\ \medskip forthcoming in \textit{Management Science\/}}

\usepackage{graphicx}
\usepackage{comment}

\newtheorem{theorem}{Theorem}

\newtheorem{definition}{Definition}
\newtheorem{lemma}{Lemma}
\newtheorem{proposition}{Proposition}

\newtheorem{example}{Example}
\newtheorem{observation}{Observation}

\def\calb{\mathcal{B}}
\def\calc{\mathcal{C}}   \def\calf{\mathcal{F}}
  \def\calp{\mathcal{P}}
 \def\calm{\mathcal{M}}

\DeclareMathOperator{\rank}{rank}
\def\pieq{\underline{\pi}}
\def\hn{\hat{\nu}}
\def\hs{\hat{\sigma}}

\def\calm{\mathcal{M}}

\newcommand{\be}{\begin{equation}}
\newcommand{\ee}{\end{equation}}
\newcommand{\bes}{\begin{equation*}}
\newcommand{\ees}{\end{equation*}}
\newcommand{\bea}{\begin{eqnarray}}
\newcommand{\eea}{\end{eqnarray}}
\newcommand{\beas}{\begin{eqnarray*}}
\newcommand{\eeas}{\end{eqnarray*}}

\newcommand{\abs}[1]{\left \vert#1\right \vert}

\renewcommand{\v}{\varphi_{\triangleright}}
\newcommand{\vp}{\varphi_{\triangleright'}}
\newcommand{\tr}{\triangleright}
\newcommand{\trp}{\triangleright'}

\def \ovf {\overline{f}}
\def \unf {\underline{f}}
\def \he {{0}}
\def \ve {{r_{u}}}
\def \cals {\calm_S}

\renewenvironment{proof}[1][Proof]{\textbf{#1.} }{\  \rule{0.5em}{0.5em}}

\begin{document}
\maketitle
\begin{abstract}
A \textit{priority system\/} has traditionally been the protocol of choice for the allocation of scarce life-saving resources
during public health emergencies. Covid-19 revealed the limitations of this allocation rule. 
Many argue that priority systems abandon ethical
values such as equity by discriminating against disadvantaged communities. We show
that a restrictive feature of the traditional priority system largely drives these limitations. 
Following \textit{minimalist market design\/}, 
an institution design paradigm that integrates research and policy efforts, 
we formulate pandemic allocation of scarce life-saving resources as a new application of market design.  
Interfering only with the restrictive feature of the priority system to address its shortcomings, 
we formulate a \textit{reserve system\/} as an alternative allocation rule. 
Our theoretical analysis develops a general theory of reserve design. We relate
our analysis to debates during Covid-19 and describe the impact of our paper on policy and practice.
\end{abstract}


\newpage

\onehalfspacing

\section{Introduction}

Agencies responsible for public health and emergency preparedness
regularly design guidelines to allocate scarce life-saving resources in crisis situations.
These situations range from wartime triage medicine to public
health emergencies, such as influenza pandemics and Covid-19.
Items in short supply include vaccines, ventilators, and anti-viral treatments.

How to implement a rationing system for life-saving resources during a crisis presents a complicated
question rife with ethical concerns.   Rationing guidelines typically
start by articulating several ethical principles.  These principles include
equity, which is fair distribution of benefits and burdens; utilitarianism,
which is maximizing welfare; reciprocity, which is respecting contributions others have made in the past;
instrumental valuation, which is respecting contributions others could make in the future;
solidarity, which is fellowship with other members of society; non-discrimination, which is requiring
that certain individual characteristics such as gender, race, and age play no role in allocation protocols.
Guidelines also emphasize procedural values, such as accountability, reasonableness, and transparency.\footnote{
Examples of this framework are in \citet{who:07}, \citet{prehn/vawter:08}, \citet{nejm:20}, and \citet{daley:20}, among others.}

After articulating these ethical principles, guidelines describe how to operationalize them with an allocation rule.
Traditionally, the most common rule suggested  in these guidelines is a \textit{priority system\/}
in which patients are placed into a single priority order, and allocation 
is made in order of priority.\footnote{For example, 2018 US Centers for Disease Control Influenza Vaccine
Allocation guidelines place patients into one of four tiers based on
(1)  provision of homeland and national security, (2) provision of health care and community
support services, (3) maintenance of critical infrastructure, and (4) membership of the general population \citep{cdc:18}.}
In some cases, the priority order is derived from an objective scoring method, resulting in a  \textit{priority point system}. 
This practice is especially prominent for rationing of ICU beds and ventilators,
where 19 states in the US base the priority score on the Sequential Organ Failure Assessment (SOFA) score, a
measure of organ dysfunction \citep{piscitello:20}.\footnote{The European Society of Intensive Care Medicine devised the
SOFA score at a consensus meeting in October 1994 in Paris, France \citep{vincent/etal:96}.
Each of six organ systems---lungs, liver, brain, kidneys, blood clotting, and blood pressure---is independently 
assigned a score of 1 to 4.  The SOFA score sums
these six scores, and sicker patients are assigned higher scores.
While not initially designed as a prognostic score, some subsequent research supports
its use for that end \citep{jones/trzeciak/kline:09}.}

The Covid-19 pandemic has spurred renewed interest in medical rationing guidelines
and has revealed several important limitations of the existing allocation mechanisms.
Whether it is rationing of ventilators, antiviral drugs or vaccines,
a common theme in many debates is that existing guidelines and allocation mechanisms have given up on certain values.
For example, advocates for disadvantaged groups criticize priority point systems that use the SOFA score.
They argue that these criteria are discriminatory for they fail to acknowledge pre-existing discrimination in access to health care \citep{schmidt:20}.
Similarly, disability advocates argue against rationing plans based solely on survival probabilities
because they are inherently discriminatory for certain types of disabled patients.
Some even reject a detailed triage protocol in favor of random selection \citep{neeman:20}.

In this paper, we report on our integrated research and policy efforts on design of pandemic medical resource allocation 
protocols during Covid-19. In these efforts we followed an institution design paradigm called \textit{minimalist market design}  \citep{sonmez:23}, 
and succeeded in informing policy in several jurisdictions. 
Under this framework, the main tasks of a design economist are the following:
\begin{enumerate}
\item Identify the mission of the institution. 
What are the policy objectives of policymakers, system operators, and other stakeholders? 
\item Assess whether the existing institution (in our application the priority system) is a good choice for the design objectives. 
\item If the existing institution is not a good choice, then find the root causes of the failures, and design an 
alternative  institution that corrects them by only interfering with the root causes of the failures. 
\item In case there are multiple ``minimalist'' designs through tasks 1-3, present a comprehensive analysis 
of these competing institutions. 
\end{enumerate}

At the outset of the pandemic, we observed that many shortcomings of existing protocols 
are directly tied to the restriction of allocation rules to the priority system.  A priority system's single priority order for all units
impedes its ability to represent a variety of ethical considerations.
As an alternative, we formulated and proposed the \textit{reserve system\/} by directly addressing  
the root cause of the shortcomings of the priority system.  

In a reserve system, units are divided into multiple categories, with each representing an ethical value
or a balance of multiple ethical values. Rather than relying on a single priority order to allocate all units,
a category-specific priority order prioritizes individuals for units in each category.
This heterogeneity allows a rationing system to accommodate the desired ethical values without
needing to aggregate them into a single metric or into a strict lexicographic hierarchy.

In some applications, minimalist market design prescribes a unique institution. Some examples are 
the US Army's branching system for the cadets  \citep{greenberg/pathak/sonmez:23} and Supreme Court judgments 
in India for affirmative action policies for government jobs \citep{sonmez/yenmez22}. 
In these applications, the fourth task above under minimalist market design is redundant. 
In our application, in contrast, a reserve system is a large class of allocation rules which have the
flexibility to accommodate a rich set of normative principles. 
However, when individuals can receive units from multiple categories, some of these normative implications
may be hard to observe even for experts, and policymakers may risk undermining 
policy objectives by choosing the wrong reserve system. 
Consequently, we present a comprehensive analysis of the reserve system in the formal part of our paper. 

The rest of our paper is organized as follows. In Section \ref{sec:minimalist}, we detail how minimalist market design 
shaped our integrated research and policy efforts for pandemic allocation of scarce life-saving resources during Covid-19 pandemic. 
In Section \ref{sec:model}, we present a general model and analysis of the reserve system. This section also includes 
a detailed discussion of the earlier literature on reserve systems, and relates our analytical results to earlier ones in the literature. 
In Section \ref{field}, we discuss the policy impact of our paper during and after Covid-19 pandemic. 
We conclude in Section \ref{conclusion}, and  relegate resource-dependent design considerations, 
a theory of smart reserve systems, and the proofs of our formal results to the Online Appendix. 

\section{A Case Study in Minimalist Market Design: Pandemic Medical Resource Allocation} \label{sec:minimalist}

\textit{Minimalist market design} \citep{sonmez:23} is a framework that integrates research and policy efforts to influence
the design of real-life allocation mechanisms.  
Our paper and policy work on the allocation of scarce life-saving resources during the Covid-19 pandemic is an exercise  
in this institution design paradigm.

Traditionally, economists have rarely, if ever, been part of 
triage committees or task forces that design pandemic medical resource allocation systems \citep{pathak/sonmez/unver-jama:20}. 
One reason for this lack of representation is the following: Whereas the primary normative focus of welfare economics largely 
revolves around preference \textit{utilitarianism\/} \citep{li2017ethics}, bioethics and public health scholars and practioners frown 
upon a narrow focus on this principle. 
Preference utilitarianism is also the dominant normative objective for mainstream approaches in design economics.  
In contrast, the objectives of policymakers, system operators  and other stakeholders---which may 
include non-utilitarian principles---take central stage in \textit{minimalist market design\/}.  
This cautious and respectful approach to institution design is at the heart of the  policy impact 
of our paper---reported in Section  \ref{field}---during and after the Covid-19 pandemic.

\citet{sonmez:23} identifies three main steps, and in some applications, a fourth step to achieve policy goals within this framework.  
We begin by briefly reviewing these steps and how they fit into our integrated research and policy efforts during  Covid-19. 

The first step under the minimalist market design is identifying the main objectives of the policymakers, stakeholders, and authorities that operate the system. The historical evolution of pandemic resource allocation guidelines for influenza and other highly contagious respiratory diseases holds key clues regarding these objectives, as do bioethics and law literature on the topic. We first identify and document these objectives as fulfilling multiple ethical criteria in a balanced manner. 

The second step is determining whether the institution in place satisfies these objectives. 
In our application of pandemic medical resource allocation, various forms of the 
priority system turn out to be the protocol of choice.  These systems fail to satisfy key design objectives as we discuss and document below.

The third step is proposing an alternative to the current institution through a \textit{minimally invasive} intervention. 
To do so, it is important to identify and address 
the \textit{root causes\/} of the inconsistency between objectives and practice.  
Reliance on a single priority list to formulate the claims of individuals on the scarce resources is the root cause of the failures of 
existing pandemic rationing protocols. 
In this respect, we formulate and advocate for the \textit{reserve system\/}, which differs from the priority system in this key aspect only, and fulfills the design objectives for the allocation of  various scarce life-saving resources (e.g. ventilators, ICU beds, therapeutics, vaccines) through multiple priority lists. 

In some settings, a unique minimalist intervention may eliminate the inconsistency 
between the objectives of the stakeholders and the practice. 
In those cases, the three main steps of minimalist market design prescribe a unique intervention.
In other applications, however, some of the primary objectives of the stakeholders may  be incompatible with others.\footnote{A well-known 
example in market design literature is the incompatibility between the normative principles of \textit{Pareto efficiency\/} and \textit{no justified envy}
in the context of school choice \citep{balinski/sonmez:99, abdulkadiroglu/sonmez:03}.}
In such applications, a design economist may need to formulate compromises between these objectives.  
Finally, in other cases, multiple minimalist interventions  may eliminate the  inconsistency 
between the primary objectives and the practice. 
The fourth step of the minimalist market design may be potentially valuable in these applications, especially in applications  
``in which issues of social, racial and distributive justice are particularly salient'' \citep{hitzig2020}.   
Pandemic medical resource allocation is one such application.

The goal of  the fourth step in minimalist market design is maintaining \textit{informed neutrality\/} \citep{li2017ethics} between 
reasonable normative principles  in design proposals. 
Our formal analysis in Section \ref{sec:model} pertains to this aspect of our application. 
Since our proposed reserve system has many flexible ways to fulfill multiple ethical principles, 
we develop new theory to interpret and characterize this flexibility. 
This allows us to propose alternative solutions to policymakers with different tradeoffs
and helps them to avoid various unintended consequences.  

In the next four subsections, we discuss how each of these steps were carried out 
in the context of pandemic medical resource allocation, 
and how our minimalist approach to institution design helped us achieve policy goals and shape aspects of 
Covid-19 resource allocation guidelines or procedures virtually in real-time with the progress of this paper.

\subsection{Ethical Principles for Allocation of Scarce Life-Saving Resources} 

In the early days of the Covid-19 pandemic, it became clear that medical systems in many countries would be overwhelmed in providing care for patients. The most urgently needed life-saving resources, such as intensive care unit (ICU) beds and ventilators, were insufficient to meet expected demand, even in countries with advanced health systems. New vaccines and therapeutics had to be developed to combat the SARS-CoV-2 virus. Once developed, the initial demand for these resources would likely surpass the supply, which was bounded by production capacity. As a result, rationing of such resources would be inevitable. 

The rationing of ICU beds and ventilators was already underway in Northern Italy \citep{rosenbaum2020facing} when leading bioethicists laid out principles of the design of triage protocols in an issue of the \textit{New England Journal of Medicine} (available online in March 2020). In this issue, \citet{nejm:20} urges healthcare providers to use multiple ethical principles in the design of triage protocols rather than a single criterion. Because subjective value judgments could lead to different weights for each principle, this paper argues for adopting transparent, fair, and consistent allocation methods determined with the consensus of affected parties, clinicians, patients, public officials, and others. \citet{daley:20} also advocates using multiple principles in determining rationing protocols. 

Different advocates put forward specific principles including the  \textit{utilitarian principle\/} of saving most lives or saving most life years,  \textit{instrumental value}, i.e., healthcare professionals and front-line workers who will be important in helping society cope with the pandemic should get priority, the \textit{life-cycle principle} in which younger people should be prioritized over older patients due to a fairness notion, which states each person should have a similar expected duration of life, \textit{equal treatment of equals} whereby similar patients should have an equal chance for the resources, and \textit{reciprocity} in which patients who have done good deeds for the society in the past should have priority. 

While these papers mention these principles in detail explicitly, they are largely silent about their operationalization with a specific allocation protocol. We have to look into the history of pandemic triage guidelines in the US and older bioethics literature to find what allocation procedures were actually proposed or adopted.

\subsection{Priority System  for Allocation of Scarce Life-Saving Resources} \label{sec:priority}

Prior to our work, the most common allocation mechanism for pandemic medical rationing
was a version of a \textbf{priority system}.  
Under this system, individuals are priority-listed based on various criteria reflecting the ethical values guiding the allocation of the 
scarce medical resource, and all units   are allocated based on this single priority order.

In some applications, most notably for the allocation of ventilators and ICU beds,
the underlying priority order is obtained through a monotonic scoring function.
This refinement of a priority system is called a \textbf{priority point system}.
Under this system, each ethical value is also represented with a monotonic function.
Values are then integrated with an additive formula, which produces an aggregate point score for each patient.
The claims of patients on scarce life-saving resources are determined based on their point scores,
where a lower score may be associated with a higher claim or a lower claim depending on the application.
Sometimes priority scores may be coarsened into multiple tiers, and all patients in the same tier may have the same claim.
Tie-breaking within a tier is typically based on clinical criteria or lotteries.

A single-principle point system is a priority point system based on only one ethical value.
The 2015 New York State Ventilator Guideline \citep{nys:15} is a prominent example.  In the system,
eligible patients are ordered based on a measure of  estimated mortality risk called the \textit{SOFA score\/}, which is re-evaluated every 48 hours.
In cases of excess demand among members of a given priority tier,
New York and other proposals recommend random allocation--- a lottery---among equal-priority patients \citep{nys:15,nejm:20}.

Several bioethicists and clinicians criticize single-principle priority point systems solely based on the SOFA score for ignoring multiple ethical values.
These critics emphasize the need to integrate a variety of ethical values and they
advocate for a \textit{multi-principle} approach (e.g., \citealp{white:09} and \citealp{maryland:17}).
\citet{white:09} describes a multi-principle priority point system
where several ethical values are placed on a
numerical scale and summed up across ethical values to arrive at a single number.
Variants of the system shown in Table 1 are widespread as the leading multi-principle
priority point system for ventilators,\footnote{\citet{kamp:20} reports that several hundred US hospitals had adopted this system at the
outset of the Covid-19 pandemic.} and more than half of US states adopted either a single- or multiple-principle priority point system 
during the Covid-19 pandemic \citep{whyte:20}.

\medskip

\begin{center}
\includegraphics[scale=0.8]{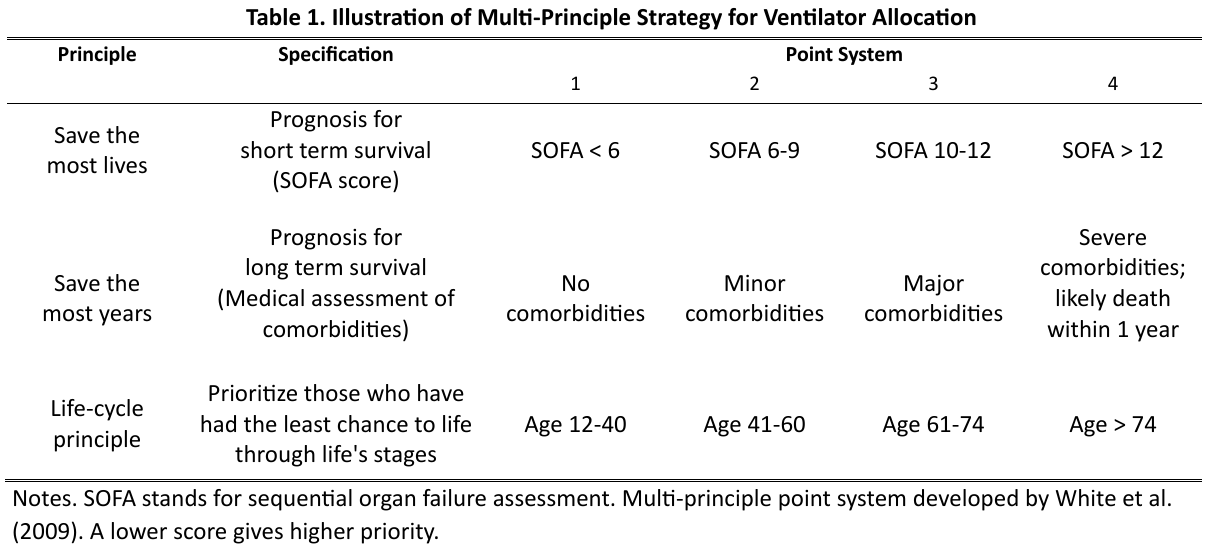}
\end{center}

While practical,  priority  systems have a number of important limitations.
A priority system struggles to integrate different ethical values because some values are
incommensurable.  Incommensurable values do not share a common
standard of measurement and cannot be compared with one another.  
For example, the values of saving the most lives and the life-cycle principle
in Table 1 are scaled to numeric values, even though there is no natural scaling.
In addition, many ethical values have implications for group composition,
and a priority point system (or a priority system in general) lacks the flexibility to accommodate these considerations.
In many cases, these challenges have led to the exclusion of some of ethical values all together.
We elaborate on several of these points next, focusing on examples of the debates on rationing of ventilators
and ICU beds. Since virtually all states with guidelines recommend priority point systems, we
present their shortcomings, but many of our points directly apply to the broader class of priority systems in general.

 \subsubsection{Failure to Represent the Desired Ethical Values}\label{failure}

A priority point system requires that ethical values be mapped to a single linear order.  However, some ethical principles involve claims of patients which cannot be represented with a monotonic function.
One example is group-based policies, such as those related to social justice issues regarding providing access to certain groups of individuals, such as historically socio-economically disadvantaged groups, at proportions equal to or higher than their percentage representation in the population. As we will discuss in Section \ref{sec:implementation}, concerns
for access for these groups has been an important desired feature of medical resource allocation mechanisms during the Covid-19 pandemic for therapeutics and vaccines. 
This shortcoming of priority systems is exacerbated due to the fact that considerations based on group composition cannot be represented with a single function that relies only on individual attributes.

When constructing priority points and incorporating multiple ethical values, a priority point system norms or scales different and potentially incommensurable ethical principles into one dimension.  These challenges are like the usual ones associated with aggregating social alternatives
into a single ordering based on multiple inputs -- a situation that involves ``comparing apples to oranges.''
The debate on how rationing guidelines should
compare claims of children versus adults illustrates this issue.
Initial Massachusetts guidelines state that indicators that feed into scores for adults are not
reliable for children \citep{mass:20}. They explain that ``scoring systems that are meaningful for adult critical
care patients do not apply to pediatric patients or newborns.''
As a result, the Massachusetts guidelines use a different scoring system for children.
However, their point system then uses a single priority point system
to evaluate all patients together.  This decision ends up comparing the point scores of children with those of adults.

Third, the fact that all resources are ordered using a single uniform priority
order can result in the exclusion of certain ethical values.  A prominent example
appears in the debate about prioritizing essential personnel.
Many groups argue that essential personnel, and especially frontline healthcare workers,
should receive priority allocation of scarce resources  under triage scenarios.
This view is also strongly endorsed by medical ethicists based on the backward-looking principle of \textit{reciprocity} and
the forward-looking principle of \textit{instrumental value} \citep{nejm:20}.
Nevertheless, states such as Minnesota and New York 
had to give up on this consideration in the past, largely due to  concerns about the extreme scenarios where no units
may remain for the rest of the society.
The Minnesota Pandemic Working group reasons that \citep{minnesota:10}:
\begin{quote}
\textsf{... it is possible that they [key workers] would use most, if not all,
of the short supply of ventilators; other groups systematically would be deprived access.}
\end{quote}
The New York State Task Force recognized the need to provide ``insurance''
for frontline health workers, but 
 ultimately decided against such a priority  \citep{nys:15}:
\begin{quote}
\textsf{Expanding the category of privilege to include all the workers listed above may mean that
only health care workers obtain access to ventilators in certain communities. This approach
may leave no ventilators for community members, including children; this alternative was unacceptable to the Task Force.}
\end{quote}
For both states, the committees abandoned the ethical values of reciprocity and instrumental value because
of the limitation of priority systems.  In a priority system, providing preferential access to any group for any 
portion of the resources means giving preferential treatment for all units.

Fourth, a single priority order may struggle to integrate the principle of non-exclusion.
This principle is the idea that every patient, no matter
his or her circumstances, should have some hope of obtaining a life-saving resource.
In the March 2020 Alabama rationing plan, individuals with severe or profound mental
disabilities were considered ``unlikely candidates for ventilator support.''\footnote{After
widespread backlash, this plan was withdrawn on April 9, 2020.}   Washington state
guidelines recommend that hospital patients with ``loss of reserves in energy, physical ability, cognition
and general health'' be switched to outpatient or palliative care \citep{fink:20}.
In a priority system coupled with excess demand for available resources by the higher-priority groups---even without any explicit exclusion of certain types of individuals---some patients
in a lower-priority group would never be treated during a shortage.

 \subsubsection{Implementation Issues in Existing Guidelines} \label{sec:implementation}

These conceptual challenges with a priority point system are reflected in actual design challenges in
several guidelines. We describe three examples.

Massachusetts Crisis Standards of Care guidance for the Covid-19 pandemic
was developed in April 2020 by a committee consisting of
medical experts and ethicists \citep{ma-cc:20}.
The guidelines provided an adaption of the system described in Table 1 without
the life-cycle consideration. However,
after precisely spelling out the priority order with a table of numbers for each dimension, the document states:
\begin{quote}
\textsf{Individuals who perform tasks that are vital to the public health response, including all those whose work directly to support the provision of care to others, should be given heightened priority.}
\end{quote}
This clause provides no further description of how heightened priority is
to be implemented.  This
lack of transparency contrasts with the level of
precision regarding other ethical principles, and may
reflect their inability to arrive at consensus given the underlying priority point system.\footnote{Illustrating
this tension, revised MA
guidelines issued on October 20, 2020 no longer include any reference for heightened priority for these essential personnel groups \citep{ma-cc2:20}.}

Crisis standards of care guidelines initially developed in Pittsburgh
use a similar adaptation of
the system described in Table 1.  They offer a vague description
of tie-breakers:
\begin{quote} 
\textsf{In the event that there are ties in priority scores between patients, life-cycle considerations will be used as a tiebreaker, with priority going to younger patients, who have had less opportunity to live through life's stages. 
In addition, individuals who perform tasks that are vital to the public health 
response specifically, those whose work directly supports the provision of acute care to others will
also be given heightened priority (e.g., as a tiebreaker between identical priority scores)}
\end{quote}
In their adaptation of Table 1, the designers saw
saving the most lives as more justified than either the life-cycle principle or the instrumental value principle.
However, the guidelines did not choose between these two latter ethical values in the event of tie-breaking.

The third example is from the June 6, 2020 update to the Arizona allocation
framework \citep{arizona:20}.  This document also offers a table
prioritizing patients based on SOFA scores and whether a patient
is expected to live or die within one or five years despite successful treatment
of illness.  It then warns that ``a situation could arise where limited resources
are needed by two or more patients with the same triage priority scores''
in which case ``additional factors \textit{may} be considered as priorities.''
Among the list of additional factors are whether patients are pediatric patients,
first responders or health care workers, single caretakers for minors
or dependent adults, pregnant, or have not had an opportunity to experience
life stages.   There is no further detail on how multiple tie-breakers would be implemented.

Beyond these specific updates to guidelines during the Covid-19 pandemic, there are also concerns that incomplete
descriptions have rendered such guidelines ineffective in other settings.
During the 2004 shortage of the influenza vaccine, \citet{schoch-spana:05} state that CDC guidelines were too general and broad. Specifically,
\begin{quote}
\textsf{Local providers thus faced gaps in the local supply of inactivated vaccine as well
as the absence of \textit{a priori} prioritization standards relevant to initial and evolving local
conditions.  Practitioners and local and state health authorities throughout the
U.S. faced a similar predicament.}
\end{quote}
 Despite these vagaries, some state departments of health penalized clinicians if protocols were not followed.
 For example, \citet{lee:04} describes that Massachusetts threatened a fine or prison time for whoever violates the CDC order on distribution
during the 2004 flu shot shortage. The requirement to follow an incompletely specified system places clinicians in a difficult position.

\subsection{Reserve System as a Minimalist Alternative to Priority System}

Challenges presented in Sections \ref{failure} and \ref{sec:implementation} mainly stem from one limiting
feature of a priority  system: it relies on a single priority ranking of patients that is identical for all units. 
Thus, the root cause of its failure is the forced use of a uniform priority criteria for allocation of each and every unit of
the scarce medical resource. 
This limitation is directly addressed under a reserve system which allows for heterogeneity of patient claims over different units. 

A \textbf{reserve system} has the following three main parameters:
\begin{enumerate}
\item a division of all units into multiple segments referred to as  
\textit{reserve categories},\footnote{This division is for accounting purposes only, and it does not attach a specific unit to a category.}
\item number of units in each of these categories, and
\item a \textit{priority order\/} of individuals for each of reserve category.
\end{enumerate}
Observe that a priority system is a single-category reserve system. 
This aspect of the reserve system makes it easier to communicate with policymakers
and experts in bioethics and public health who are familiar with the priority system. 

Reserve categories can differ either based on the groups to receive higher priority or
the combination of ethical principles to be invoked. Thus, categories can directly or indirectly incorporate different ethical principles advocated for the rationing of 
life-saving resources during the  Covid-19 pandemic (e.g., see \citealp{nejm:20} and \citealp{daley:20}). As a result, the reserve system minimally differs 
from the more restrictive priority system by allowing category-specific criteria to prioritize individuals. 

Various forms of the reserve system have been the procedure of choice  
in the allocation of various scarce resources, especially when compromises need to be reached among different segments of society. In many settings, reserve categories and the number of resources reserved for each category were determined after years of community engagement. 
For example, the most extensive affirmative action program in the world takes place in India in the allocation of state jobs and public school seats. After decade-long community discussions, a reserve system was adopted for affirmative action (see 1979 Mandal Commission Report for the summary of the efforts) and was formulated in 1992 Indra Sawhney v. Union of India Supreme Court case \citep{sonmez/yenmez22}. Chicago public schools also adopted a reserve system in admission to highly selective high schools after US federal law prohibited the use of racial criteria. Instead, 
five reserve categories were adopted based on socioeconomic group specifications \citep{dur/pathak/sonmez:20}. 

The priority order of patients for each reserve category also incorporates information on \textit{eligibility criteria}. 
If every individual is eligible only through one category, 
then these categories can be processed in parallel to determine who will receive the resources, and the reserve system is uniquely identified. 
However, in most applications, many individuals will be eligible through multiple reserve categories. For example, essential workers, such as healthcare professionals or front-line workers, can be eligible to be assigned an ICU bed through the essential-worker category and also through an unreserved category that is open to all patients. In such cases, it is important to specify which category will be used to satisfy the demand.  
This brings us to the fourth step of the minimalistic market design.

\subsection{Designing Reserve Systems for Rationing of Life-Saving Resources}

When  category membership overlaps (i.e., when at least one individual is eligible to receive
units from two or more categories), the three main parameters of the reserve system allow for multiple  ways 
to operationalize a reserve system.  
Handling this subtle aspect of the reserve system and communicating it to policymakers and other stakeholders requires extra care.  
On the one hand, this overlapping nature of category memberships brings an additional degree of freedom in the design of reserve categories, 
allowing for a wider range of policies. 
On the other hand, this aspect of a reserve system may also 
be a potential source of loopholes in rationing guidelines and thus may result in unintended consequences 
(cf. \citealp{dur/kominers/pathak/sonmez:18, pathak/rees-jones/sonmez:20, pathak/rees-jones/sonmez:20b, sonmez/yenmez22}). 
Therefore, before a reserve system can be recommended and utilized for the rationing of scarce pandemic resources, 
it is important to understand the implications of this extra degree of freedom  
to help policymakers make informed decisions on which of these systems better serve their policy objectives. 
That is, the role of the fourth step of the minimalistic market design is to illustrate the
range of possibilities, while still maintaining \textit{informed neutrality\/}  \citep{li2017ethics} between reasonable normative principles.

Starting with its introduction to market design literature with \citet{hafalir2013}, 
virtually all studies on reserve systems have focused on a special setting 
where category-specific priorities are uniquely determined by an underlying baseline priority order together with  category membership.  
Our application on pandemic medical resource allocation, in contrast, calls for a more flexible structure on category-specific priority orders. 
Therefore, to  maintain informed neutrality between reasonable normative principles in 
recommending reserve systems for various settings in  scarce medical resource allocation,  in Section \ref{sec:model} 
we develop a general theory of reserve systems based on 
a richer system  of priority orders.

 \section{A General Theory of Reserve Systems}\label{sec:model}

Reserve systems are common 
in applications where competing interest groups fail to agree on allocation criteria.  
If each individual is eligible to receive units only
from one of these categories, then the theory of priority systems directly extend to this most basic form of reserve systems. 
However, in most applications at least some of the individuals are eligible to receive units from multiple categories, thereby
resulting in multiplicities in implementation of  ``categorized'' priority systems. 
In this section, we develop a general theory of reserve systems, with particular emphasis on
the analytical structure and the distributional implications of this multiplicity.

The earlier literature on reserve systems considers a special setting, where all categories share a common baseline priority order, 
although various protected groups receive preferential treatment or exclusive access in some of the categories. 
In contrast, we develop a theory of reserve systems 
where no cross-category structure is imposed on priority orders.\footnote{Our analyses in Section \ref{sec:baseline} 
and Appendix \ref{sec:smart} are two exceptions to this statement.} 
  
Our model allows priority orders at various categories  
to be completely independent  from each other, although they do not need to be.
Since it may not be possible to capture priorities for all
ethical considerations though adjustments of a single baseline priority order, 
this level of generality is necessary for our main application.\footnote{This more general structure is 
not only needed for our main application of medical resource allocation, but also in other applications. 
For example in Germany, priority for college seats depends on applicant waiting time for (up to) 20\% of the seats, on their performance in 
high school leaving exams for (up to) 20\% of the seats, and on college-specific criteria for the rest of the seats (\citealp{westkamp2013}). } 
To the best of our knowledge, reserve systems have not been studied in this generality before.

\subsection{The Primitives}

Despite the generality of our model,  the terminology is tailored to the main application of pandemic medical resource rationing.
There is a set of \textbf{patients} $I$ and  $q$ identical \textbf{medical units} to allocate.
There is a set of \textbf{reserve categories} $\mathcal{C}$.
For every category $c\in \calc$, $r_c$ units are \textbf{reserved} so that $\sum_{c \in \calc} r_c=q$.
It is
important to emphasize that individual units are not associated with the categories in our model.
The phrase ``$r_c$ units are reserved'' does not mean specific units are set aside for category $c$. Rather, it
means that for the purposes of accounting, a total of unspecified $r_c$ units are attached to category $c$.

For every category $c \in \calc$, there is a linear \textbf{priority order} $\pi_c$ over the set
of patients $I$ and $\emptyset$. This priority order represents the relative claims of the patients on
units in category $c$ as well as their  eligibility for those units.
For every category $c \in \calc$ and patient $i \in I$,
we say that patient $i$ is \textbf{eligible} for category $c$  if
\[ i \mathrel{\pi_c} \emptyset. \]
Given priority order $\pi_c$, we represent its weak order
by $\underline{\pi}_c$. That is,  for any $x,y \in I \cup \{\emptyset\}$,
$$ x  \mathrel{\underline{\pi}_c} y \qquad \iff \qquad x=y \; \; \mbox{ or }\; \; x \mathrel{\pi_c} y .$$
For our main application of pandemic rationing,  $\pi_c$ orders patients based on the
balance of ethical principles guiding the allocation of units in category $c$.

\subsection{The Outcome: A Matching}

A \textbf{matching}
$\mu: I \rightarrow \calc \cup \{\emptyset\}$ is a function
that maps each patient either to a category or to $\emptyset$ such that
$\abs{\mu^{-1}(c)}\leq r_c$ for every category $c\in \calc$.
For any patient $i \in I$, $\mu(i)=\emptyset$ means that the patient
does not receive a unit and $\mu(i) = c \in \calc$ means that the patient receives a unit reserved
for category $c$. Let $\calm$ denote the set of matchings.

Our formulation of the outcome of a reserve system involves 
a subtle but important modeling choice. 
In models where there is a single type of good and each individual has unit demand, 
the traditional way to describe the outcome of a reserve system is through
a \textit{choice rule\/}, which indicates the set of individuals who are awarded a unit  for any
set of applicants. Since each individual is indifferent between all units of the good, this formulation 
is often seen sufficient for analysis. However, for our purposes it is not. 

One of the primary appeals of the reserve system is its ability to facilitate compromises between 
various interest groups by allocating scarce resources through multiple categories.  
While individuals are indifferent between all units of the scarce good, they have potentially different claims on
units from different categories. And consequently merely specifying  who receives a unit may not always
be sufficient. For various analytical exercises, it is also important to specify through which
category individuals receive their units.\footnote{As an illustration, consider affirmative action in India where a percentage of government jobs are reserved for
historically discriminated groups such as Scheduled Castes (SC).  
By  the Supreme Court judgment \textit{Indra Sawhney (1992)\/}, 
units reserved for SC cannot be awarded to candidates from other groups even if there are not sufficiently many applicants from SC.  
Moreover, by the same judgment, any member of SC who is awarded one of these SC-reserve units needs to have lower merit score than 
any one who receives any unreserved unit (cf. \citealp{sonmez/yenmez22}). 
These legal requirements cannot be verified by simply specifying who receives a unit. 
It is also necessary to specify through which category the unit is received.}
Hence, we deviate from the prior formulation and let an outcome embed information about the category.

For any matching $\mu \in \calm$ and any subset of patients $I'\subseteq I$, let $\mu(I')$ denote the
set of patients in $I'$ who are matched with a category under matching $\mu$. More formally,
\[\mu(I')=\big\{i\in I':\mu(i)\in \calc\big \}.\]
These are the patients in $I'$ who are matched (or equivalently who are assigned units) under matching $\mu$.

\subsection{Axioms}

In real-life applications of our model, it is important to allocate the units to qualified individuals without wasting any
and abiding by the priorities governing the allocation of these units.
We next formulate this idea through three axioms:

\begin{definition} A matching  $\mu \in \calm$  \textbf{complies with eligibility requirements} if,
for any $i \in I$ and $c \in \calc$,
\[ \mu(i) = c \quad \implies \quad  i \mathrel{\pi_c} \emptyset.
\]
\end{definition}

Our first axiom formulates the idea that units in any category should be awarded only to eligible individuals.
For most applications of  rationing of life-saving resources, any patient who is eligible for one category can also
be eligible for any category. In those applications, if a patient is ineligible for all categories, then this patient can as well be  dropped
from the set of individuals. Hence, compliance with eligibility requirement vacuously holds
in applications when all individuals are eligible for all categories. 
This includes most applications in pandemic resource allocation, although there can be examples where
certain individuals may be eligible for only some of the categories.

\begin{definition}  A matching  $\mu \in \calm$ is \textbf{non-wasteful} if,
for any  $i\in I$ and $c \in \calc$,
\[ i \mathrel{\pi_c} \emptyset \quad \mbox{ and } \quad \mu(i) = \emptyset \quad \implies \quad \abs{\mu^{-1}(c)} = r_c.\]
\end{definition}

Our second axiom formulates the idea that no unit should go idle as long as there is an eligible individual to award it.
That is, if a unit remains idle, then there should not be any  unmatched individual who is eligible for the unit.
In those applications of  rationing vital resources where each patient is eligible for all units,
non-wastefulness corresponds to either matching all the units or all the patients.

\begin{definition} A matching  $\mu \in \calm$ \textbf{respects priorities} if,
for any $i, i'  \in I$ and $c \in \calc$,
\[ \mu(i) = c \; \mbox{ and } \;  \mu(i') = \emptyset  \quad \implies \quad i \mathrel{\pi_c} i'.
\]
\end{definition}

Our last axiom formulates the idea that for each category, the units should be allocated
based on the priority order of individuals in this category.

As far as we know, in every real-life application of a reserve system, each of these three axioms is either explicitly
or implicitly required. Hence, we see these three axioms as minimal requirements for a reserve system.
In the next two sections we present two characterizations of matchings that satisfy these axioms,
one through a notion akin to competitive equilibria and the second based on the celebrated deferred acceptance algorithm
 of \citet{gale/shapley:62}.

\subsection{Cutoff Equilibria}
\label{sec:char-cutoff}

In many real-life applications of reserve systems, the outcome is often publicized
through a system that identifies the lowest priority individual that qualifies for admission for each category.
This representation makes it straightforward to verify that an allocation was computed
following the announced policy because an individual can compare her priority to the announced cutoffs.
Often a cardinal representation of the priority order, such as a merit score or a lottery number, is used
to identify these ``cutoff'' individuals. 
This observation motivates the following equilibrium notion.

For any category $c \in \calc$,   a {\bf cutoff} $f_c$ is an element of $I \cup \{\emptyset\}$
such that $f_c \mathrel{\underline{\pi}_c} \emptyset$.
Cutoffs in our model play the role of prices for exchange or production economies, where
the last condition corresponds to prices being non-negative in those economies.
We refer to a list of cutoffs $f=(f_c)_{c\in \calc}$ as a {\bf cutoff vector}. Let $\calf$ be the set of cutoff vectors.

Given a cutoff vector $f \in \calf$, for any patient $i\in I$, define  the {\bf budget set} of patient $i$ at cutoff vector $f$ as
$$\calb_i(f)=\left \{c \in \calc \; : \; i \mathrel{\underline{\pi}_{c}} f_{c} \right \}.$$
A {\bf cutoff equilibrium} is a pair consisting of  a cutoff vector and a matching $(f,\mu) \in \calf \times \calm$ such that
\begin{enumerate}
\item For every patient $i \in I$,
\begin{enumerate}
 \item $\mu(i) \in \calb_i(f) \cup \{\emptyset\}$, and
  \item $ \calb_i(f) \not= \emptyset \quad \implies \quad \mu(i)  \in \calb_i(f).$
\end{enumerate}
\item For every category $c \in \calc$, $$|\mu^{-1}(c)|<r_c \quad \implies \quad f_c=\emptyset.$$
\end{enumerate}
A cutoff equilibrium is an analogue of a competitive equilibrium for a reserve system. A cutoff vector-matching pair is a cutoff equilibrium if
\begin{enumerate}
   \item each patient who has a non-empty budget set is matched with a category in her budget set, and each patient who has an empty budget set remains unmatched, and
   \item each category which has not filled its quota under this matching has cutoff $\emptyset$.
 \end{enumerate}
 The first condition corresponds to ``preference maximization within budget set'' and the second condition corresponds to
 the ``market clearing condition.''

Our first characterization result gives an equivalence  between cutoff equilibrium matchings
and matchings that satisfy our basic axioms.

\begin{theorem}
\label{thm:cutoff}
For any matching $\mu \in \calm$ that complies with eligibility requirements, is non-wasteful, and respects priorities,
there exists a cutoff vector
$f \in \calf$  that supports the pair $(f,\mu)$ as a cutoff equilibrium.
Conversely,  for any cutoff equilibrium  $(f,\mu) \in \calf \times \calm$,
matching $\mu$ complies with eligibility requirements, is non-wasteful, and respects priorities.
\end{theorem}

There can be multiple equilibrium cutoff vectors that
support a matching at cutoff equilibria. Next, we explore the structure of equilibrium cutoff vectors.

For any matching $\mu \in \calm$ and category $c\in \calc$, define
\begin{align}
\ovf^\mu_c = \left\{ \begin{array}{ll}
\min_{\pi_c} \mu^{-1}(c)  & \mbox{if } |\mu^{-1}(c)|=r_c  \\
\emptyset  & \mbox{otherwise}
 \end{array}\right. \qquad \mbox{and, }  \label{eq:max-cutoff}
 \end{align}
\begin{align}
\unf^\mu_c = \left\{ \begin{array}{ll}
\min_{\pi_c} \Big \{ i \in \mu(I) : i \mathrel{\pi_c} \max_{\pi_c} \big(I \setminus \mu(I)\big) \cup \{\emptyset\}  \Big \} & \mbox{if } \max_{\pi_c} \big(I \setminus \mu(I)\big) \cup \{\emptyset\}  \not=\emptyset  \\
\emptyset  & \mbox{otherwise}
 \end{array}\right. . \label{eq:min-cutoff}
 \end{align}
 Here,
 \begin{itemize}
\item  $\ovf^\mu_c$ identifies
\begin{itemize}
\item the lowest $\pi_c$-priority patient who is matched with category $c$ if all category-$c$ units
 are exhausted under $\mu$,  and
 \item $\emptyset$ if there are some idle category-$c$  units under $\mu$,
\end{itemize}
\end{itemize}
\noindent whereas
\begin{itemize}
\item  $\unf^\mu_c$ identifies
\begin{itemize}
\item the lowest $\pi_c$-priority patient with the property that every weakly higher $\pi_c$-priority patient
than her is matched under $\mu$
if some category-$c$ eligible patient is unmatched under $\mu$, and
\item $\emptyset$ if all category-$c$ eligible patients are matched under $\mu$.
\end{itemize}
\end{itemize}
Let $\mu$ be any matching that respect priorities.  By construction,
\[ \ovf^\mu_c \, \mathrel{\underline{\pi}_c}  \, \unf^\mu_c \qquad \mbox{ for any } c\in\calc.
\]
Our next result characterizes the set of cutoff vectors.

\begin{lemma} \label{prop:eq-cutoff-vector-structure}
Let  $\mu \in \calm$ be a matching that complies with eligibility requirements, is non-wasteful, and respects priorities.
Then the pair $(g,\mu)$ is a cutoff equilibrium if, and only if,
\[ \ovf^\mu_c   \, \mathrel{\underline{\pi}_c} \, g_c \, \mathrel{\underline{\pi}_c} \,  \unf^\mu_c \qquad \mbox{ for any } \; c\in\calc.
\]
\end{lemma}

An immediate corollary to Lemma \ref{prop:eq-cutoff-vector-structure} is that for each cutoff equilibrium matching $\mu$,
$\ovf^\mu = (\ovf^\mu_c)_{c\in \calc}$ is a \textbf{maximum equilibrium cutoff  vector}
and $\unf^\mu = (\unf^\mu_c)_{c\in \calc}$ is a \textbf{minimum equilibrium cutoff  vector}.

Of these equilibrium cutoff vectors, the first one has a clear  economic interpretation.
The maximum equilibrium cutoff of a category  indicates the \textit{selectivity\/} of this particular category.
The higher the maximum cutoff is the more competitive it becomes to receive a unit through this category.
This is also the cutoff that is typically announced in real-life applications of reserve systems due to its clear interpretation.
The interpretation of the minimum equilibrium cutoff of a category is more about the entire matching  than the category itself,
and in some sense it is artificially lower than the maximum equilibrium cutoff due to individuals who are matched with other categories.
All other equilibrium cutoffs between the two are also artificial in a similar sense.
Therefore, for much of our analysis, we focus on the maximum equilibrium cutoff vector.

\subsection{Characterization via Deferred Acceptance Algorithm}\label{sec:char}

Although Theorem \ref{thm:cutoff} gives a full characterization of matchings that satisfy our three axioms,
it leaves open the question of how to find such a matching. In this section, we present  a procedure to
construct all such matchings utilizing the celebrated  deferred-acceptance algorithm by \citet{gale/shapley:62}.

Consider the following hypothetical many-to-one matching market.
The two sides of the market are the set of patients $I$ and the set of categories $\calc$.
Each patient $i \in I$ can be matched with at most one category, whereas each category $c \in \calc$ can be
matched with as many as $r_c$ patients.
Category $c$ is endowed with the linear order $\pi_c$ that is specified in the primitives of the original rationing problem.

Observe that in our hypothetical market, all the primitives introduced so far  naturally follows
from the primitives of the original problem.
The only primitive of the hypothetical market  that is somewhat ``artificial'' is the next one:

Each patient $i \in I$ has a strict preference relation $\succ_i$ over the set $\calc \cup \{\emptyset\}$, such that,
for each patient $i \in I$,
\[ c \; \succ_i \; \emptyset   \quad \iff \quad
\mbox{patient $i$ is eligible for category $c$}.
\]
While in the original problem a patient is indifferent between  all units (and therefore all categories as well),
in the hypothetical market she has strict preferences between the categories.
This ``flexibility'' in the construction of the hypothetical market is the basis of our main characterization.

For each patient $i \in I$, let $\calp_i$ be the  set of all preferences constructed in this way, and
let $\calp = \times_{i \in I} \calp_i$.

Given a preference profile $\succ\, =(\succ_i)_{i \in I}$,
the individual-proposing deferred acceptance algorithm  (DA)  produces a matching as follows.

\medskip

	\begin{quote}
        \noindent{}{\bf Individual Proposing Deferred Acceptance Algorithm ($\mathbf{DA}$)}

		\noindent{}{\bf Step 1:}
            Each patient in $I$ applies to her most preferred category among categories for which she is eligible.
            Suppose that $I_c^1$ is the set of patients who apply to
            category $c$. Category $c$ tentatively assigns
            applicants with the highest priority according to $\pi_c$ until
            all patients in $I_c^1$ are chosen or all $r_{c}$ units
            are allocated, whichever comes first, and permanently rejects the rest. If there are no rejections, then stop.
	
		\noindent{}{\bf Step k:}
			Each patient who was rejected in Step k-1 applies to her next
			preferred category among categories for which she is eligible,
			if such a category exists. Suppose
			that $I_c^k$ is the union of the set of patients who were tentatively
            assigned to category $c$ in Step k-1 and the set of patients who
            just applied to category $c$. Category $c$ tentatively assigns patients
            in $I_c^k$ with the highest priority according to $\pi_c$ until all
            patients in $I_c^k$ are chosen or all $r_{c}$ units are allocated, whichever
            comes first, and permanently rejects the rest. If there are no rejections,
            then stop.\smallskip

            The algorithm terminates when there are no rejections, at which point all tentative
            assignments are finalized.
	\end{quote}

A matching $\mu \in \calm$  is called \textbf{DA-induced} if it is  the outcome of
DA for some preference profile $\succ \, \in \calp$.

We are ready to present our next  result:

\begin{proposition}\label{thm:characterization}
A matching complies with eligibility requirements, is non-wasteful, and respects priorities
if, and only if, it is DA-induced.
\end{proposition}

Not only is this result a second characterization of matchings that satisfy our three basic axioms,
it also provides a concrete procedure to calculate all such matchings.
Equivalently, Proposition \ref{thm:characterization} provides us with a procedure to derive all cutoff equilibria.
This latter interpretation of our characterization leads us to a refinement of cutoff equilibrium matchings explored in our next section.

\subsection{Sequential Reserve Matching} \label{sec:sequential}

An interpretation of the DA-induced matchings is helpful to motivate in focusing a subset of these matchings.
Recall that the hypothetical two-sided matching market constructed above relies on an
artificial preference profile $(\succ_i)_{i \in I}$ of patients over categories.
What this corresponds to under DA is that any patient $i$ is considered for
categories that deem her eligible in sequence, following the ranking of these categories under her artificial preferences $\succ_i$.
Unless there is a systematic way to construct these preferences, it may be difficult to motivate
adopting this methodology for real-life applications.
For example, if a patient $i$ is considered first for an unreserved category and then for an essential personnel category,
whereas another patient $j$ with similar characteristics is considered for them in the reverse order,
it may be difficult to justify this practice.
That is, a potentially large set of matchings
satisfy our three axioms, but not all are necessarily obtained through an intuitive procedure.
This may be a challenge especially in the context of medical rationing, since \textit{procedural fairness\/}
is also an important ethical consideration in this context.
Procedural fairness is the main motivation for our focus in a subset of these matchings.

In many real-life applications of reserve systems, institutions process reserve categories sequentially
and allocate units associated with each category one at a time using its category-specific priority order.
We next formulate matchings obtained in this way and relate them to our characterization in Proposition \ref{thm:characterization}.

An \textbf{order of precedence} $\triangleright$ is a linear order over the set of categories $\calc$.
For any two categories $c,c' \in \calc$,
$$ c \mathrel{\triangleright} c' $$
means that category-$c$ units are to be allocated before category-$c'$ units.
In this case, we say category $c$ \textbf{has higher precedence} than category $c'$.
Let $\Delta$ be the set of all orders of precedence.

For a given order of precedence $\triangleright \in \Delta$, the induced \textbf{sequential reserve matching}
$\varphi_\triangleright$, is a matching that is constructed as follows:

 \begin{quote}

Suppose categories are ordered under $\triangleright$ as \[c_1 \mathrel{\triangleright} c_2 \mathrel{\triangleright} \hdots \mathrel{\triangleright} c_{\abs{\calc}}.\]
Matching $\varphi_\triangleright$ is found sequentially in $\abs{\calc}$ steps:

\noindent{}\textbf{Step 1:} Following their priority order under $\pi_{c_1}$,
 the highest priority $r_{c_1}$ category-$c_1$-eligible patients in $I$ are
matched with category $c_1$.
If there are fewer than $r_{c_1}$ eligible patients in $I$ than all of these eligible patients are matched with category $c_1$.
Let $I^1$ be the set of patients matched in Step 1.

\noindent{}\textbf{Step k:}  Following their priority order under $\pi_{c_k}$,
the highest priority $r_{c_k}$ category-$c_k$-eligible patients in $I \setminus \cup_{k'=1}^{k-1} I^{k'}$
are matched with category $c_k$.  If there are less than $r_{c_k}$  eligible patients in $I \setminus \cup_{k'=1}^{k-1} I^{k'}$
then all of these eligible patients are matched with category $c_k$. Let $I^k$ be the set of patients matched in Step k.
 \end{quote}

Given an order of precedence $\triangleright \in \Delta$, the induced
sequential reserve matching complies with eligibility requirements, is non-wasteful, and it respect priorities.
Thus, it is DA-induced by Proposition \ref{thm:characterization}.
Indeed  it corresponds to a specific DA-induced matching.

\begin{proposition}\label{prop:seq.equiv}
Fix an order of precedence  $\triangleright \in \Delta$. Let
the preference profile $\succ^{\triangleright} \in \calp$ be such that,
for each patient $i \in I$ and pair of categories $c, c' \in \calc$,
\[  c \; \succ^{\triangleright}_i \; c' \quad \iff \quad   c \; \mathrel{\triangleright} \; c'.
\]
Then the sequential reserve matching $\varphi_\triangleright$ is DA-induced from
the preference profile $\succ^{\triangleright}$.
\end{proposition}

We conclude this section with a comparative static result regarding the maximum equilibrium cutoff vectors
supporting sequential reserve matchings:

\begin{proposition} \label{prop:seq-reserve-comp-stat-cutoffs}
 Fix two distinct categories $c,c' \in \calc$ and a pair of orders of precedence   $\triangleright,\triangleright' \in \Delta$ such that:
\begin{itemize}
\item $c' \; \triangleright \; c$,
\item $c \; \triangleright' \; c'$, and
\item for any $\hat{c} \in \calc$ and $c^* \in \calc\setminus \{c, c'\}$
\[ \hat{c} \; \triangleright \; c^* \quad \iff \quad \hat{c} \; \triangleright' \; c^*. \]
\end{itemize}
That is, $\triangleright'$ is obtained from $\triangleright$ by only changing the order of $c$ with its immediate predecessor $c'$.
Then,
\[ \ovf^{\vp}_c \mathrel{\underline{\pi}_c} \,  \ovf^{\v}_c. \]
\end{proposition}

Recall that the maximum equilibrium cutoff for a category is indicative of how selective the category is.
Therefore,
the earlier a category is processed under a sequential reserve matching the more selective it becomes
by Proposition \ref{prop:seq-reserve-comp-stat-cutoffs}.
This result is intuitive because the earlier a category is processed, the larger is the set of patients who compete
for these units in a set-wise inclusion sense.

\subsection{Reserve Systems under a Baseline Priority Order} \label{sec:baseline}

In many real-life applications of reserve systems,
there is a baseline priority order  $\pi$ of individuals.
Starting with \citet{hafalir2013}, the earlier market design literature on reserve
systems exclusively considered this  environment.
This priority order may depend on scores in a standardized exam, a random lottery, or arrival time of application.
For ventilator, ICU, antiviral or monoclonal antibody allocation, it may depend on SOFA scores described in Section \ref{sec:priority}.
This baseline priority order is used to construct the priority order for each of the reserve categories,
although each category except one gives preferential treatment to a  specific subset of individuals.
For example, in our main application these could be essential personnel or persons from disadvantaged communities.
In this section, we focus on this subclass of reserve systems and present an analysis of reserve matching on this class.
In contrast to Section \ref{sec:model} where our analysis is more general than earlier literature 
and contributions are more conceptual than technical, 
analysis in this section represents our deeper technical contributions in relation to earlier literature.

To formulate this subclass of reserve systems,  
we designate a \textbf{beneficiary group} $I_c \subseteq I$ for each category $c \in \calc$.
It is assumed that  all patients in its beneficiary group are
eligible for a category. That is, for any $c \in \calc$ and $i\in I_c$,
\[ i \; \pi_c \; \emptyset. \]

There is an all-inclusive category $u \in \calc$, called the  \textbf{unreserved} (or \textbf{open}) category,
which has  all patients as its set of beneficiaries and endowed with the baseline priority order.
That is,
\[ I_u= I  \quad \mbox{ and  } \quad \pi_u = \pi. \]

Any other category $c \in \calc \setminus \{u\}$,  referred to as a \textbf{preferential treatment} category,
has a more exclusive  set $I_c \subsetneq I$ of beneficiaries
and it is endowed with a priority order $\pi_c$ with the following structure: for any pair of patients $i, i' \in I$,
\begin{align*}
 i\in I_c\quad \mbox{ and } \quad  i' \in I \setminus I_c\quad  & \implies \quad i \mathrel{\pi_c} i', \\
i,i' \in I_c  \quad \mbox { and } \quad i \mathrel{\pi}  i' \quad & \implies \quad i \mathrel{\pi_c} i', \mbox{ and }\\
i,i' \in I \setminus I_c \quad \mbox { and } \quad i \mathrel{\pi}  i' \quad & \implies \quad i \mathrel{\pi_c} i'.
\end{align*}
Under $\pi_c$, beneficiaries of category $c$ are prioritized over patients who are not,
but otherwise their relative priority order is induced by the baseline priority order $\pi$.

Let $I_g$,  referred to as the set of \textbf{general-community patients},
be the set of patients who are each a beneficiary of the  unreserved category only:
\[I_g = I \setminus \cup_{c\in \calc \setminus \{u\}} I_c.\]

In particular, two types of such problems have widespread applications.

We say that a priority profile $(\pi_{c})_{c\in \calc}$ has \textbf{soft reserves} if, for any category $c \in \calc$ and  any patient $i \in I$,
\[ i \mathrel{\pi_{c}} \emptyset. \]
Under a soft reserve system all individuals are eligible for all categories.
This is the case, for example, in our main application of pandemic resource allocation.

We say that a priority profile $(\pi_c)_{c\in \calc}$ has \textbf{hard reserves} if, for any preferential treatment category
$c \in \calc \setminus \{u\}$,
\begin{enumerate}
\item $ i \; \mathrel{\pi_c} \; \emptyset$ for any of its beneficiaries  $i \in I_{c}$, whereas
\item $ \emptyset \; \mathrel{\pi_{c}} \; i$ for any patient  $i \in I \setminus I_{c}$ who is not a beneficiary.
\end{enumerate}
Under a hard reserve system,  only the beneficiaries of a preferred treatment category are eligible for
units in this category. This is the case, for example, in H-1B visa allocation in the US.

Allocation rules based on sequential reserve matching are used in many practical applications.
While an aspect often ignored in practical applications,
it is important to pay attention to the choice of the order of precedence in these problems,
for it has potentially significant distributional implications.
In this subsection, we focus on sequential reserve matching under soft reserves,
as this case likely is the relevant case for our main application of pandemic rationing.\footnote{The case of hard reserves is 
much simpler to analyze and offers more general results. See Theorem 2 in  \citet{pathak/sonmez/unver/yenmez:20a}.}

We already know from Proposition \ref{prop:seq-reserve-comp-stat-cutoffs} that the later a category is processed,
the less competitive it becomes. A natural follow-up question is whether this also means that the beneficiaries of
this category necessarily benefits from  this comparative static exercise.
The answer would be straightforward, if each patient was a beneficiary of a single category.
But this is not the case in our model, because even if each patient is a beneficiary of at most one preferential treatment category,
they are also each a beneficiary of the unreserved category. Indeed, even if that was not the case, unless the reserves are hard
non-beneficiaries may still be matched with units from preferential treatment categories.
So the answer to this question is not an immediate implication of Proposition \ref{prop:seq-reserve-comp-stat-cutoffs}.
Under some assumptions such as
when there is only one preferential treatment category  \citep{dur/kominers/pathak/sonmez:18},
this question is already answered in the affirmative.
However, as we present in the next example, this is not always the case.

\begin{example} \label{ex:cat6} Suppose there are $q=6$ medical units to be allocated in total. There are six categories: the unreserved category $u$ and five preferential treatment categories $c,c',c^*,\hat{c},\tilde{c}$  and each category has a single unit capacity.

Suppose there are seven patients. All patients are beneficiaries of the unreserved category $u$:
$$
I_u=\{i_1,i_2,i_3,i_4,i_5,i_6,i_7\}.
$$
The beneficiaries of preferential treatment categories $c$, $c^*$, and $\tilde{c}$ are given as  $$I_{c}=\{i_1,i_3,i_6\}, \quad I_{c^*}=\{i_2,i_5\}, \quad I_{\tilde{c}}=\{i_4,i_7\},$$
while there are no beneficiaries of preferential treatment categories $c'$ and $\hat{c}$: $I_{c'}=\emptyset$ and $I_{\hat{c}}=\emptyset$. There are also no general-community patients: $I_g=\emptyset$.  Suppose $\pi$, the baseline priority order of patients, is given as $$i_1 \; \mathrel{\pi} \; i_2 \; \mathrel{\pi} \; i_3 \; \mathrel{\pi} \; i_4 \; \mathrel{\pi} \;  i_5 \; \mathrel{\pi} \; i_6 \mathrel{\pi} \; i_7.$$ Also assume that all patients are eligible for all preferential treatment categories besides the unreserved category $u$.

We consider two sequential reserve matchings based on the following two orders of precedence:
$$ \tr :  \qquad c' \; \mathrel{\tr} \; c \; \mathrel{\tr} \; c^* \; \mathrel{\tr} \; \hat{c} \; \mathrel{\tr} \; \tilde{c} \; \mathrel{\tr} \; u,$$
and
$$ \trp : \qquad  c \; \mathrel{\trp} c' \; \mathrel{\trp} \; c^* \; \mathrel{\trp} \;  \hat{c} \; \mathrel{\trp} \; \tilde{c} \; \mathrel{\trp} \; u. $$

In the following table, we demonstrate the construction of the two induced  sequential reserve matchings by processing their mechanics
in parallel:

\begin{center}
\begin{tabular}{c|cc|ccc}
 & \multicolumn{2}{c|}{Order of Precedence $\tr$} & \multicolumn{2}{c}{Order of Precedence $\trp$} \\ \hline
  Step & Category & Patient &  Category & Patient \\ \hline
$1$ & $c'$ & $i_1$ & $c$ & $i_1$ \\
$2$ & $c$ & $i_3$ & $c'$ & $i_2$ \\
$3$ & $c^*$ & $i_2$ & $c^*$ & $i_5$ \\
$4$ & $\hat{c}$ & $i_4$ & $\hat{c}$ & $i_3$ \\
$5$ & $\tilde{c}$ & $i_7$ & $\tilde{c}$ & $i_4$ \\
$6$ & $u$ & $i_5$ & $u$ & $i_6$
\end{tabular}
\end{center}
Thus the two sequential reserve matchings match the patients
$$\v=\{i_1,i_2,i_3,i_4,i_5,i_7\} \qquad \mbox{and} \qquad \vp=\{i_1,i_2,i_3,i_4,i_5,i_6\}.$$
In this problem category-$c'$ and $\hat{c}$ units are treated as if  they are of unreserved category $u$, as these two categories do not have any beneficiaries in the problem. We use the baseline priority order $\pi$ to match them.

Under the first order of precedence $\tr$, the highest $\pi$-priority patient $i_1$, who is also a category-$c$ beneficiary, receives the first unit, which is reserved for category $c'$. As a result, $i_3$, who is the next category-$c$ beneficiary, receives the only category-$c$ unit.  In the end, units associated with categories $c^*$ and $\tilde{c}$ are matched with their highest and lowest priority beneficiaries $i_2$ and $i_7$, respectively. The highest priority beneficiary of $\tilde{c}$, patient $i_4$, receives the category-$\hat{c}$ unit, which is processed like the unreserved category and before $\tilde{c}$.  Hence, the lowest priority beneficiary of category $c$, $i_6$ remains unmatched as the last unit, which is reserved for the unreserved category, goes to $i_5$. Thus,
$$\v(I_c)=\{i_1,i_3\}$$
is the set of matched category-$c$ beneficiaries.

Under the second order of precedence $\trp$ that switches the order of $c$ and $c'$, the selectivity of category $c$ increases as it is processed earlier: the highest priority category-$c$ patient $i_1$ receives its unit instead of $i_3$. This leads to the units associated with categories $c^*$ and $\tilde{c}$ being matched with their lowest and highest priority beneficiaries $i_5$ and $i_4$, respectively -- this is a switch of roles for these categories with respect to $\tr$. This is because the highest-priority beneficiary of $c^*$, patient $i_2$, is now matched with category $c'$, which is processed like the unreserved category before $c^*$. This enables the lowest priority beneficiary of category $c$, patient $i_6$, to be matched with the unreserved category as she is prioritized higher than $i_7$ under the baseline priority order. Hence,
 $$\vp(I_c)=\{i_1,i_3,i_6\}$$ is the set of matched category-$c$ beneficiaries.

 Thus, $$\v(I_c) \subsetneq \vp(I_c)$$ although category $c$ is processed earlier under $\trp$ than under $\tr$.
 \end{example}\medskip

Example \ref{ex:cat6} shows that earlier positive comparative static results in the literature fail to extend in
our more general model.  
Also  observe that
our negative example holds even though each patient is a beneficiary of at most one preferential treatment category.
Nevertheless, a positive result holds  provided that there are at most five categories
and each patient is a beneficiary of no more than one preferential treatment category.

\begin{proposition} \label{prop:comp-stat-soft-5}
Assuming (i) there are at most five categories, and (ii) each patient is a beneficiary of at most one preferential treatment category,
consider a soft reserve system induced by a baseline priority order.
Fix a preferential treatment category $c \in \calc \setminus \{u\}$,
another category $c' \in \calc \setminus \{c\}$, and a pair of orders of precedence   $\triangleright,\triangleright' \in \Delta$ such that:
\begin{itemize}
\item $c' \; \triangleright \; c$,
\item $c \; \triangleright' \; c'$, and
\item for any $\hat{c} \in \calc$ and $c^* \in \calc\setminus \{c, c'\}$,
\[ \hat{c} \; \triangleright \; c^* \quad \iff \quad \hat{c} \; \triangleright' \; c^*. \]
\end{itemize}
That is, $\triangleright'$ is obtained from $\triangleright$ by only changing the order of $c$ with its immediate predecessor $c'$.
Then,
\[ \varphi_{\triangleright'}(I_c) \subseteq  \varphi_{\triangleright}(I_c). \]
\end{proposition} \bigskip

Sequential reserve systems play a  central role in practical applications including in 
applications for pandemic medical resource allocation reported in Section \ref{field}. 
Nevertheless, even in the basic version of the  problem considered in this section, 
sequential reserve systems may have some limitations when beneficiaries of preferential categories overlap (cf. \citealp{sonmez/yenmez22}). 
For instance,  they may be wasteful  for the case of  hard reserves as  Example \ref{ex:non-wastefulness-problem} in Appendix \ref{sec:smart} shows. 
While soft reserves are preferable for allocation of most life-saving resources,  in Appendix \ref{sec:smart}
we further introduce a theory of non-wasteful reserve systems for this version of the problem. 
Rather than processing the preferential treatment categories sequentially, 
processing them in parallel in a \textit{smart\/} way turns out to be the key aspect of these systems.

\subsection{Related Theoretical Literature} \label{sec:rel-lit}

Our formal analysis of reserve systems contributes to literature in matching market design focused on distributional issues.

For the special setting with baseline priorities we also consider in Section \ref{sec:baseline}, 
\citet{hafalir2013} introduce reserve systems to the matching market design literature. 
In addition to an open category where the priority order is the same as the baseline priority order, 
in their analysis there is  a single preferential treatment category. 
 \citet{hafalir2013} further assume that the preferential treatment category is processed prior to the open category. 
 This version of a reserve system is  called the \textit{minimum guarantee\/} policy, whereas 
 the version  where the open category is processed prior to the preferential treatment category, later introduced in 
 \citet{dur/kominers/pathak/sonmez:18}, is called the \textit{over-and-above} policy. 
 
The concept of \textit{order of precedence\/} used in sequential reserve systems 
is formally introduced by  \citet{kominersandsonmez2016}, albeit in a more general form for individual units (or slots).  
For the basic version of the model considered in  \citet{hafalir2013}, 
\citet{dur/kominers/pathak/sonmez:18} present the following comparative statics result:
The later a  preferential treatment unit is processed,  the weakly better it is
for its beneficiaries. A corollary of this result is that, the later the preferential category is processed under a sequential reserve system, the
weakly better it is  its beneficiaries. 
Proposition \ref{prop:comp-stat-soft-5} shows that this result extends for as long as there are at most four preferential treatment categories
(and thus five categories in total). 
However, by Example \ref{ex:cat6}, the result fails when there are five or more  preferential treatment categories. 

Unlike Proposition \ref{prop:comp-stat-soft-5} and our results on smart reserve systems in Appendix \ref{sec:smart}, 
Theorem \ref{thm:cutoff} and Propositions   \ref{thm:characterization}-\ref{prop:seq-reserve-comp-stat-cutoffs}
assume no cross-category structure on priority orders. As such, these results have no direct antecedents in the literature.

While Theorem \ref{thm:cutoff} is novel and our paper is the first one to formally introduce the notion of cutoff equilibrium for reserve systems,
the use of this notion is widespread in real-life applications of reserve systems.
In particular, the outcomes of reserve systems are often announced together with the cutoffs that support them.
Examples include admission to exam high schools in Chicago  \citep{dur/pathak/sonmez:20}, assignment of government positions in
India \citep{sonmez/yenmez22},  college admissions in Brazil \citep{aybo16},
and H-1B visa allocation in the US for years 2006-2008 \citep{pathak/rees-jones/sonmez:20}. For the first three of these applications the cutoffs are given in terms of
exam or merit scores, whereas for the last application the cutoffs are given in terms of the date of visa application receipt.
While the concept of cutoff equilibrium for reserve systems is novel to our paper, cutoffs are used
in simpler matching environments in the absence of distributional considerations (see, e.g., \citealp{balinski/sonmez:99} and
\citealp{azevedo/leshno:14}).

In addition to presenting a characterization of outcomes that satisfy three basic axioms,
Proposition \ref{thm:characterization} also  provides a procedure to calculate all cutoff matchings. While our characterization itself is novel,
the use of the deferred acceptance algorithm to derive a specific cutoff matching is not.
This technique to obtain cutoff levels for reserve systems has been employed in various real-life applications, including in
school choice algorithms of Boston Public Schools \citep{dur/kominers/pathak/sonmez:18} and
Chile \citep{correa:19}, and college admissions algorithm for Engineering Colleges in India \citep{baswana:18}.

Our characterizations in Theorem  \ref{thm:cutoff} and Proposition \ref{thm:characterization} use three simple axioms.
As such, the resulting matchings can fail to be Pareto efficient in some applications.
This failure has to do with rather mechanical and inflexible assignment of agents to categories as presented by
Example \ref{ex:non-wastefulness-problem} in Appendix \ref{sec:smart}. This can be mitigated by filling reserves in a ``smart'' way.
The class of  smart reserve matching algorithms in Appendix \ref{sec:smart} does precisely this
and through maximal utilization of reserves always generates a  Pareto efficient matching.
The smart reserve matching algorithm is a generalization of the meritorious horizontal choice rule by \citet{sonmez/yenmez22}, 
who introduced the idea under exclusively ``minimum guarantee'' type reserves.

In addition to above-discussed papers which directly relate to our analysis,
several  others  examine allocation under various  forms of distributional constraints such
as upper quotas and regional quotas.
Some notable papers along these lines include \citet{abdulkadiroglu2005}, \citet{biro2010}, \citet{kojima2012},
\citet{ehlers2014}, \citet{echeniqueandyenmez2015}, \citet{kamadaandkojima2015, kamadaandkojima2017, kamadaandkojima2018},  
\citet{bo2016}, \citet{dogan2016, dogan:17}, \citet{goto/kojima/kurata/tamura/yokoo:17},  \citet{fragiadakisandtroyan2017}, \citet{combe:18}, \citet{kojima/tamura/yokoo:18}, \citet{tomoeda:18},
\citet{ehlers/morrill:19}, and \citet{imamura:20}.

Finally, our paper also introduces medical rationing into the market design literature. 
Follow-up research on this application of market design 
include \citet{aziz/brandl:21}, \citet{delacretaz:21}, \citet{dur/morrill/phan:21}, \citet{grigoryan:21} and \citet{Akbarpour/Budish/Dworczak/Kominers:23}.

\section{Policy Impact}\label{field}

A few days after the initial draft of this paper 
was circulated, the Boston Globe
reported that revised guidelines for ventilator rationing in Massachusetts give
preferences to medical personnel \citep{kowalczyk:20}.  Given the prior debates in Minnesota
and New York, we were curious to see how this priority was implemented.
The new April 2020 Massachusetts state guidelines made \textit{no} preference for medical personnel,
and were a version of White's priority point system \citep{mass:20}.  The Boston
Globe picked up on ambiguous language about a possible preference for medical personnel
in this document, despite the fact that the final document did not include preference
for medical personnel. 

Upon seeing this article, our research team contacted Professor Dr. Robert Truog, who was mentioned in the Boston
Globe article as a member of the working group and was one of the scholars who had called for new
rationing guidelines in \citet{daley:20}, which inspired our initial work on this topic.  
Robert Truog,  who is the Director of the Center for Bioethics at Harvard Medical School, convened a meeting and also invited  legal scholar
Professor Govind Persad and Professor Dr. Douglas White.  
Govind Persad is a faculty member at the University of Denver Sturm College of Law, and he
is the second author of \citet{nejm:20}, the main paper that inspired our current work.
Douglas White is a
Critical Care Professor and the Director of the Program on Ethics and Decision Making in Critical Illness at the University of Pittsburgh Medical Center (UPMC), 
and he has co-developed the multi-principle priority point system (cf. \citealp{white:09}), which is
the main allocation procedure proposed in the bioethics literature using the multiple ethical principle framework.   
After the meeting, we teamed up with these three experts
to advocate for the reserve system to broader bioethics and medical communities 
\citep{sonmez-chest:21}. Our new collaborators encouraged us to use the alternative name of ``categorized priority system''  to emphasize
the close connection between our proposed reserve system and the traditional priority system.  
To the best of our knowledge, a reserve system has neither been proposed nor implemented in the field for 
pandemic rationing of life-saving resources prior to the Covid-19 pandemic.

Around the same time we also started collaborating with University of Pennsylvania bioethicist Professor Harald Schmidt and 
Massachusetts General Hospital Clinical  Care Physician Dr. Emily Rubin. 
Schmidt was interested in utilizing the reserve system to mitigate disparities in healthcare access  \citep{Schmidt/Pathak/Sonmez/Unver:2020}. 
Schmidt became a major proponent of the reserve system, and as we discuss in Section \ref{sec:NASEM}, he 
played a key role in the adoption of the reserve system in several states during the Covid-19 vaccine rollout which started in December 2020. 
Rubin  was a member of the working group for determining state rationing guidelines in Massachusetts. 
As we discuss in Section \ref{sec:MA-theraupetics}, we later worked with her in designing and deploying a reserve system 
for allocation of monoclonal antibody therapies in the state \citep{rubin/etal:chest2021}.  

Following our interaction with various interdisciplinary groups and public health officials during the Covid-19 pandemic, 
a reserve system has  been recommended or deployed in several  jurisdictions. 
We outline various design considerations for different life-saving resources that we learnt and developed through these interactions in Appendix \ref{app:design}.  
In all these collaborations, the close connection between the reserve system and the traditional  priority system 
made it possible for our team to communicate the strengths  of our proposal to various stakeholders in a simple and transparent way. 
Many jurisdictions  were caught unprepared for the challenges posed by the Covid-19 pandemic. 
The flexibility offered with the policy levers in reserve system, i.e. the selection of the reserve categories and  the allocation criteria along with the 
number of units in these categories, made it easier to reach compromises in several committees and task forces. 

We next report various outreach activities we conducted during Covid-19 and 
policy impact of our efforts  in several jurisdictions due to these activities.

\subsection{Antiviral Therapy Allocation in Pennsylvania}

Gilead's Remdesivir was one of the first antiviral therapies granted emergency use authorization
by the US Food and Drug Administration.  Following this approval in May 2020, UPMC devised a new procedure to allot its allotment of Remdesivir, obtained
from the federal government.   Douglas White led the effort to construct reserve categories through community 
engagement, and three reserve categories were established based on whether a patient
is from a hardest hit region (based on the Area Deprivation Index), whether the patient
is an essential worker (using the Pennsylvania state definition), and whether the patient is 
extended to die within one-year (based on various clinical indicators).  
The system was implemented as a weighted lottery given projections on total demand, and more than 300 patients went
through the procedure.  Our team
provided software to implement the rationing system.  \citet{whiteCCM:22} provides additional details.

The UPMC system was held up as a model policy for the allocation of scarce anti-viral medications.
The Pennsylvania Commonwealth later recommended it statewide \citep{Pennsylvania:2020}.

\subsection{NASEM Framework for Equitable Vaccine Allocation} \label{sec:NASEM}

By late summer of 2020, the focus shifted from ventilator and antiviral therapy rationing to the upcoming
vaccine rollout.  Starting with the beginning of the pandemic, there was an intense debate about equitable vaccine allocation. 
 
In July 2020, the U.S. CDC and NIH commissioned the National Academies of Sciences, Engineering, 
and Medicine (NASEM) to formulate  recommendations on the equitable allocation of Covid-19 vaccines.
In September 2020,  NASEM's committee of distinguished experts solicited public  comments on a discussion draft 
of their Framework for Equitable Allocation of Covid-19 Vaccine  \citep{nasem:20}.

In written and oral comments, our collaborator Harald Schmidt inquired about the mechanism to prioritize members of hard-hit communities.  
In preparation for this contingency and in collaboration with Schmidt, weeks earlier we circulated a white paper 
illustrating how easily a traditional tiered priority system can be ``modified'' as a reserve system to include equity as a goal though 
an index of social vulnerability \citep{pathak/etal:20}. 
This precise formulation through minimalist market design  
was brought to the attention of the committee as a possible mechanism to embed equity in their framework.
 Around the same time, \citet{persad/peek/emanuel:20} endorsed our proposed reserve system for vaccine rollout in a \textit{JAMA\/} viewpoint. 
 Following the terminology adopted in \citet{sonmez-chest:21}, they described a reserve system as a ``categorized priority system,'' writing:
\bigskip

\begin{minipage}{0,9\textwidth}
\textsf{``Dividing the initial vaccine allotment into priority access categories and using medical criteria to prioritize within each category is a promising approach. For instance, half of the initial allotment might be prioritized for frontline health workers, a quarter for people working or living in high-risk settings, and the remainder for others. Within each category, preference could be given to people with high-risk medical conditions. Such a categorized approach would be preferable to the tiered ordering previously used for influenza vaccines, because it ensures that multiple priority groups will have initial access to vaccines.''}
\end{minipage}

\bigskip

In October 2020, NASEM published their final Framework \citep{nasem:20b}.  Following the recommendation in \citet{persad/peek/emanuel:20}  and
using the same formulation in \citet{pathak/etal:20}, the NASEM Framework 
recommended a 10 percent over-and-above reserve for hard-hit areas, where
hard-hit is measured by the CDC's Social Vulnerability Index (SVI).  Their stated
justification emphasized the flexibility of a reserve system:

\bigskip

\begin{minipage}[c]{0,9\textwidth}
\textsf{``The committee does not propose an approach in which, within each phase, all vaccine is first given to people in high SVI areas. Rather the committee proposes that the SVI be used in two ways. First as previously noted, a reserved 10 percent portion of the total federal allocation of Covid-19 vaccine may be reserved to target areas with a high SVI (defined as the top 25 percent of the SVI distribution within the state).''}
\end{minipage}

\bigskip

NASEM's recommendations, together with those from the  CDC's Advisory Committee on Immunization Practices,
were made available to states and other jurisdictions in charge of vaccine distribution.  

\subsection{Symposium on Vaccine Allocation and Social Justice}

Despite the NASEM recommendation, by the beginning of December 2020,
only Tennessee adopted a formal reserve system for the upcoming vaccine rollout. 
At this point, we co-organized a symposium directed at state policymakers and academics on ethical aspects of vaccine allocation \citep{ariadne:2020}. 
Our co-organizers were Ariadne Labs, a joint healthcare advocacy center of Brigham and Women's Hospital and Harvard T. H. Chan School of Public Health. 
We invited several state officials and also the top vaccine official of Tennessee, Dr. Michelle Fiscus, to explain the adoption process of their system.      

During the symposium, we developed a better understanding of the needs of states and other jurisdictions in operationalizing a vaccine allocation system and acquainted with different state officials, who made presentations on how their task forces were working to come up with a vaccine allocation plan or participated as listeners. 

The symposium led to two important direct policy developments. Our symposium partner Ariadne Labs worked with Massachusetts to develop a vaccine allocation guideline featuring a reserve system. Our team had several meetings with California officials and introduced them to the idea of a reserve system and how to operationalize it. 
Subsequently, Massachusetts and California each adopted a reserve system for the vaccine rollout in December 2020 and March 2021 respectively.  
By May 2021, more than a dozen states and jurisdictions adopted the reserve system
during various phases of the vaccine rollout \citep{Schmidt-NatureMedicine:2021}. 
We next elaborate on the allocation frameworks in some of these jurisdictions.

\subsection{Vaccine Allocation Plans at Various Jurisdictions}

In December 2020, the federal government
began allocation of doses of Pfizer and Moderna vaccine to state, tribal, local and territorial public health agencies in proportion to their population.

 In the vaccine allocation context, a reserve is often formulated as a community reserve, in which extra units
are awarded to a particular location, and the authorities at the location are in charge of distribution to individuals.  A popular reserve category
is a hard-hit reserve defined by geography, where units are awarded above and beyond the population proportion of that geographic area.
Table 2 contains examples from several states and juristictions.\footnote{Additional details on these and other plans can be found
at https://www.covid19reservesystem.org/}

\begin{table}
\includegraphics[scale=0.65]{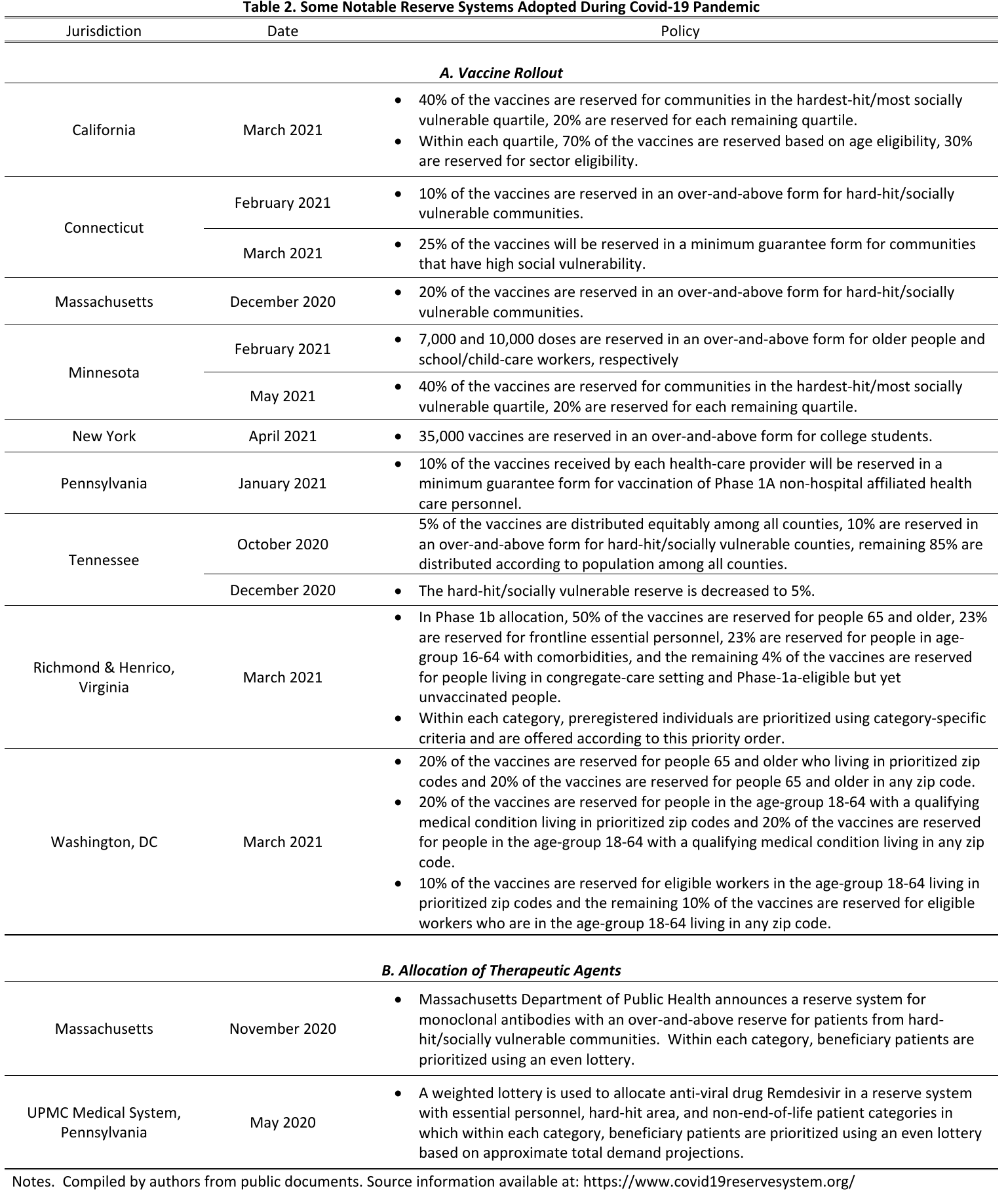}
\end{table}

In October 2020, as mentioned earlier,
Tennessee became the first state to adopt a reserve system for Covid-19 vaccine allocation
\citep{tn:20}. The initial plan has the following reserve categories:
5\% of vaccines will be distributed equitably among all 95 counties,
10\% of vaccines will be reserved by the State for use in targeted areas with high vulnerability to morbidity and mortality from the virus,
and 85\% will be  distributed among all 95 counties based upon their populations.
Reserve sizes were subsequently revised for  specific Covid-19 vaccines
in response to the differences in vaccine storage requirements \citep{tn:20b}.

In December 2020, Massachusetts released their vaccine allocation plan.
The plan describes that 20\% of vaccines will be reserved 
for communities that have experienced a disproportionate Covid-19 burden and high social vulnerability.
The state adopted an over-and-above implementation, allocating additional
units to hard-hit areas \citep{ma:20,biddinger:20}.
Within a community, the allocation is based on Phase prioritization.

Several other jurisdictions also embedded 
equity in their vaccine rollout through over-and-above reserves for their hard-hit communities, 
but other forms of reserve systems also found applications in the field. 
For example, in February 2021, Connecticut announced a 25\% community-level reserve 
for communities with high social vulnerability, formulated as a minimum guarantee, and in January, Pennsylvania started to use a 10\% minimum guarantee reserve for healthcare personnel who are not affiliated with any hospital system.

In March 2021, California announced a particularly elaborate reserve system in which 40\% of vaccines are 
reserved for communities in the first quartile of the Healthy Places Index, 
an index of socioeconomic disadvantage, and 20\% are reserved for communities in each of the other quartiles.
Within each quartile, 30\% of the vaccines reserved for the quartile are allocated based on age eligibility, and 70\% of the vaccines reserved for the quartile 
are allocated based on sector eligibility.  It is  noteworthy  to emphasize that California's reserve system utilizes  category-specific prioritization criteria. 
As shown in Table 2, 
other jurisdictions that adopted reserve systems utilizing category-specific prioritization criteria include Richmond, Virginia and Washington DC.

Three other plans in the table are noteworthy for showcasing
the flexibility of a reserve system.  First, while most reserve systems come in an over-and-above form, in 
February 2021, Connecticut announced a 25\% community-level reserve for communities with high social vulnerability, formulated as a minimum guarantee.
Second, Richmond City and Henrico County Health Districts in Virginia proposed several reserve 
categories  \citep{Richmond-2:2021, Richmond-1b:2021}, some with category-specific prioritization criteria.
Third, Washington DC's plan also uses the flexibility of a reserve with categories
based on age, worker type, qualifying medical conditions and certain prioritized zip codes as shown in the table. 

Since Richmond City and Henrico County plan uses the full flexibility of our model with different category-specific priorities, we give it in detail here (although the plan partially specifies the category-specific priorities): \medskip 

\noindent{\bf Richmond and Henrico Reserve System for Phase 1b Covid-19 Vaccine Rollout}
Categories, their share of Covid-19 vaccines, and factors that affect category-specific priorities were specified by the health districts as follows:
\begin{enumerate}
	\item Phase 1a and Congregate Care (4\% of units) 
	\item Adults age 65+ (50\% of units)
	\begin{itemize}
		\item Age (older residents have higher priority)
		\item Race and ethnicity (Black, Hispanic/Latinx, and American Indian or Alaska Native residents have higher priority)
		\item Burden of disease in the area where a person lives 
		\item Social  Vulnerability Index (SVI) Score  of the area where a person lives
	\end{itemize}
	\item Frontline Essential Workers (23\% of units)
	\begin{itemize}
		\item Age (older residents have higher priority)
		\item Race and ethnicity (Black, Hispanic/Latinx, and American Indian or Alaska Native residents have higher priority)
		\item Burden of disease in the area where a person lives 
		\item Social  Vulnerability Index (SVI) Score  of the area where a person lives
	\end{itemize}
	\item People ages 16-64 with Comorbidities (23\% of units)
	\begin{itemize}
		\item Age (older residents have higher priority)
		\item Race and ethnicity (Black, Hispanic/Latinx, and American Indian or Alaska Native residents have higher priority)
		\item Socioeconomic status (residents who are under and uninsured have higher priority
	\end{itemize}

\end{enumerate}

\subsection{Monoclonal Antibody Allocation in Massachusetts} \label{sec:MA-theraupetics}

In November 2020, the FDA granted an emergency use authorization
for monoclonal antibody treatments.
The Massachusetts Department of Public Health assembled a committee of doctors,
health center representatives, and ethicists to devise a plan for equitable distribution of doses delivered to Massachusetts.
Dr. Emily Rubin, a member of the committee who had been following our progress since May 2020,  
inquired about whether our proposed reserve system can be deployed and 
operationalized.  
Following our interaction, the December guideline recommended the use of a reserve system for within-hospital allocation
 \citep{ma-mAb:20}.  Patients are in placed into two tiers:
(1) patients who are age $\geq$ 65 and patients aged 18 and older with BMI $\geq$ 35,
and (2) all other eligible patients.

At each time interval (defined either as once or twice a day), there are two serial allocations for treatment spots. The first allocation of 80\% of 
available doses, known as the ``open allocation,'' will be available to all patients \citep{ma-mAb:20}. The second allocation of 20\% 
of available doses are available to patients who live in a census tract with SVI $>$ 50\% or 
in a city or town with a 7-day average Covid-19 incidence rate in the top quartile as reported in the most recent MA DPH weekly COVID report (``hardest-hit'').
For the open 80\%,  allocation takes place in order
of tier, with lottery tie-breaking within tier.
For the 20\% reserve, allocation is first
for patients from hardest-hit areas.
For these patients, allocation takes place in order of tier, with lottery tie-breaking within tier.  If units remain in the 20\% reserve, then they
are allocated to patients who are eligible for open units in order of tier,
with lottery tie-breaking. \citet{rubin/etal:chest2021} reports the experience of the Mass General Brigham health system with the implementation of the resulting reserve system  for allocation of monoclonal antibody therapies.

\subsection{Post-pandemic Policy Impact: Crisis Care Guidance in Oregon}

After the pandemic, some states formed advisory committees to evaluate and reform their 
crisis standards of care which will be adopted in a public health emergency or an overwhelming disaster. 
Oregon is one of these states. 

In May 2023, Oregon Health Authority (OHA) requested public input for
a preliminary framework prepared by the Oregon Resource Allocation Advisory Committee \citep{ORAAC:23}. 
In this document, the committee presented six criteria under consideration for crisis care triage
and discuss how they might be used in a stand-alone fashion or combined as part of a multi-criteria approach.

As their recommended methodology to implement multiple criteria, the committee recommended the reserve system:
\bigskip

\begin{minipage}[c]{0,9\textwidth}
\textsf{``In a major health emergency situation, ease of implementation is a feature that
needs to be taken seriously. Models such as those described below can all be
implemented readily via a methodological approach known as Categorized
Priority System (sometimes also called a Reserve System). During the Covid-19
pandemic, systems that combined factors such as survivability, level of
disadvantage and essential worker status were successfully developed for
purposes including allocating vaccines, tests and treatments. Custom-made, 
free-of-charge software has been developed to facilitate implementation.''}
\end{minipage}

\bigskip

As can be seen from this recommendation, two of the factors that contributed to the committee's endorsement of 
reserve system are its practicality and the free-of-charge software we made publicly available  for its implementation.\footnote{The free software is available from the first three co-authors upon request; its demo video can be seen at https://www.covid19reservesystem.org/software.}

\subsection{Other Endorsements for the Reserve System in Bioethics, Critical Care Medicine and Law Literatures}
Following our efforts in recommending or designing reserve systems for different settings, many members of
the medical community and civil rights advocates have also been receptive to the adoption of this methodology.

Some retrospective studies assessed the effectiveness of field applications of certain allocation methods in overcoming pre-existing racial inequities in access to healthcare. \citet{wu2022disparities} shows that during the Covid-19 pandemic at a large healthcare system, Black and Hispanic patients were less likely to receive monoclonal antibody treatment. This was a discrepancy that could not be explained through the refusal of the patients to take the treatment. They recommend exploring the use of a reserve system in the future to correct such disparities. 
Using a large data set from 208 US hospitals in 2014 and 2015,  \citet{wasserman2020setting} shows that  SOFA scores were associated with an overestimated mortality among Black patients compared with White patients, and this was associated with a structural disadvantage for Black patients in crisis standards of care guidelines. They also recommend the adoption of reserve systems to overcome the problems associated with biased SOFA-score-based prioritization using social disadvantage-based reserve categories.

The Editorial, \citet{makhoul2021reserve}, in response to \citet{rubin/etal:chest2021} characterizes a reserve system as a method that is backed by growing field evidence and is a pragmatic framework for equitable medical resource allocation, citing the success of its use during the Covid-19 pandemic. They also note that

\bigskip

\begin{minipage}[c]{0,9\textwidth}
\textsf{``As reserve systems become more prevalent, it is important to acknowledge and understand the psychological effects on participants. Not only do reserve systems enable policymakers to allocate resources equitably, but they also signal to participants that expert judgment has been used to design a system for maximal societal benefit. Participants eligible for prioritized categories (eg, patients from high-SVI zip codes) may feel more adequately safeguarded.''}
\end{minipage}

\bigskip

Finally,  emphasizing the policy roles of its ``minimum guarantee'' and ``over-and-above'' variants, 
\cite{hellman/nicholson:21} advocates for reserve system as a ``middle road'' between saving the most lives and equity
for citizens with disability. 

\bigskip

\begin{minipage}[c]{0,9\textwidth}
\textsf{``The reserve can be operationalized in either of two ways. It can provide a `boost' for a group, 
giving it extra resources, as might be justified in the case for health care workers. Or it can function 
as a “protective measure” to ensure that members of a group are not left out altogether. 
The order in which the reserve category is processed determines whether it functions as a boost or a protective measure. 
If the reserve category is processed first, it functions as a protective measure; if it is process second, 
it functions as a boost. In our view and consistent with the ADA’s [The Americans with Disabilities Act's] 
emphasis on equal rather than superior opportunity, and in their view also, the reserve for disability should function not as a boost but as a protective measure.\\[0.3cm]
While a discussion of mathematics of the reserve system is beyond the scope of this paper, 
what is important to understand is that this model provides an algorithm that enables policymakers to allocate resources according to multiple principles–one that can be fashioned in advance and applied much like the algorithms in 
SOFA-type scoring. For our purposes, this method provides a middle road between the goals of saving the most lives and ensuring that benefits and harms are distributed fairly''}
\end{minipage}

\medskip

\section{Conclusion} \label{conclusion}

In two different ways, this paper is about compromise between considerations that pull in different directions.\footnote{We are grateful
to an anonymous referee for this insightful observation.}
The first way is in the context of the specific market design application formulated in this paper. How can various competing ethical 
values be balanced to allocate scarce and vital life-saving resources in a disaster setting? The second way is methodological. 
How can the role of a design economist in advancing scholarly research in  a given application be balanced with 
the role of a design economist in developing practical and realistic tools which have a shot at influencing policy?
We answer both questions in this paper together by proposing a reserve system as a flexible allocation rule to balance
various ethical values and as a procedure that only minimally differs from the traditional priority system.  

As a compromise between multiple ethical principles in a pandemic medical resource allocation setting, we formulate and advocate for 
a reserve system where each ethical principle is represented with a reserve category.  
The underlying ethical principle at any category $c$ gives rise to a category-specific priority order $\pi_c$ of individuals for category $c$, 
which is then used to allocate the units reserved for category $c$. 
When there is only one ethical value, the reserve system reduces to the traditional priority system which had been
prevalent in the field. During the Covid-19 pandemic, this close connection gave our team a distinct advantage to communicate our ideas
with experts in medical ethics and officials in healthcare. 
By correctly identifying some of the challenges they face due to limitations of the priority system and addressing those
with minimal interference, we found that many in these communities are receptive to our ideas. 
This broader strategy of institution redesign, minimalist market design \citep{sonmez:23}, has recently been effective 
in other settings as well such as the 2021 reform of the US Army's branching system \citep{greenberg/pathak/sonmez:23}. 
There is, however, one important aspect of the current application that differentiates it from the Army application, 
and it is this difference that motivated our formal analysis in Section \ref{sec:model}.  

In the case of the 2021 reform of the US Army's branching system,  \citet{greenberg/pathak/sonmez:23} show that there is
a unique mechanism that satisfies the Army's design objectives. Thus, the paradigm of minimalist market design 
has a unique prescription for the Army's branching application.  
In contrast, in our current application, our proposed reserve system refers to a broad class of allocation rules. 
While many elements of this class can be used to overcome the failures of the traditional priority system, 
different elements of the class can have disparate distributional implications. 
This richness can be both a blessing by giving policymakers additional flexibility in the design and
a curse by causing unintended consequences due to the adoption of a wrong element of the broad class.  
Indeed, in settings where reserve categories correlate due to a baseline priority order, such unintended consequences 
and resulting policy challenges have been prevalent in the field 
(cf. \citealp{dur/kominers/pathak/sonmez:18, pathak/rees-jones/sonmez:20, pathak/rees-jones/sonmez:20b, sonmez/yenmez22}).   
Our formal analysis in Section \ref{sec:model} provides 
policymakers and other stakeholders with the analytical tools which enable them to choose a reserve system
that truly reflects their policy objectives. 

Our paper has shown that the field of market design has matured to a stage where it can provide real-time guidance 
for society during a crisis. It has also demonstrated that minimalist market design is an effective paradigm to communicate the ideas of design economists with 
policymakers and experts in other fields when they are most useful in a timely manner.

\bibliographystyle{aea}
\bibliography{Pandemic-MS-v58-July2023-ArXiv}

\newpage

\begin{center}
\textbf{\Large Appendix}
\end{center}

\setcounter{page}{1}
\renewcommand{\thepage}{A.\arabic{page}}
\renewcommand{\thefigure}{A.\arabic{figure}}
\renewcommand{\thetable}{A.\arabic{table}}
\renewcommand{\thetheorem}{A.\arabic{theorem}}
\renewcommand{\thelemma}{A.\arabic{lemma}}
\renewcommand{\theexample}{A.\arabic{example}}
\renewcommand{\theassumption}{A.\arabic{assumption}}

\appendix

\section{Resource-Dependent Design Considerations} \label{app:design}

In this appendix, we offer some thoughts about how to implement a reserve system
might depend on the resource that is rationed.
In Section \ref{sec:model}, we model pandemic rationing as a one-shot static reserve system.
Several vital resources, however, must be rationed during a pandemic as patients in need present.
Hence, it is important to formulate how our static model can be operationalized in an application where patient arrival
and allocation are both dynamic. The adequate formulation depends on the specific characteristics of the rationed resource.
Most notably, answers to the following two questions factor in the implementation details:
\begin{enumerate}
\item  \textit{Is the resource fully consumed upon allocation or is it durable, utilized over a period, and can it be re-allocated?}
\item \textit{Is there immediate urgency for allocation?}
\end{enumerate}
Since most guidelines are on rationing of vaccines, ventilators, ICU beds, and anti-viral drugs or treatments,
we focus our discussion on these four cases.

\subsection{Vaccine Allocation} \label{subsec:vaccine}

A unit of a vaccine is consumed upon allocation and reallocation of the unit is not possible.
Moreover, there is no immediate urgency to allocate a vaccine. Hence, a large number of units can be allocated simultaneously.
Therefore, vaccine allocation is an application of our model where our proposed reserve system can be implemented on a static basis
as vaccines become available.

This is, however, not the only reasonable way a reserve system can be operationalized for vaccine allocation.
In the United States, there is a tradition of distributing influenza vaccines at local pharmacies or healthcare providers on a  first-come-first-serve basis.
This practice can be interpreted as a single-category special case of a reserve system where the priorities are based on
the time of arrival. This practice can easily be extended to any sequential reserve matching system with multiple categories
where the baseline priorities are determined by the time of arrival.  Under this dynamic implementation of a reserve system,
as a patient arrives to a healthcare provider she is allocated a vaccine  as long as there is availability in a
category for which she is a beneficiary.  If there are multiple such categories, the patient is assigned a
unit from the category that has the highest precedence under the sequential reserve matching.
While  many have criticized first-come-first-serve allocation because of biases it induces based on access to health care (e.g., \citealp{kinlaw:07}),
reserve categories can be designed to mitigate these biases, even if priority is first-come-first-serve within each category.
For example, there can be a reserve category for patients from rural areas.
There is an important precedent for using a reserve system in this dynamic form.
Between 2005-2008,  H-1B immigration visas in the US were allocated through a reserve system with
general and advanced-degree reserve categories
where priority for each category was based on the application arrival time \citep{pathak/rees-jones/sonmez:20}.

\subsection{Ventilator/ICU Bed Allocation}

Since the relevant characteristics of ventilators and ICU beds are identical in relation to our model, the
implementation of reserve systems for these resources will be similar. Therefore, we present the details of their implementation
together. For simplicity in this subsection, we refer the resource in short supply as a ventilator.

A ventilator is durable and can be reassigned once its use by its former occupant is completed.
Moreover, there is always urgency in allocation of this vital resource.
These two features make direct static implementation of a reserve system impractical;
implementation always has to be dynamic.
One important observation on ventilator allocation is key to formulate the implementation:
since a ventilator is durable and assigned to a patient for a period, it can be interpreted as a good
which is allocated at each instant. During the course of using a ventilator, a patient's clinical situation and her priority for
one or more categories  may change.
Therefore, with the arrival of each new patient, the allocation of all units has to be
reevaluated.
As such, the following additional ethical and legal consideration has an important bearing on
the design of a reserve system:
\begin{itemize}
\item[3.] \textit{Can a patient be removed from a ventilator once she is assigned?}
\end{itemize}

There is widespread debate on this issue in the United States.  \citet{piscitello:20} describes
25 states with protocols that discuss the ethical basis of re-assigning ventilators.
As of June 2020, the majority of guidelines support ventilator withdrawal.
If a ventilator
can be withdrawn, the design is simpler (and effectively identical to static implementation with each new arrival). While patient data needs to be updated through the duration of ventilator use,
no fundamental adjustment is needed for the design of the main parameters of the reserve system.
Of course in this scenario, it is possible that the category of the unit occupied by the patient may change over time.
For example, a patient may initially be assigned a unit from the general category even though
she has  sufficiently high priority for multiple categories such as the general category and essential personnel category.
At a later time, she may only have high enough priority for   the latter category.
In this case, the patient will continue using the ventilator although for accounting purposes she will
start consuming a unit from a different category.

If a ventilator cannot be withdrawn,  a reserve system can still be applied with a grandfathering
structure to reflect the property rights of patients who are already assigned.
In this case, the priority system has to give highest priority to occupants of the units from any category
for as long as they  can hold these units despite a change in their clinical situation
or  arrival of patients who otherwise would have higher priority for these units.

\subsection{Anti-viral Drugs or Treatments}

For anti-virals drugs and treatments, the vital resource is consumed upon allocation (as in vaccine allocation)
but there is typically urgency and allocation decisions will need to be made as patients arrive (as in ventilator allocation).
One possible dynamic implementation is based on first-come-first-serve arrival within reserve categories.  This would be akin to
the dynamic allocation scenario for ventilators with a baseline priority structure that depends on
patient arrival time as described in Section \ref{subsec:vaccine}.  Alternatively, drug assignment can be batched within pre-specified time-windows.  Drugs can then be assigned based on expectations of the number of patients in each
category over this time window.   Since drugs would be administered by a clinician, the relationship between a reserve system and cutoffs can be particularly valuable.
A clinician can simply assign the treatment to a patient if she clears the cutoff for any reserve for which she is eligible.
In fact, after the first version of our paper was circulated, our team assisted with the design of the system used at  the University of Pittsburgh Medical Center to allocate the anti-viral drug Remdesivir in May 2020 with this implementation \citep{whiteCCM:22}. 
The system had special provisions for hardest hit and essential personnel and used lotteries for prioritization.
 In November 2020, we also assisted the Massachusetts working group on the equitable
 allocation of Covid-19 therapies to help design a reserve
system for allocation of monoclonal antibodies \citep{rubin/etal:chest2021}.

\section{Smart Reserve Matching} \label{sec:smart}

Although virtually all practical applications of reserve systems are implemented through sequential reserve matching,
this class of mechanisms may suffer from an important shortcoming: they may lead to Pareto inefficient outcomes,
due to myopic processing of reserves. The following example illustrates both how this may happen,
and also motivates a possible refinement based on smart processing of reserves.

\begin{example}\label{ex:non-wastefulness-problem} Consider a hard reserve system induced by baseline priority order $\pi$.
There are two patients $I=\{i_1,i_2\}$ who are priority ordered as
$$i_1 \mathrel{\pi} i_2 $$
under the baseline priority order $\pi$.
There are two categories; an unreserved category $u$ with an all-inclusive  beneficiary set of $I=\{i_1,i_2\}$,
and  a preferential treatment category $c$ with a beneficiary set $I_c=\{i_1\}$ of a single preferential treatment patient.
Both categories have a capacity of one unit each (i.e., $r_c=r_u=1$).
Since the reserves are hard, the resulting category-specific priority orders are given as follows:
\[  i_1 \; \pi_u \; i_2 \; \pi_u \; \emptyset \qquad \mbox{and} \qquad   i_1 \; \pi_c \; \emptyset \; \pi_c \; i_2.
\]
Consider the sequential reserve matching $\varphi_{\tr}$  induced by the order of
precedence $\tr$, where
\[  u \; \tr \; c.
\]
Under matching $\varphi_{\tr}$,   first patient $i_1$ is matched with the unreserved category $u$ and
subsequently the category-$c$ unit is left idle since no remaining patient is eligible for this preferential treatment category.
Therefore,
$$ \varphi_{\tr}  =
\left(\begin{array}{cc} i_1 & i_2 \\
u & \emptyset
\end{array} \right),
$$
resulting in the set of matched patients  $\varphi_{\tr}(I) = \{i_1\}$.

Next consider the sequential reserve matching  $\varphi_{\tr'}$  induced by the order of
precedence $\tr'$, where
\[  c \; \tr' \; u.
\]
Under matching $\varphi_{\tr'}$,   first patient $i_1$ is matched with the preferential treatment category $c$ and
subsequently patient $i_2$ is matched with the unreserved category $u$.
Therefore,
\[  \varphi_{\tr'}  =\left(\begin{array}{cc} i_1 & i_2 \\
c & u
\end{array}   \right),
\]
resulting in the set of matched patients  $\varphi_{\tr'}(I) = \{i_1, i_2\}$.
Since $\varphi_{\tr}(I)  \subsetneq  \varphi_{\tr'}(I)$,  matching  $\varphi_{\tr}$ is Pareto dominated by matching $\varphi_{\tr'}$.
\end{example}

Observe that Example \ref{ex:non-wastefulness-problem} also illustrates that the cause of Pareto inefficiency is
the myopic allocation of categories under sequential reserve matchings.
Under matching  $\varphi_{\tr}$,  the more flexible unreserved unit is allocated to patient $i_1$ who is the only
beneficiary of category-$c$. This results in a suboptimal utilizations of reserves, which can be avoided with
the concept of ``smart  reserve matching'' we introduce below.

To this end,  we first introduce a new axiom, which together with non-wastefulness imply Pareto efficiency.

\begin{definition} A matching $\mu \in \calm$ is {\bf maximal in beneficiary assignment} if
$$\mu \in \underset {\nu \in \calm} {\arg \max} \; \left \vert \bigcup_{c \in \calc \setminus\{u\}} \big(\nu^{-1}(c)\cap I_c \big ) \, \right \vert .$$
\end{definition}

This axiom simply requires that the reserves should be maximally assigned to target beneficiaries to
the extent it is feasible. It precludes the myopic assignment of categories to patients since
the desirability of a matching depends on the structure of the matching as a whole rather than the individual assignments
it prescribes for each category.

It is worth noting that the inefficiency observed in Example \ref{ex:non-wastefulness-problem} is
specific to the case of hard reserves and cannot happen for soft reserves, as in our main application of pandemic rationing.
Nonetheless, maximality in beneficiary assignment is a desirable axiom in general including for soft reserves
because sub-optimal utilization of reserves may receive heightened scrutiny.  For example, consider a scenario with two preferential treatment categories,
essential personnel and disadvantaged, each with one unit of reserve.  Suppose patient A is both essential personnel and disadvantaged,
 patient B is disadvantaged, and there are several other patients who are neither.
 One possible way  to use these reserves is to assign patient A to the disadvantaged reserve, leaving no other
preferential treatment patients available for the essential personnel reserve.  In this case, the essential personnel reserve would be opened up
to patients who are neither essential personnel nor disadvantaged.  This could in turn mean only one of the reserves
is assigned to members of the target beneficiary groups.  This outcome could be seen problematic since an alternative,
which assigns patient A to the essential personnel reserve (instead of the disadvantaged reserve) and patient B to the disadvantaged reserve,
accommodates both reserves.  By imposing maximality in beneficiary assignment, we avoid this shortcoming
through a ``smart'' utilization of reserves. 

Building on this insight, we next present a polynomial-time algorithm that generates smart cutoff matchings:

\begin{quote}
  \noindent{}{\bf Smart Reserve Matching Algorithm}

 Fix a parameter  $n \in \{0,1,\hdots,r_u\}$ that represents the number of unreserved units to be processed in
 the beginning of the algorithm.\footnote{For $n=0$, this algorithm is equivalent to the meritorious horizontal algorithm
 in \citet{sonmez/yenmez22}.}
 The remaining unreserved units are to be processed at the end of the algorithm.

Fix a baseline priority order $\pi$, and for the ease of description relabel patients so that
\[i_1 \mathrel{\pi} i_2 \mathrel{\pi} \hdots \mathrel{\pi} i_{\abs{I}}.\]

\begin{enumerate}

  \item[\textbf{Step 0.}] Find a matching that is maximal and complies with eligibility requirements by temporarily deeming that a patient $i \in I$ is  eligible for a category $c \in \calc \setminus \{u\}$ if and only if $i \in I_c$, and no patient is eligible for unreserved category $u$.\footnote{This is known as a \emph{bipartite maximum cardinality matching problem} in graph theory and many \emph{augmenting alternating path algorithms} \citep[such as those by][]{hopcroft/karp:73, karzanov:73} can solve it in polynomial time.} The solution finds the maximum number of patients who can be matched with a preferential treatment category that they are beneficiaries of. Denote the number of patients matched by this matching as $n_b$.

  \item[\textbf{Step 1.}]  Let $J^u_0=\emptyset$, $J_0=\emptyset$. Fix parameters $\kappa\gg \varepsilon>0$ such that $\varepsilon<1$ and $\kappa>|I|$. For $k=1,\hdots,|I|$ we repeat the following substep given $J^u_{k-1}$, $J_{k-1}$:
  
  \textbf{Step 1.($k$).}     \begin{enumerate}

      \item[i.] if $|J^u_{k-1}|<n$ continue with (i.A) and otherwise continue with (ii).
      \begin{enumerate}

        \item[A.] Temporarily deem all patients in $J^u_{k-1}\cup\{i_k\}$ eligible only for category $u$ and all other patients eligible only for the categories in $\calc \setminus \{u\}$ that they are beneficiaries of.

        \item[B.] for every pair $(i,x) \in I\times \calc\cup\{\emptyset\}$ define a weight $W_{i,x} \in \mathbb{R}$ as follows:

        \begin{itemize}

          \item If $x\in \calc$ and $i$ is temporarily eligible for $x$ as explained in (i.A),
          	\begin{itemize}
          		\item if $i \in J^u_{k-1}\cup J_{k-1}$, then define $W_{i,x}:=\kappa$,
          		\item otherwise, define $W_{i,x}:=\varepsilon$.
          	\end{itemize}

          \item If $x\in \calc$ and $i$ is not temporarily eligible for $x$ as explained in (i.A), define $W_{i,x}:=-\varepsilon$.

          \item If $x=\emptyset$, define $W_{i,x}:=0$.
        \end{itemize}

        \item[C.] Solve the following \emph{assignment problem} to find a matching\footnote{A polynomial-time algorithm such as the \emph{Hungarian algorithm} \citep{kuhn:55} solves this problem.} 
        	$$
        		\sigma \in \underset {\mu \in \calm} {\arg \max} \sum_{i \in I}W_{i,\mu(i)}.
        	$$

        \item[D.] If $|\sigma(I)|=n_b+|J^{u}_{k-1}|+1$ then define $$J^u_{k}:=J^u_{k-1}\cup\{i_k\} \quad \mbox{and} \quad J_{k}:=J_{k-1},$$ and go to Step 1.($k+1$) if $k < |I|$ and Step 2 if $k=|I|$.

        \item[E.] Otherwise, go to (ii).
      \end{enumerate}

      \item[ii.] Repeat (i) with the exception that $i_k$ is temporarily deemed eligible only for  the categories in $\calc \setminus \{u\}$ that she is a beneficiary of in Part (ii.A). Parts (ii.B) and (ii.C) are the same as Parts (i.B) and (i.C), respectively, with the exception that weights are constructed with respect to the eligibility construction in (ii.A). Parts (ii.D) and (ii.E) are as follows:
      \begin{enumerate}
        \item[D.] If $|\sigma(I)|=n_b+|J^u_{k-1}|$ and $\sigma(i)\not=\emptyset$ for all $i \in J_{k-1}\cup\{i_k\}$, then define $$J^u_{k}:=J^u_{k-1} \quad \mbox{and} \quad J_{k}:=J_{k-1}\cup\{i_k\},$$ and  go to Step 1.($k+1$) if $k < |I|$ and Step 2 if $k=|I|$.

        \item[E.] Otherwise, $$J^u_{k}:=J^u_{k-1} \quad \mbox{and} \quad J_{k}:=J_{k-1},$$ and  go to Step 1.($k+1$) if $k < |I|$ and Step 2 if $k=|I|$.
      \end{enumerate}
    \end{enumerate}

  \item[\textbf{Step 2.}]
    \begin{enumerate}
      \item Find a matching $\sigma$ as follows:

      \begin{enumerate}
        \item Temporarily deem all patients in $J^u_{|I|}$ eligible only for category $u$, all patients in $J_{|I|}$ eligible only for the categories in $\calc \setminus \{u\}$ that they are beneficiaries of, and all other patients ineligible for all categories.

        \item Find a maximal matching $\sigma$ among all matchings that comply with the temporary eligibility requirements defined in (i).\footnote{This problem can be solved using a polynomial \emph{augmenting alternating paths algorithm} \citep[for example see][]{hopcroft/karp:73, karzanov:73}.}
      \end{enumerate}

      \item Modify $\sigma$ as follows:

      One at a time assign the remaining units unmatched in $\sigma$ to the remaining highest priority patient in $I \setminus (J^u_{|I|} \cup J_{|I|})$ who is eligible for the category of the assigned unit in the real problem  in the following order:
      \begin{enumerate}

        \item the remaining units of the preferential treatment categories in $\calc \setminus \{u\}$ in an arbitrary order, and

        \item the remaining units of the unreserved category $u$.
      \end{enumerate}
    \end{enumerate}

\end{enumerate}

  \medskip
 \end{quote}

Every matching $\sigma$ constructed in this manner is referred to as a \textbf{smart reserve matching} induced
by assigning $n$ unreserved units subsequently at the beginning of the algorithm.
Let $\cals^{n}$ be the set of all reserve matchings for a given $n$.

The idea behind the smart reserve matching algorithm can be summarized as follows: 

We sequentially process patients with respect to their priority under the baseline priority order for each $k=1,2,\hdots,|I|$. 
Step 1.($k$).i is invoked if some of the initial $n$ unreserved category units are still available and not all of them are already committed to a subset $J^u_{k-1}$ of patients among $i_1,\hdots,i_{k-1}$. 

When Step 1.($k$).i is invoked, we commit to giving an unreserved unit to $i_k$ if this assignment will not decrease the maximum number of patients who can make use of the preferential treatment categories that they are beneficiaries of, denoted as $n_b$ in the algorithm, in addition to the subset $J_{k-1}$ of patients already committed to these units among $i_1,\hdots,i_{k-1}$. Otherwise, $i_k$ gets a unit from a preferential treatment category in Step 1.($k$).ii. 

In case Step 1.($k$).i is skipped, and we directly continue with Step 1.($k$).ii. In this case, we commit to giving a preferential treatment category unit to  $i_k$ from a category she is beneficiary of if this assignment, in addition to patients in $J_{k-1}$, does not decrease the maximum number of patients who can make use of the preferential treatment category that they are beneficiaries of, $n_b$. Otherwise, we skip patient $i_k$ for now.

We invoke Step 2 after we commit $n$ unreserved units to patients in  $J^u_{|I|}$ and the maximum number of preferential treatment category units, $n_b$, to patients in $J_{|I|}$. In this step, one at a time, we give one remaining unit from an eligible category to the remaining patients according to the baseline priority order.

We have the following technical result about the sets of patients matched under smart reserve matchings:
\begin{lemma} \label{lem:smart-samepatients}
Consider either a soft reserve system or a hard reserve  system induced by a baseline priority order $\pi$. 
For any $n\in \{0,1,\hdots,r_u\}$ and any two smart reserve matchings $\sigma,\nu \in \cals^{n}$, $$\sigma^{-1}(u)=\nu^{-1}(u), \; \mbox{ and } \; \cup_{c\in \calc\setminus \{u\}}(\sigma^{-1}(c)\cap I_c)= \cup_{c\in \calc\setminus \{u\}}(\nu^{-1}(c)\cap I_c),$$ and moreover,  $$\sigma(I)=\nu(I).$$
\end{lemma}

Lemma \ref{lem:smart-samepatients} states that for a given baseline priority order $\pi$,  every smart reserve matching for a given $n$ matches the same set of patients with the unreserved category $u$, the same set of patients with preferential treatment categories in $\calc\setminus\{u\}$ that they are beneficiaries of, and thus, the same set of patients overall.

In a soft reserve system or a hard reserve system, for a given $n$, we denote the set patients matched in every smart reserve matching with $n$ unreserved units processed first as $I^n_S$. 
By Lemma \ref{lem:smart-samepatients}, for any $\sigma \in \cals^n$, $$I^n_S=\sigma(I).$$

Our next result on smart reserve matchings is as follows:

\begin{proposition}\label{prop:smart-maximal} 
Consider either a soft reserve system or a hard reserve system induced by a baseline priority order $\pi$. 
For any $n\in \{0,1,\hdots,r_u\}$, 
any smart reserve matching in $\cals^{n}$  complies with eligibility requirements, is non-wasteful, respects priorities, and is maximal in beneficiary assignment. \end{proposition}

The choice of parameter $n$  is not without a  consequence. In particular,
matchings produced by the algorithm with the lowest parameter $n=0$ and the highest parameter $n=r_u$ both have
distinctive distributional consequences with potentially important policy implications.

\begin{theorem}\label{thm:smart-cutoffs}
 Consider either a soft reserve system or a hard reserve  system induced by a baseline priority order $\pi$.
Let $\sigma_{\ve} \in \cals^{\ve}$ be a smart reserve matching when all unreserved units are assigned 
first (i.e., $n=r_u$)
and $\sigma_{\he} \in \cals^{\he}$ be a smart reserve matching when all unreserved units are assigned 
last (i.e., $n=0$). 
Then, for any  cutoff equilibrium matching $\mu \in \calm$ that is maximal in beneficiary assignment,
  $$\ovf^{\sigma_{\ve} }_u \; \underline{\pi} \;\; \ovf^\mu_u  \;\; \underline{\pi} \;\; \ovf^{\sigma_{\he} }_u.$$
\end{theorem}

  Theorem \ref{thm:smart-cutoffs} states that among all maximum equilibrium cutoff vectors that
  support maximal matchings in beneficiary assignment, the selectivity of the unreserved category is
  \begin{itemize}
   \item  the most competitive for smart reserve matchings with $n=r_u$, that is, when all unreserved
   category units are assigned in the beginning of the algorithm, and
   \item the least competitive for smart reserve matchings with $n=0$, that is, when all unreserved
   category units are assigned  at the end of the algorithm.
 \end{itemize}
 
We believe Theorem \ref{thm:smart-cutoffs} can have important role in policy. The most widespread
real-life applications of sequential reserve systems  are in one of the two forms  ``over-and-above''  or ``minimum guarantee.''
Theorem \ref{thm:smart-cutoffs} shows that both forms have their counterparts in smart reserve matching systems.
Adopting a smart reserve matching instead of a sequential reserve matching therefore need not interfere with existing policies.

On a technical note,  Theorem \ref{thm:smart-cutoffs} is silent about the qualifications of maximum equilibrium cutoffs of 
preferential treatment categories supporting smart reserve matchings. While keeping the maximality in beneficiary assignment, 
the patients who are matched with preferential treatment  categories are uniquely determined under the two extreme smart 
reserve matchings (and for any given $n$), respectively. However, there can be multiple ways of assigning these patients 
to different preferential treatment categories. Thus, the maximum equilibrium cutoffs of preferential treatment categories are not uniquely determined for these matchings.

\section{Proofs} \label{app:proofs} \label{app:proofs1}
\subsection{Proofs of the Results in Section \ref{sec:char-cutoff}}
\newcommand{\vI}{\mu}
\newcommand{\vpI}{\mu'}

\newcommand{\vIR}{\varphi_{\triangleright}^{\tilde{I}}}
\newcommand{\vpIR}{\varphi_{\triangleright'}^{\tilde{I}}}
\renewcommand{\v}{\varphi_{\triangleright}}
\renewcommand{\vp}{\varphi_{\triangleright'}}
\renewcommand{\tr}{\triangleright}
\renewcommand{\trp}{\triangleright'}
\renewcommand{\th}{\hat{\tr}}
\newcommand{\thp}{\hat{\tr}'}
\newcommand{\vh}{\varphi_{\th}^{\tilde{I}}}
\newcommand{\vhp}{\varphi_{\thp}^{\tilde{I}}}

\noindent{}\begin{proof}[Proof of Theorem \ref{thm:cutoff}] \medskip

\noindent{}\textit{\bf Part 1.} Suppose matching $\mu \in \calm$ complies with eligibility requirements, is non-wasteful, and respects priorities. We construct a cutoff vector $f \in\calf$ as follows: For each category $c \in \calc$, define
$$ f_c = \left \{ \begin{array}{ll}
   \min_{\pi_c} \mu^{-1}(c) & \qquad \mbox{if } |\mu^{-1}(c)|=r_c, \\
   \emptyset & \qquad \mbox{otherwise}.
 \end{array} \right .
$$
Fix a category $c \in\calc$. If $|\mu^{-1}(c)|=r_c $ then $f_c \in \mu^{-1}(c)$ by construction. Since $\mu$ complies with eligibility requirements, then $f_c \mathrel{\pi}_c \emptyset$. On the other hand, if $|\mu^{-1}(c)|<r_c $, then $f_c = \emptyset$. Therefore, in either case $f_c \mathrel{\pieq_c} \emptyset$. We showed that $f \in \calf$, i.e., it is a well-defined cutoff vector.

Next, we show that $(f,\mu)$ is a cutoff equilibrium. Condition 2 in cutoff equilibrium definition is immediately satisfied because if for any $c \in \calc$, $|\mu^{-1}(c)|<r_c$, then $f_c=\emptyset$ by construction.

We next show that Condition 1 in cutoff equilibrium definition is also satisfied in two parts.
Let $i \in I$.
\begin{itemize}
\item[(a)] We show that $\mu(i) \in \calb_i(f)\cup\{\emptyset\}$. If $\mu(i)=\emptyset$ then we are done. Therefore, suppose $\mu(i)=c$ for some $c \in \calc$.  Two cases are possible: \begin{itemize}
\item If $|\mu^{-1}(c)|=r_c$, then $f_c =\min_{\pi_c} \mu^{-1}(c)$, and hence $i \mathrel{\pieq_c} f_c$. Thus, $c \in \calb_i(f)$.
\item If $|\mu^{-1}(c)|<r_c$, then $f_c=\emptyset$ by construction. Since $\mu$ complies with eligibility requirements, $i \mathrel{\pi_c} f_c=\emptyset$. Thus, $\mu(i) \in \calb_i(f)$.
\end{itemize}

\item[(b)] We show that $\calb_i(f)\not=\emptyset \implies \mu(i)\in \calb_i(f)$. Suppose $\calb_i(f)\not=\emptyset$; but to the contrary of the claim, suppose that  $\mu(i)\not\in \calb_i(f)$. By Condition 1(a) in the definition of a cutoff equilibrium, $\mu(i)=\emptyset$. Let $c\in \calb_i(f)$.  Since $\mu$ respects priorities,   then for every $j \in \mu^{-1}(c)$ we have $j \mathrel{\pi_c} i$. If $|\mu^{-1}(c)|=r_c$, then by construction, $f_c \in \mu^{-1}(c)$, and hence, $f_c \mathrel{\pi_c} i$, contradicting $c \in \calb_i(f)$. We conclude that $|\mu^{-1}(c)|<r_c$. Then by construction, $f_c=\emptyset$. Since $c \in \calb_i(f)$, $i \mathrel{\pi_c} f_c=\emptyset$. These two statements together with $\mu(i)=\emptyset$ contradict non-wastefulness of $\mu$. Thus, $\mu(i)\in \calb_i(f)$.
\end{itemize}
Hence, we showed that $(f,\mu)$ is a cutoff equilibrium.
\medskip

\noindent{}\textit{\bf Part 2.} Conversely, suppose  pair $(f,\mu) \in \calf \times \calm$ is a cutoff equilibrium. We will show that matching $\mu$ complies with eligibility requirements, is non-wastefulness, and respects priorities.\medskip

\noindent{}\textit{Compliance with eligibility requirements:} Consider a patient $i \in I$. Since by Condition 1(a) of cutoff equilibrium definition $\mu(i)\not=\emptyset$ implies $\mu(i)\in \calb_i(f)$, we have $i \mathrel{\pieq_c} f_c$. Since the cutoff satisfies $f_c \mathrel{\pieq_c} \emptyset$, by transitivity of $\pi_c$, $i \mathrel{\pi_c} \emptyset$. Therefore, $\mu$ complies with eligibility requirements.\medskip

\noindent{}\textit{Non-wastefulness:} Let $i \in I$ be such that $\mu(i)=\emptyset$ and $i \mathrel{\pi_c} \emptyset$ for some $c \in \calc$. We show that $|\mu^{-1}(c)|=r_c$. Then by Condition 1(a) of the definition of a cutoff equilibrium for $(f,\mu)$, we have $\calb_i(f)=\emptyset$. In particular $c \notin \calb_i(f)$. Then $f_c \mathrel{\pi_c} i$, implying that $f_c \mathrel{\pi_c} \emptyset$ and hence $|\mu^{-1}(c)|=r_c$. Thus, $\mu$ is non-wasteful. \medskip

\noindent{}\textit{Respect of Priorities:} Let patient $i \in I$ be such that for some category $c\in \calc$, $\mu(i)=c$ while for some patient $j \in I$, $\mu(j)=\emptyset$.  We show that $i \mathrel{\pi_c} j$, which will conclude that matching $\mu$ respects priorities. By Condition 1(b) of cutoff equilibrium definition, $\calb_j(f)=\emptyset$. In particular, $f_c \mathrel{\pi_c} j$. Since $\mu(i)=c$, by Condition 1(a) of cutoff equilibrium definition, $c \in \calb_i(f)$, implying that $i \mathrel{\pieq_c} f_c$. By transitivity of $\pi_c$, $i \mathrel{\pi_c} j$.
\end{proof} \bigskip

\noindent{}\begin{proof}[Proof of Lemma \ref{prop:eq-cutoff-vector-structure}]
We prove the lemma in three claims. Let matching $\mu \in \calm$ comply with eligibility requirement, be non-wasteful, and respect priorities. \medskip

\noindent{}\textbf{Claim 1.} $\ovf^\mu$ is the maximum equilibrium cutoff vector supporting $\mu$, i.e., ($\ovf^\mu,\mu)$ is a cutoff equilibrium and for every cutoff equilibrium $(f,\mu)$, $\ovf^\mu_c \mathrel{\pieq_c} f_c$ for every $c\in \calc$.  \medskip

\noindent{}\emph{Proof.} We prove the claim in two parts.

\noindent{}\underline{Part 1.}We show that $(\ovf^\mu,\mu)$ is a cutoff equilibrium:

We restate the definition of $\ovf^\mu$ given in Equation (\ref{eq:max-cutoff}) in the main text: For every $c\in \calc$,
\begin{align*}
\ovf^\mu_c = \left\{ \begin{array}{ll}
\min_{\pi_c} \mu^{-1}(c)  & \mbox{if } |\mu^{-1}(c)|=r_c  \\
\emptyset  & \mbox{otherwise}
 \end{array}\right. .  \end{align*}
By this definition $\ovf^\mu \in \calf$. Moreover, Condition 2 in the definition of a cutoff equilibrium is trivially satisfied.

We show that Condition 1(a) holds next. Let $i \in I$. If $\mu(i)=\emptyset$ then Condition 1(a) is satisfied for $i$. Suppose $\mu(i)=c$ for some $c \in \calc$. We have $i \mathrel{\pieq_c} \min_{\pi_c}\mu^{-1}(c)$. Moreover $i \mathrel{\pi_c}\emptyset$, as $\mu$ complies with eligibility requirements. Thus, $i \mathrel{\pieq_c} \ovf^\mu_c \in \{\emptyset, \min_{\pi_c}\mu^{-1}(c)\}$, and hence, $\mu(i) \in \calb_i(\ovf^\mu)$, showing Condition 1(a) is satisfied.

Finally, we show that Condition 1(b) is satisfied. We prove its contrapositive. Let $i \in I$ be such that $\mu(i)\not \in \calb_i(\ovf^\mu)$. Thus, $\mu(i)=\emptyset$ by Condition 1(a). Let $c \in \calc$. If $|\mu^{-1}(c)|<r_c$, then   $\ovf^\mu_c = \emptyset \mathrel{\pi_c} i$ by non-wastefulness of $\mu$. If $|\mu^{-1}(c)|=r_c$, then  $j \mathrel{\pi}_c i$ for every $j \in \mu^{-1}(c)$ as $\mu$ respects priorities; thus,
$\ovf^\mu_c = \min_{\pi_c} \mu^{-1}(c) \mathrel{\pi_c} i$.
In either case, we have  $c \not \in \calb_i(\ovf^\mu)$. Thus, we get $\calb_i(\ovf^\mu)=\emptyset$, showing that Condition 1(b) also holds for $(\ovf^\mu,\mu)$, and hence, completing the proof that
$(\ovf^\mu,\mu)$ is a cutoff equilibrium.
\medskip

\noindent{}\underline{Part 2.} Let $(f,\mu)$ be a  cutoff equilibrium. We prove that $\ovf^\mu_c \mathrel{\pieq_c} f_c$ for every $c\in \calc$:

Suppose, for contradiction, that there exists some category $c \in \calc$ such that $f_c \mathrel{\pi_c} \ovf^\mu_c$. Then $|\mu^{-1}(c)| =r_c$ as $(f,\mu)$ is a cutoff equilibrium and  $f_c \mathrel{\pi_c} \emptyset$, which follows from the fact that   $\ovf^\mu_c \mathrel{\pieq_c} \emptyset$.  Thus, $\ovf^\mu_c=\min_{\pi_c} \mu^{-1}(c) \mathrel{\pi_c} \emptyset$ by definition and $\mu$ complying with eligibility requirements. Then for the patient  $i=\ovf^\mu_c$, $\mu(i)=c \not \in \calb_i(f)$ contradicting that $(f,\mu)$ is a cutoff equilibrium. Thus, such a category $c$ does not exist, and hence, $\ovf^\mu$ is the maximum cutoff equilibrium vector supporting matching $\mu$. $\diamond$ \medskip

\noindent{}\textbf{Claim 2.} $\unf^\mu$ is the minimum equilibrium cutoff vector supporting $\mu$, i.e., $(\unf^\mu,\mu)$ is a cutoff equilibrium and for every cutoff equilibrium $(f,\mu)$, $f_c \mathrel{\pieq_c} \unf^\mu_c$ for every $c\in \calc$.  \medskip

\noindent{}\emph{Proof.} We prove the claim in two parts. \medskip

\noindent{}\underline{Part 1.} We show that $(\unf^\mu,\mu)$ is a cutoff equilibrium:

We restate $\unf^\mu$ using its definition in Equation (\ref{eq:min-cutoff}):
for every $c\in \calc$,
\begin{align*}
\unf^\mu_c = \left\{ \begin{array}{ll}
\min_{\pi_c} \Big \{ i \in \mu(I) : i \mathrel{\pi_c} x_c  \Big \} & \mbox{if } x_c \not=\emptyset  \\
\emptyset  & \mbox{otherwise}
 \end{array}\right.
 \end{align*}
 where $x_c$ is defined as
 $$x_c=\max_{\pi_c} \big (I\setminus \mu(I)\big )\cup \{\emptyset\}.$$
 For every $c \in \calc$, since $x_c \mathrel{\pieq_c} \emptyset$, we have
 $ \min_{\pi_c} \big \{ i \in \mu(I) : i \mathrel{\pi_c}x_c  \big \} \mathrel{\pieq}_c \emptyset$, if $x_c\not=\emptyset$. Hence, $\unf^\mu_c \mathrel{\pieq_c} \emptyset$ showing that $\unf^\mu\in \calf$. \medskip

 We show that the conditions in the definition of a cutoff equilibrium are satisfied by $(\unf^\mu,\mu)$. \medskip

 \noindent{}\emph{Condition 2.} Suppose $|\mu^{-1}(c)|<r_c$ for some $c \in \calc$. For any $i \in I\setminus \mu(I)$ we have $\emptyset \mathrel{\pi}_c i$ by non-wastefulness of $\mu$. Thus, $x_c=\emptyset$. This implies $\unf^\mu_c=\emptyset$ by its definition. Hence, Condition 2 is satisfied.\medskip

\noindent{}\emph{Condition 1(a).} Let $i \in I$. If $\mu(i)=\emptyset$ then Condition 1(a) is satisfied for $i$. Suppose $\mu(i)=c$ for some $c \in \calc$. We have $i \mathrel{\pieq_c} \ovf^\mu_c$, since we showed in Claim 1 that $(\ovf^\mu,\mu)$ is a cutoff equilibrium. Two cases are possible about $\mu^{-1}(c)$:
\begin{itemize}
\item If $|\mu^{-1}(c)|<r_c$: we showed in proving Condition 2 that $x_c = \emptyset$, thus, $\ovf^\mu_c=\unf^\mu_c=\emptyset$. Since $i \mathrel{\pi_c} \emptyset$, $c=\mu(i) \in \calb_i(\unf^\mu)$ showing that Condition 1(a) holds for $i$.

\item If $|\mu^{-1}(c)|=r_c$: Then $\ovf^\mu_c=\min_{\pi_c} \mu^{-1}(c) \mathrel{\pi_c} x_c$:
as otherwise
\begin{itemize}
\item if $x_c \in I$, then $\mu(x_c)=\emptyset$ (by definition of $x_c$) and yet $c \in \calb_{x_c}(\ovf^\mu)$,
a contradiction to $(\ovf^\mu,\mu)$ being a cutoff equilibrium;
\item if $x_c=\emptyset$, then (i) $x_c \mathrel{\pi}_c \ovf^\mu_c \in I$ contradicts $\ovf^\mu$ being a cutoff vector, and  (ii) $x_c=\ovf^\mu_c $ contradicts $|\mu^{-1}(c)|=r_c$. Thus we cannot have $x_c \mathrel{\pieq_c} \ovf^\mu_c$ in this case either.
\end{itemize}
Thus, $$\ovf^\mu_c \mathrel{\pieq_c} \min_{\pi_c}\{i\in I: i \mathrel{\pi_c} x_c\}=\unf^\mu_c.$$
Then $$i \mathrel{\pieq_c} \ovf^\mu_c  \mathrel{\pieq_c} \unf^\mu_c,$$
implying $c=\mu(i) \in \calb_i(\unf^\mu)$ and showing that Condition 1(a) holds for $i$.\medskip
\end{itemize}

\noindent{}\emph{Condition 1(b).} Let $i \in I$ be such that $\calb_i(\unf^\mu)\not=\emptyset$. Let $c \in \calb_i(\unf^\mu)$.
\begin{itemize}
\item if $x_c=\emptyset$: Then $i \mathrel{\pi_c} \emptyset=\unf^\mu_c$. By definition of $x_c$, $i \in \mu(I)$, i.e., $\mu(i)\not=\emptyset$.
\item if $x_c\not=\emptyset$: Then $i \mathrel{\pieq_c} \unf^\mu_c \mathrel{\pi_c} x_c$, which in turn implies that $\mu(i)\not=\emptyset$ by definition of $\unf^\mu_c$ and $x_c$.
\end{itemize}
In either case, by Condition 1(a),  $\mu(i) \in \calb_i(\unf^\mu)$. Thus, Condition 1(b) is satisfied for $i$.

These conclude proving that $(\unf^\mu,\mu)$ is a cutoff equilibrium.\medskip

\noindent{}\underline{Part 2.} Let $(f,\mu)$ be a cutoff equilibrium. We prove that $f_c \mathrel{\pieq_c} \unf^\mu_c$ for every $c\in \calc$:

Suppose to the contrary of the claim $\unf^\mu_c \mathrel{\pi_c} f_c$ for some $c \in \calc$. Now, $\unf^\mu_c$ is a patient, because $f_c \mathrel{\pieq_c} \emptyset$ by the definition of a cutoff vector. By definition of $\unf^\mu_c$, $x_c\not=\emptyset$ and  $\unf^\mu_c \mathrel{\pi_c} x_c$. We cannot have $x_c \mathrel{\pieq_c} f_c$, as otherwise, we have $c \in \calb_{x_c}(f)$; however, by definition  of $x_c$, $\mu(x_c)=\emptyset$, contradicting $(f,\mu)$ is a cutoff equilibrium. Thus $f_c \mathrel{\pi_c} x_c$. Since $x_c$ is eligible for $c$, $f_c \in I$. Furthermore, $f_c \in \mu(I)$, since $c \in \calb_{f_c}(f)$ and $(f,\mu)$ is a cutoff equilibrium. Therefore, $f_c \in \{j \in \mu(I) : j \mathrel{\pi_c} x_c\}$. Since $\unf^\mu_c = \min_{\pi_c} \{j \in \mu(I) : j \mathrel{\pi_c} x_c\}$, we have $f_c \mathrel{\pieq_c} \unf^\mu_c$, contradicting $\unf^\mu_c \mathrel{\pi_c} f_c$. Therefore, such a category $c$ cannot exist, and hence, $\unf^\mu$ is the minimum equilibrium cutoff vector supporting $\mu$.  $\diamond$ \medskip

\noindent{}\textbf{Claim 3.} For any given two cutoff equilibria  $(f,\mu) $ and $(g,\mu)$ such that $f_c \mathrel{\pieq_c} g_c$ for every $c\ \in \calc$,  the pair $(h,\mu)$ is also a cutoff equilibrium where  $h \in \calf$ satisfies  for every $c\ \in \calc$,
$f_c \mathrel{\pieq_c} h_c \mathrel{\pieq_c} g_c$.\medskip

\noindent{}\emph{Proof.} We can obtain cutoff vector $h$ from  $g$  after a sequence of repeated applications of the following operation: Change the cutoffs of one of the categories $c\in \calc$ of an input vector $f' \in \calf$ so that its cutoff \emph{increases} by one patient and gets closer to $h_c$ than $f'_c$. We start with $f'=g$ to the sequence.  We show that each iteration of this operation results with a new equilibrium cutoff vector $g'$ supporting $\mu$ and we use this $g'$  as the input of the next iteration of the operation. Since the outcome vector gets closer to $h$ at each step, the last cutoff vector of the sequence is $h$ by finiteness of the patient set, and thus, $(h,\mu)$ is a cutoff equilibrium:

Suppose $c'\in \calc$ is such that $h_{c'} \mathrel{\pi_{c'}} g_{c'}$.  We prove that for cutoff vector $g'\in \calf$ such that $g'_{c}=g_{c}$ for every $c \in \calc\setminus \{c'\}$ and $g'_{c'}=\min_{\pi_{c'}}\{i \in I : i \mathrel{\pi_{c'}} g_{c'}\}$, $(g',\mu)$ is a cutoff equilibrium. It is straightforward to show that $g' \in \calf$.
Observe also that $\calb_i(g')=\calb_i(g)$ for every $i \in I \setminus\{j\}$ where $j = g_{c'}$. Three cases are possible regarding $j$:
\begin{itemize}
\item If $j \not \in I$: $j = \emptyset$.
\item If $j \in I$ and $\mu(j)=c'$: Category   $c' \in \calb_j(f)$ as $(f,\mu)$ is a cutoff equilibrium. However, $f_{c'} \mathrel{\pieq_{c'}} h_{c'} \mathrel{\pi_{c'}} j=g_{c'}$, contradicting that  $c' \in \calb_j(f)$. Therefore, this case cannot happen.
\item If $j \in I$ and $\mu(j)\not=c'$: Observe that $\mu(j)\not=\emptyset$, as $c' \in \calb_j(g)$ and $(g,\mu)$ is a cutoff equilibrium. Moreover, we have $\mu(j)\in \calb_j(g)$, in turn together with $\mu(j)\not=c'$ implying that $\mu(j) \in  \calb_j(g')$ as $g'_{\mu(j)}=g_{\mu(j)}$.
\end{itemize}
These and the fact that $(g,\mu)$ is a cutoff equilibrium (specifically its Condition 1(b)) show that $\mu(i) \in \calb_i(g')$ for every $i\in I$ such that $\calb_i(g')\not =\emptyset$, proving Condition 1(b) holds in the definition of cutoff equilibrium for $(g',\mu)$.

Since $(g,\mu)$ is a cutoff equilibrium (specifically Conditions 1(a) and 1(b) of the definition) imply that for every $i\in I$, $\mu(i)=\emptyset \iff \calb_i(g)=\emptyset$. Therefore, we have $\mu(i)=\emptyset$ for every $i \in I$ such that $\calb_i(g') =\emptyset$, because $\calb_i(g')\subseteq \calb_i(g)=\emptyset$, where the set inclusion follows from the fact that the cutoffs have weakly increased for each category under $g'$. This and Condition 1(b) that we showed above imply that for all $i \in I$, $\mu(i)\in \calb_i(g')\cup\{\emptyset\}$. Thus, Condition 1(a) in the definition of a cutoff equilibrium is also satisfied by $(g',\mu)$.

We show Condition 2 is also satisfied proving that for every $c \in C$, $g'_c=\emptyset$ if $|\mu^{-1}(c)|<r_c$ to conclude that $(g',\mu)$ is a cutoff equilibrium. Suppose $|\mu^{-1}(c)|<r_c$ for some $c \in \calc$. If $c\not=c'$, then $g'_c=g_c=\emptyset$, where the latter equality follows from $(g,\mu)$ being a cutoff equilibrium (specifically its Condition 2). If $c=c'$, $f_{c'}=\emptyset \mathrel{\pi_{c'}} g_{c'}$, where the first equality follows from $(f,\mu)$ being a cutoff equilibrium (specifically its Condition 2). This contradicts. $g \in \calf$. So $c\not=c'$, completing the proof. $\diamond$
\end{proof}
\bigskip

\subsection{Proof of the Result in Section \ref{sec:char}}
\noindent{}\begin{proof}[Proof of Proposition \ref{thm:characterization}]
\medskip

\noindent \textbf{\textit{Sufficiency}:} We first prove that any DA-induced matching complies with eligibility requirements, is non-wasteful, and respects priorities. Let $\succ \in \calp$ be a preference profile of patients over categories and $\emptyset$. Suppose $\mu \in \calm$ is DA-induced from this preference profile.
\medskip

\noindent
\textit{Compliance with eligibility requirements:} Suppose that $\mu(i)=c$ for some $c \in \calc$. Then $i$ must apply to $c$ in a step of the DA algorithm, and hence, $c \mathrel{\succ_i} \emptyset$. By construction of $\succ_i$, this means $i \mathrel{\pi_c} \emptyset$. Therefore, matching $\mu$ complies with eligibility requirements.

\medskip

\noindent
\textit{Non-wastefulness:} Suppose that $i\mathrel{\pi_c} \emptyset$ and
$\mu(i)=\emptyset$ for some category $c\in \calc$ and patient $i \in I$. By construction of $\succ_i$, $c \mathrel{\succ_i} \emptyset$ because $i$ is eligible for $c$. As patient $i$ is unmatched in $\mu$, she applies to $c$ in some step of the DA algorithm. However,  $c$ rejects $i$ at this or a later step. This means, $c$ should have been holding at least $r_c$ offers from eligible patients at this step. From this step on, $c$ always holds $r_c$ offers and eventually all of its units are assigned: $\abs{\mu^{-1}(c)}=r_c$. Hence, matching  $\mu$ is non-wasteful. \medskip

\noindent
\textit{Respecting priorities:} Suppose that $\mu(i) =c$ and $\mu(i') = \emptyset$ for two patients $i,i' \in I$ and a category $c \in \calc$.
For every category $c'\in \calc$, $\pi_{c'}$ is used to choose eligible patients at every step of the DA algorithm.  Therefore, $\mu(i) =c$ implies  $i \mathrel{\pi_c} \emptyset$. Since $\mu(i')=\emptyset$, then it must be either
because $\emptyset \mathrel{\pi_c} i'$ or because $i \mathrel{\pi_c} i'$. In the first
case, we get $i \mathrel{\pi_c} i'$ as well because $\mathrel{\pi_c}$ is transitive.
Therefore, matching $\mu$ respects priorities.
\medskip

\noindent
\textbf{\textit{Necessity:}} We now prove that any matching $\mu \in \calm$ with the three stated properties
is DA-induced from some preference profile. We construct a candidate  preference profile $\succ \in \calp$ as follows:
\begin{itemize}

\item Consider a patient $i\in \mu^{-1}(c)$ where $c\in \calc$.
Since $\mu$ complies with eligibility requirements, $i$ must be eligible for category $c$. Let $i$
rank category $c$ first in $\succ_i$. The rest of the ranking in $\succ_i$ is arbitrary as long as all eligible categories are ranked above the empty set.

\item Consider an unmatched patient $i\in \mu^{-1}(\emptyset)$. Let $i$ rank categories
in any order in $\succ_i$ such that only eligible categories are ranked above the empty set.
\end{itemize}

We now show that $\mu$ is DA-induced from preference profile $\succ$. In the induced DA algorithm under $\succ$, for every category $c'\in \calc$, patients
in $\mu^{-1}(c')$ apply to category $c'$ first. Every unmatched patient $j \in \mu^{-1}(\emptyset)$ applies to her
first-ranked eligible category according to $\succ_j$, if there is any. Suppose $c\in \calc$ is this category. Since $\mu$ respects priorities, $j$ has a lower priority than any patient in $\mu^{-1}(c)$, who also applied to $c$ in Step 1. Furthermore, since $\mu$ is non-wasteful, $\abs{\mu^{-1}(c)}=r_c$ (as there are unmatched eligible patients for this category, for example $j$).
Therefore, all unmatched patients in $\mu$ are rejected at the first step of the DA algorithm. Moreover, for every category $c' \in \calc$, all patients in $\mu^{-1}(c')$ are tentatively accepted by category $c'$ at the end of Step 1.

Each unmatched patient in $j \in \mu^{-1}(\emptyset)$ continues to apply according to $\succ_j$ to the other categories at which
she is eligible. Since $\mu$ respects priorities and is non-wasteful,
she is rejected from all categories for which she is eligible one at a time, because each of these categories $c \in \calc$ continues to tentatively hold patients $\mu^{-1}(c)$ from Step 1 who all have higher priority than $j$ according to $\pi_c$, as $\mu$ respects priorities. Moreover, by non-wastefulness of $\mu$, $\abs{\mu^{-1}(c)}=r_c$, as there are unmatched eligible patients (for example $j$) under $\mu$.

As a result, when the algorithm stops,
the outcome is such that, for every category $c'\in \calc$, all patients in $\mu^{-1}(c')$ are
matched with $c'$. Moreover, every patient in $\mu^{-1}(\emptyset)$ remains unmatched at the end. Therefore, $\mu$ is DA-induced from the constructed
patient preferences $\succ$.
\end{proof}
\bigskip

\subsection{Proofs of the Results in Section \ref{sec:sequential}}

\noindent{}\begin{proof}[Proof of Proposition \ref{prop:seq.equiv}]
Let $\triangleright \in \Delta$ be a precedence order and $\varphi_\triangleright$ be the
associated sequential reserve matching. We show that $\varphi_\triangleright $  is DA-induced from  preference profile $\succ^\triangleright=\left(\succ^{\triangleright}_{i}\right)_{i \in I}$.

For every patient $i\in I$, consider another strict preference relation
$\succ'_i$ such that all categories are ranked above the empty set and,
furthermore, for any $c,c'\in \calc$,
\[c\mathrel{\succ'_i}c' \iff c\mathrel{\triangleright}c'.\]

Note that the relative ranking of two categories for which $i$ is eligible is the same
between $\succ^\triangleright_i$ and $\succ'_i$.

We use an equivalent version of the DA algorithm as the one given in the text. Consider a Step $k$: Each patient $i$ who is not tentatively accepted currently by a category applies to the best category that has not rejected her yet according to $\succ'_i$. Suppose
			that $I_c^k$ is the union of the set of patients who were tentatively
            assigned to category $c$ in Step $k-1$ and the set of patients who
            just proposed to category $c$. Category $c$ tentatively assigns eligible patients
            in $I_c^k$ with the highest priority according to $\pi_c$ until all
            patients in $I_c^k$ are chosen or all $r_{c}$ units are allocated, whichever
            comes first, and permanently rejects the rest.

Since for any category $c \in \calc$ and any patient $i\in I$ who is ineligible for category $c$, $\emptyset \mathrel{\pi_c} c$, the
outcome of the DA algorithm when the preference profile is $\succ^\triangleright$
and $\succ'=(\succ'_i)_{i\in I}$ are the same.

Furthermore, when the preference
profile is $\succ'$, the DA algorithm works exactly like the sequential reserve procedure that is used to construct $\varphi_\triangleright$. We show this by induction. Suppose $\triangleright$ orders categories as $c_1 \mathrel{\triangleright} c_2 \mathrel{\triangleright} \hdots \mathrel{\triangleright}c_{|\calc|}$.
As the inductive assumption, for $k>0$, suppose for categories $c_1,\hdots,c_{k-1}$,  the tentative matches  at the end of Step $k-1$ and final matches at the end  under the DA algorithm from $\succ'$ are identical to their matches in sequential reserve matching $\varphi_\triangleright$.

We next consider Step $k$ of the DA algorithm from $\succ'$. Only patients who are rejected from category $c_{k-1}$ apply in Step $k$ of the DA algorithm and they all apply to category $c_k$. Then $c_k$ uses its priority order $\pi_{c_k}$ to tentatively accept the $r_{c_k}$ highest-priority eligible applicants (and if there are less than $r_{c_k}$ eligible applicants, all eligible applicants), and rejects the rest. Observe that since every patient who is not tentatively accepted by a category
$c_1,\hdots,c_{k-1}$ applied to this category in Step $k$,  none of these patients will ever apply to it again; and by the inductive assumption no patient who is tentatively accepted in categories $c_1,\hdots,c_{k-1}$ will ever be rejected, and thus, they will never apply to $c_k$, either. Thus, the tentative acceptances by $c_k$ will become permanent at the end of the DA algorithm. Moreover, this step is identical to Step $k$ of the sequential reserve procedure under precedence order $\triangleright$ and the same patients are matched with category $c_k$ in $\varphi_\triangleright$. This ends the induction.

Therefore, we conclude that $\varphi_\triangleright$ is DA-induced from patient preference profile  $\succ^\triangleright$.
\end{proof}\bigskip

\noindent{}\begin{proof}[Proof of Proposition \ref{prop:seq-reserve-comp-stat-cutoffs}]
  Let $J^\tr,J^{\trp} \subseteq I$ be the sets of patients remaining just before category $c$ is processed under the sequential reserve matching procedure induced by $\tr$ and $\trp$, respectively. Since $c$ is processed earlier under $\trp$ and every other category preceding $c$ and $c'$ under $\tr$ and $\trp$ are ordered in the same manner order,  $J^{\tr} \subseteq  J^{\trp}$.  Two cases are possible:
  \begin{enumerate}
  \item If $|\vp^{-1}(c)|<r_c$: Then $J^{\tr} \subseteq  J^{\trp}$ implies $|\v^{-1}(c)|<r_c$. Therefore, by Equation (\ref{eq:max-cutoff}), $\ovf_c^{\vp}=\emptyset=\ovf_c^{\v}$.

  \item If $|\vp^{-1}(c)|=r_c$: Then $J^{\tr} \subseteq  J^{\trp}$ implies, $$\ovf_c^{\vp}=\min_{\pi_c} \vp^{-1}(c) \mathrel{\pieq_c} \min_{\pi_c} \v^{-1}(c),$$  where the first equality follows by Equation (\ref{eq:max-cutoff}). By the same equation, $\ovf_c^{\v} \in \{\emptyset,\min_{\pi_c} \v^{-1}(c)\}$ and by the definition of a cutoff vector, $\ovf_c^{\vp} \mathrel{\pieq_c} \emptyset$. Thus,   $$\ovf_c^{\vp}\mathrel{\pieq_c}\ovf_c^{\v}.$$
  \end{enumerate}
  \end{proof}\bigskip

\subsection{Proof of Result in Section \ref{sec:baseline}} \label{app:proofs3} \bigskip 

In this subsection,  we first show some lemmas that we will use in the proof of Proposition
\ref{prop:comp-stat-soft-5}, the only result in Section \ref{sec:baseline}.

Fix a soft reserve system induced by the baseline priority order $\pi$. Suppose each patient is a beneficiary of at most one preferential treatment category.

First, we introduce some concepts.

We introduce function $\tau:I \rightarrow (\calc\setminus\{u\}) \cup \{\emptyset\}$ to denote the preferential treatment category that a patient is beneficiary of, if there is such a category. That is, for any patient $i\in I$, if $i\in I_c$ for some $c \in \calc\setminus\{u\}$, then $\tau(i)=c$, and  if $i \in I_g$, i.e., $i$ is a general-community patient, then $\tau(i)=\emptyset$.

For a category $c^*\in \calc$, a set of patients $\tilde I \subseteq I$, and a patient $i\in \tilde I$,
let $\rank(i;\tilde I,\pi_{c^*})$ denote the rank of $i$ among patients in $\tilde I$ according to $\pi_{c^*}$.

We consider \emph{incomplete} orders of precedence. For a given subset of categories
$\calc^* \subseteq \calc$, we define an \textbf{order of precedence on $\calc^*$} as a linear order on $\calc^*$.
Let $\Delta(\calc^*)$ be the set of orders of precedence on $\calc^*$.

We extend the definition of sequential reserve matchings to cover incomplete precedence orders and match a subset of patients $\tilde{I}\subseteq I$ as follows: A \textbf{sequential reserve matching induced by $\tr \in\Delta(\calc^*)$ over $\tilde{I}$} is the outcome of the sequential
reserve procedure which processes only the categories in $\calc^*$ in the order of $\tr$ to match only the patients in $\tilde{I}$ and leaves all
categories in $\calc \setminus \calc^*$ unmatched and patients in $I\setminus \tilde I$ unmatched. Let $\vIR$ denote this matching.

\begin{lemma}\label{lem:2cat}
Suppose that $\tilde I\subseteq I$, $c\in \calc \setminus \{u\}$, and $\tr,\trp \in \Delta(\{u,c\})$ are such that
\begin{itemize}
\item $\tr$ is given as $u \tr c$,
\item $\trp$ is given as $c \trp u$,
\item $I(2)=\tilde I \setminus \vI(\tilde I)$ where $\vI=\vIR$,
\item $I'(2)=\tilde I\setminus \vpI(\tilde I)$ where $\vpI=\vpIR$, and
\item $\vI(\tilde I_c)\subsetneq \tilde I_c$.
\end{itemize}
Then the following results hold:
\begin{enumerate}
\item $|I(2)\setminus I'(2)|=|I'(2)\setminus I(2)|$,
\item $I'(2)\setminus I(2)\subseteq \tilde I_c$,
\item $I(2)\setminus I'(2)\subseteq \tilde I \setminus \tilde I_c$,
\item if $i\in I(2)\setminus I'(2)$ and $i'\in I'(2)$, then $i \mathrel{\pi} i'$, and
\item if $i'\in I'(2)\setminus I(2)$ and $i \in I(2)\cap I_c$, then $i' \mathrel{\pi} i$.
\end{enumerate}
\end{lemma}

\noindent{}\begin{proof}[Proof of Lemma \ref{lem:2cat}]
The first statement in Lemma \ref{lem:2cat} holds because under soft reserves
every patient is eligible for every category, which implies that
$|\vI(\tilde I)|=|\vpI(\tilde I)|$.
As a result, $|\vI(\tilde I)\setminus \vpI(\tilde I)|=|\vpI(\tilde I)\setminus \vI(\tilde I)|$, which is equivalent to
$|I'(2)\setminus I(2)|=|I(2)\setminus I'(2)|$ since $\vI(\tilde I) \setminus \vpI(\tilde I)=I'(2)\setminus I(2)$
and $\vpI(\tilde I) \setminus \vI(\tilde I)=I(2)\setminus I'(2)$.

The second statement in Lemma \ref{lem:2cat} holds because if $i\in \vI^{-1}(u)$,
then $\rank(i;\tilde I,\pi)\leq r_u$. Therefore, $i\in \vpI(\tilde I)$.
Furthermore, every $i\in \vI^{-1}(c)$ is a category-$c$ patient since there exists $j\in \tilde I_c$
such that $j \notin \vI(\tilde I)$. As a result, we get
\[\tilde I_c \supseteq \vI^{-1}(c) \supseteq \vI(\tilde I) \setminus \vpI(\tilde I) = I'(2) \setminus I(2).\]

To prove the third statement in Lemma \ref{lem:2cat}, suppose for contradiction that there
exists $i\in I(2)\setminus I'(2)$
such that $i\in \tilde I_c$. Therefore, $i\in \vpI(\tilde I)\setminus \vI(\tilde I)=I(2)\setminus I'(2)$. By the
first statement in Lemma \ref{lem:2cat}, $|I'(2)\setminus I(2)|=|I(2)\setminus I'(2)| \geq 1$
because $I(2)\setminus I'(2)$ has at least one patient. By the second
statement in Lemma \ref{lem:2cat}, $I'(2)\setminus I(2)\subseteq \tilde I_c$. Therefore, there exists
$i'\in I'(2)\setminus I(2)=\vI(\tilde I)\setminus \vpI(\tilde I)$
such that $i'\in \tilde I_c$. Since $i\in \vpI(\tilde I)$, $i'\notin \vpI(\tilde I)$, and $\tau(i)=\tau(i')$, we get
\[i \mathrel{\pi} i'.\]
Likewise, $i'\in \vI(\tilde I)$, $i\notin \vI(\tilde I)$, and $\tau(i)=\tau(i')$ imply
\[i' \mathrel{\pi} i.\]
The two displayed relations above contradict each other.

The fourth statement in Lemma \ref{lem:2cat} is true because for every
$i\in I(2)\setminus I'(2)=\vpI(\tilde I)\setminus \vI(\tilde I)$ we know that
$i\notin \tilde I_c$ by the third statement in Lemma \ref{lem:2cat}. Since $\vI(\tilde I_c)\subsetneq \tilde I_c$, there are at least $r_{c}$ patients in $\tilde I_c$. Therefore, $\vpI^{-1}(c)\subseteq \tilde I_c$, which implies that
$i\in \vpI^{-1}(u)$. Since
$i'\in I'(2)$ is equivalent to $i'\notin \vpI(\tilde I)$, we get $i\mathrel{\pi} i'$.

The fifth statement in Lemma \ref{lem:2cat} follows from $i,i'\in \tilde I_c$, $i'\in \vI(\tilde I)$, and $i \notin \vI(\tilde I)$.
\end{proof}

\begin{lemma}\label{lem:2catalt}
Suppose that $c,c'\in \calc \setminus \{u\}$ are
different categories. Let $\tilde I \subseteq I$ and $\tr,\trp \in \Delta(\{c,c'\})$ be such that
\begin{itemize}
\item $\tr$ is given as $c' \tr c$,
\item $\trp$ is given as $c \trp c'$,
\item $I(2)=\tilde I\setminus \vI(\tilde I)$ where $\vI=\vIR$,
\item $I'(2)=\tilde I\setminus \vpI(\tilde I)$ where $\vpI=\vpIR$, and
\item $\vI(\tilde I_c)\subsetneq \tilde I_c$.
\end{itemize}
Then the following results hold:
\begin{enumerate}
\item $|I(2)\setminus I'(2)|=|I'(2)\setminus I(2)|$,
\item $I'(2)\setminus I(2)\subseteq \tilde I_c$,
\item $I(2)\setminus I'(2)\subseteq \tilde I\setminus \tilde I_c$,
\item if $i\in I(2)\setminus I'(2)$ and $i'\in I'(2)$, then $i \mathrel{\pi} i'$, and
\item if $i'\in I'(2)\setminus I(2)$ and $i \in I(2)\cap \tilde I_c$, then $i' \mathrel{\pi} i$.
\end{enumerate}
\end{lemma}

\noindent{}\begin{proof}[Proof of Lemma \ref{lem:2catalt}]
If $|\tilde I_{c'}|\geq r_{c'}$, then $\vI(\tilde I)=\vpI(\tilde I)$ and, therefore, $I(2)=I'(2)$. Then
all the statements in Lemma \ref{lem:2catalt} hold trivially. Suppose that $|\tilde I_{c'}|<r_{c'}$
for the rest of the proof.

The first statement in Lemma \ref{lem:2catalt} follows as in the proof of the first statement in Lemma \ref{lem:2cat}.

The second statement in Lemma \ref{lem:2catalt} holds because if $i\in \vI^{-1}(c')$ and $i\in \tilde I_{c'}$, then
$i\in \vpI(\tilde I)$ since $|\tilde I_{c'}|<r_{c'}$. If $i\in \vI^{-1}(c')$ and $i\notin \tilde I_{c'}$, then
$\rank(i;\tilde I\setminus \tilde I_{c'},\pi)\leq r_{c'}-|\tilde I_{c'}|$.
As a result $i\in \vpI(\tilde I)$. These two statements imply that $\vI^{-1}(c')\subseteq \vpI(\tilde I)$.
Furthermore, every $i\in \vI^{-1}(c)$ is a category-$c$ patient since there exists $i\in \tilde I_c$
such that $i\notin \vI(\tilde I)$. As a result, we get that
\[I'(2) \setminus I(2)=\vI(\tilde I) \setminus \vpI(\tilde I) = \vI^{-1}(c) \setminus \vpI(\tilde I) \subseteq \vI^{-1}(c) \subseteq \tilde I_c.\]

The proof of the third statement in Lemma \ref{lem:2catalt} is the same as the proof of the
third statement in Lemma \ref{lem:2cat}.

The fourth statement in Lemma \ref{lem:2catalt} is true because for every
$i\in I(2)\setminus I'(2)=\vpI(\tilde I)\setminus \vI(\tilde I)$ we know that
$i\notin \tilde I_c$ by the third statement in Lemma \ref{lem:2catalt}. Moreover, $\vpI^{-1}(c) \subseteq \tilde I_c$, as there exists $j \in \tilde I_c$ such that $j\not\in\vI(\tilde I)$, which implies that there are at least $r_c$ category-$c$ patients. This implies $i\in \vpI^{-1}(c')$.
Furthermore, $i\notin \tilde I_{c'}$ because $\tilde I_{c'}\subseteq \vI(\tilde I)$, which follows from
$|\tilde I_{c'}|<r_{c'}$.  Consider
$i'\in I'(2)$. Then $i' \notin \vpI(\tilde I)$, which implies that
$i\mathrel{\pi} i'$ because $\vpI(i)=c'$, $\tau(i)\neq c'$, and $\vpI(i')=\emptyset$.

The proof of the fifth statement in Lemma \ref{lem:2catalt} is the same as the proof of the fifth statement in Lemma \ref{lem:2cat}.
\end{proof}

\begin{lemma}\label{lem:3cat}
Suppose that $c \in \calc \setminus \{u\}$ and $c',c^*\in \calc \setminus \{c\}$
are different categories. Let $\tilde I \subseteq I$ and $\tr,\trp \in \Delta(\{c,c',c^*\})$ be such that
\begin{itemize}
\item $\tr$ is given as $c' \tr c \tr c^*$,
\item $\trp$ is given as $c \trp c' \trp c^*$,
\item $I(3)=\tilde I\setminus \vI(\tilde I)$ where $\vI=\vIR$,
\item $I'(3)=\tilde I\setminus \vpI(\tilde I)$ where $\vpI=\vpIR$, and
\item $\vI(\tilde I_c) \subsetneq \tilde I_c$.
\end{itemize}
Then the following results hold:
\begin{enumerate}
\item $|I(3)\setminus I'(3)|=|I'(3)\setminus I(3)|$,
\item $I'(3)\setminus I(3)\subseteq \tilde I_c$,
\item $I(3)\setminus I'(3)\subseteq \tilde I\setminus \tilde I_c$, and
\item If $i'\in I'(3)\setminus I(3)$ and $i \in I(3)\cap \tilde I_c$, then $i' \mathrel{\pi} i$.
\end{enumerate}
\end{lemma}

\noindent{}\begin{proof}[Proof of Lemma \ref{lem:3cat}]
The first statement in Lemma \ref{lem:3cat} follows as in the proof of the
first statement in Lemma \ref{lem:2cat}.
Likewise, the fourth statement in Lemma \ref{lem:3cat} follows as in
the proof of the fifth statement in Lemma \ref{lem:2cat}.

To show the other two statements, we use Lemma \ref{lem:2cat} and Lemma \ref{lem:2catalt}.
Let $\hat{\tr},\hat{\tr}' \in \Delta(\{c,c'\})$ be such that $$\hat{\tr}: \quad c' \mathrel{\hat{\tr}} c \qquad \mbox{and} \qquad \hat{\tr}': \quad c \mathrel{\hat{\tr}'} c'.$$ Let $I(2)=\tilde I \setminus \vh(\tilde I)$ and $I'(2)=\tilde I \setminus \vhp(\tilde I)$. Then $I(3)=I(2) \setminus \vI^{-1}(c^*)$ and
$I'(3)=I'(2) \setminus \vpI^{-1}(c^*)$.

For both precedence orders $\tr$ and $\trp$ under the sequential reserve matching procedure, consider the beginning of the third step, at which category $c^*$ is processed.
For $\tr$, the set of available patients is $I(2)$. For $\trp$, the set of available
patients is $I'(2)$. If $I(2)=I'(2)$, then all the statements hold trivially because in this case we get $I(3)=I'(3)$.
Therefore, assume that $I(2)\neq I'(2)$. For every precedence order, $r_{c^*}$ patients with the highest priority
with respect to $\pi_{c^*}$ are chosen.

We consider each patient chosen under $\tr$
and $\trp$ for category $c^*$ one at a time in sequence with respect to the priority order $\pi_{c^*}$. For both precedence orders there are $r_{c^*}$ patients matched with $c^*$ because
$\vI(\tilde I_c) \subsetneq \tilde I_c$. Let $i_k$ be the $k^{\mbox{\scriptsize th}}$ patient chosen under
$\tr$ for $c^*$ and $i'_k$ be the $k^{\mbox{\scriptsize th}}$ patient chosen under $\trp$ for $c^*$
where $k=1,\ldots,r_{c^*}$.
Let $J_k$ be the set of patients
available when we process $\tr$ for the $k^{\mbox{\scriptsize th}}$ patient and $J'_k$ be the set of patients
available when we process $\trp$ for the $k^{\mbox{\scriptsize th}}$
patient where $k=1,\ldots,r_{c^*}$. For $k=1$, $J_k=I(2)$ and $J'_k=I'(2)$.
By definition, $J_{k+1}=J_k \setminus \{i_k\}$ and $J'_{k+1}=J'_k \setminus \{i'_k\}$.
We show that
\begin{enumerate}
\item[(a)] $J'_k\setminus J_k\subseteq \tilde I_c$,
\item[(b)] $J_k \setminus J'_k \subseteq \tilde I\setminus \tilde I_c$, and
\item[(c)] if $i\in J_k \setminus J'_k$ and $i'\in (J_k\cap J'_k)\setminus \tilde I_{c^*}$, then $i \mathrel{\pi_{c^*}} i'$.
\end{enumerate}
by mathematical induction on $k$. These three claims trivially hold for $k=1$ by Statements 2, 3, and 4 in Claims 1 and 2.

Fix $k$. In the inductive step, assume that  Statements (a), (b),  and (c) hold for $k$. Consider $k+1$. If $J_{k+1}=J'_{k+1}$, then the statements trivially hold. Assume that $J_{k+1}\neq J'_{k+1}$ which implies that
$J_{\ell} \neq J'_{\ell}$ for $\ell=1,\ldots,{k}$. There are four cases depending on which sets $i_k$ and $i'_k$ belong to. We consider each case separately.

\medskip
\noindent
\textbf{Case 1:} $i_k\in J_k\setminus J'_k$ and $i'_k\in J'_k \cap J_k$. Then
\[J'_{k+1}\setminus J_{k+1} = (J'_k \setminus \{i'_k\}) \setminus (J_k \setminus \{i_k\})=J'_k\setminus J_k\]
and
\[J_{k+1}\setminus J'_{k+1}=(J_k \setminus \{i_k\}) \setminus (J'_k \setminus \{i'_k\})=
\big((J_k \setminus J'_k) \setminus \{i_k\}\big) \cup \{i'_k\}.\]

By Statement (a) of the inductive assumption for $k$, $J'_{k+1} \setminus J_{k+1}=J'_k \setminus J_k \subseteq \tilde I_c$. Therefore, Statement (a) holds for $k+1$.

As $J'_k \not=J_k$ and $|J_k|=|J'_k|$ because of the soft-reserves condition, $J'_k \setminus J_k$ has at least one category-$c$ patient. Moreover, this patient has higher priority with respect
to $\pi_{c^*}$ than any other category-$c$ patient in $J'_k \cap J_k$ because
the former is chosen under $\tr$ while the later is not chosen under $\tr$.
Therefore, $i'_k$ cannot be a category-$c$ patient.
As a result, $J_{k+1} \setminus J'_{k+1} \subseteq \tilde I \setminus \tilde I_c$, so Statement (b) holds for $k+1$.

To show Statement (c) for $k+1$, observe that $J_{k+1}\setminus J'_{k+1}=((J_k\setminus J'_k) \setminus \{i_k\}) \cup \{i'_k\}$
and $J_{k+1 }\cap J'_{k+1}=(J_k\cap J'_k) \setminus \{i'_k\})$. Therefore, Statement
(c) for $k+1$ follows from Statement (c) for $k$ and the fact that
$i'_k \mathrel{\pi_{c^*}} i$ for any $i\in J_{k+1} \cap J'_{k+1}$.

\medskip
\noindent
\textbf{Case 2:} $i_k \in J_k\setminus J'_k$ and $i'_k\in J'_k \setminus J_k$. Then,
\[J'_{k+1}\setminus J_{k+1} = (J'_k \setminus \{i'_k\}) \setminus (J_k \setminus \{i_k\})=(J'_k\setminus J_k)
\setminus \{i'_k\}\]
and
\[J_{k+1} \setminus J'_{k+1} = (J_k \setminus \{i_k\}) \setminus (J'_k \setminus \{i'_k\})=
(J_k \setminus J'_k) \setminus \{i_k\}.\]
Therefore, $J'_{k+1}\setminus J_{k+1} \subseteq \tilde I_c$ and $J_{k+1}\setminus J'_{k+1} \subseteq \tilde I\setminus \tilde I_c$ by Statements (a) and (b) for $k$, respectively, implying
Statements (a) and (b) for $k+1$.

To show Statement (c) for $k+1$, observe that
$J_{k+1}\setminus J'_{k+1}=(J_k\setminus J'_k) \setminus \{i_k\}$
and $J_{k+1 }\cap J'_{k+1}=J_k\cap J'_k$. Therefore, Statement
(c) for $k+1$ follows from Statement (c) for $k$ trivially.

\medskip
\noindent
\textbf{Case 3:} $i_k \in J_k\cap J'_k$ and $i'_k \in J'_k \cap J_k$. In this case, $i_k=i'_k$, then
\[J'_{k+1}\setminus J_{k+1} = (J'_k \setminus \{i'_k\}) \setminus (J_k \setminus \{i_k\})=J'_k\setminus J_k\]
and
\[J_{k+1} \setminus J'_{k+1} = (J_k \setminus \{i_k\}) \setminus (J'_k \setminus \{i'_k\})=J_k \setminus J'_k.\]
Therefore, Statements (a) and (b) for $k+1$ follows from the respective statements for $k$.

To show Statement (c) for $k+1$, observe that
$J_{k+1}\setminus J'_{k+1}=J_k\setminus J'_k$
and $J_{k+1 }\cap J'_{k+1}=(J_k\cap J'_k)\setminus \{i_k\}$. Therefore, Statement
(c) for $k+1$ follows from Statement (c) for $k$ trivially.

\medskip
\noindent
\textbf{Case 4:} $i_k \in J_k\cap J'_k$ and $i'_k \in J'_k \setminus J_k$. We argue
that this case is not possible. Since $i'_k \in J'_k \setminus J_k$, $i'_k$ must be
a category-$c$ patient by Statement (a) for $k$. If $c^*=u$, then every patient
in $J_k \setminus J'_k$ has a higher priority with respect to $\pi$ than every
patient in $J_k\cap J'_k$, which cannot happen since $i_k \in J_k\cap J'_k$.
Therefore, $c^*\neq u$. Since $i'_k$ is a category-$c$ patient,
there must not be a category-$c^*$ patient in $J'_k$. By Statement (c) for $k$,
we know that every patient in $J_k\setminus J'_k$ has a higher priority
with respect to $\pi_{c^*}$ than every patient in
$(J_k \cap J'_k)\setminus I_{c^*}=J_k \cap J'_k$.
This is a contradiction to $i_k\in J_k\cap J'_k$.
Therefore, Case 4 is not possible.
\medskip

Since $I(3)=J_{r_{c^*}}$ and $I'(3)=J'_{r_{c^*}}$, Statements 2 and 3 in Lemma \ref{lem:3cat} follow from Statements (a) and (b) above, respectively.
\end{proof}

\begin{lemma}\label{lem:4cat}
Suppose that $c\in \calc \setminus \{u\}$ and $c',c^*,\tilde c \in \calc \setminus \{c\}$ are different categories. Let $\tilde I\subseteq I$ and $\tr,\trp \in \Delta(\{c,c',c^*, \tilde c\})$ be such that
\begin{itemize}
\item $\tr$ is given as $c' \tr c \tr c^* \tr \tilde c$,
\item $\trp$ is given as $c \trp c' \trp c^* \trp \tilde c$,
\item $I(4)=\tilde I\setminus \vI(\tilde I)$ where $\vI=\vIR$,
\item $I'(4)=\tilde I\setminus \vpI(\tilde I)$ where $\vpI=\vpIR$, and
\item $\vI(\tilde I_c) \subsetneq \tilde I_c$.
\end{itemize}
Then the following results hold:
\begin{enumerate}
\item $|I(4) \setminus I'(4)|=|I'(4) \setminus I(4)|$,
\item $I(4)\setminus I'(4) \subseteq \tilde I\setminus \tilde I_c$,
\item if $i'\in I'(4) \setminus I(4)$, $i' \notin \tilde I_c$, and $i\in I(4)$, then
$i' \mathrel{\pi} i$, and
\item if $i' \in I'(4) \setminus I(4)$, $i' \in \tilde I_c$, and $i\in I(4) \cap \tilde I_c$,
then $i' \mathrel{\pi} i$.
\end{enumerate}
\end{lemma}

\noindent{}\begin{proof}[Proof of Lemma \ref{lem:4cat}]
The first statement in Lemma \ref{lem:4cat} follows as in the proof of
the first statement in Lemma \ref{lem:2cat}. Likewise, the fourth statement in Lemma \ref{lem:4cat} follows as in
the proof of the fifth statement in Lemma \ref{lem:2cat}.

To prove the other two statements, we use Lemma \ref{lem:3cat}. Let $\hat{\tr},\hat{\tr}' \in \Delta(\{c,c',c^*\})$ be such that $$\hat{\tr}: \quad c' \mathrel{\hat{\tr}} c
\mathrel{\hat{\tr}} c^* \qquad \mbox{and} \qquad \hat{\tr}': \quad c \mathrel{\hat{\tr}'} c' \mathrel{\hat{\tr}'} c^*.$$ Let $I(3)=\tilde I \setminus \vh(\tilde I)$ and $I'(3)=\tilde I \setminus \vhp(\tilde I)$. Then $I(4)=I(3) \setminus \vI^{-1}(\tilde c)$ and
$I'(4)=I'(3) \setminus \vpI^{-1}(\tilde c)$.

For both precedence orders $\tr$ and $\trp$ under the sequential reserve matching procedure, consider the beginning of the fourth step, at which category $\tilde c$
is processed. For $\tr$, the set of available patients is $I(3)$.
For $\trp$, the set of available
patients is $I'(3)$. If $I(3)=I'(3)$, then $I(4)=I'(4)$ which
implies all the statements in Lemma \ref{lem:4cat}.
Therefore, assume that $I(3)\neq I'(3)$. For every precedence order,
$r_{\tilde c}$ patients with the highest priority
with respect to $\pi_{\tilde c}$ are chosen.

We consider each patient chosen under $\tr$
and $\trp$ for category $\tilde c$ one at a time in sequence with respect to the priority order $\pi_{\tilde c}$. Since
$\vI(\tilde I_c) \subsetneq \tilde I_c$, $r_{\tilde c}$ patients are matched with $\tilde c$ under both precedence
orders. Let $i_k$ be the $k^{\mbox{\scriptsize th}}$ patient chosen under
$\tr$ for $\tilde c$ and $i'_k$ be the $k^{\mbox{\scriptsize th}}$ patient chosen under $\trp$ for $\tilde c$ where $k=1,\ldots,r_{\tilde c}$.
Let $J_k$ be the set of patients
available when we process $\tr$ for the $k^{\mbox{\scriptsize th}}$ patient and $J'_k$ be the set of patients
available when we process $\trp$ for the $k^{\mbox{\scriptsize th}}$
patient where $k=1,\ldots,r_{\tilde c}$. For $k=1$, $J_k=I(3)$ and $J'_k=I'(3)$.
By definition, $J_{k+1}=J_k \setminus \{i_k\}$ and $J'_{k+1}=J'_k \setminus \{i'_k\}$.

We show that
\begin{enumerate}
\item[(a)] $J_k\setminus J'_k\subseteq \tilde I\setminus \tilde I_c$,
\item[(b)] if $\tilde c=u$, $i'\in J'_k\setminus J_k$, $\tau(i')\neq c$, and $i\in J_k$, then $i'\mathrel{\pi} i$,
\item[(c)] if $\tilde c \neq u$, $i'\in J'_k\setminus J_k$, $\tau(i')\neq c$, and $i\in J_k$, then $(J_k\cup J'_k) \cap I_{\tilde c}=\emptyset$ and $i'\mathrel{\pi} i$
\item[(d)] if $\tilde c \neq u$, then $(J'_k\setminus J_k) \cap I_{\tilde c}=\emptyset$.
\end{enumerate}
by mathematical induction on $k$. These three claims trivially hold for $k=1$ by
Statements 2 and 3 in Lemma \ref{lem:3cat}. 

Fix $k$. In the inductive step, assume that  Statements (a), (b), (c), and (d) hold for $k$.
Consider $k+1$. If $J_{k+1}=J'_{k+1}$, then the statements trivially hold. Assume that $J_{k+1}\neq J'_{k+1}$ which implies that
$J_{\ell} \neq J'_{\ell}$ for $\ell=1,\ldots,{k}$. There are four cases depending on which sets $i_k$ and $i'_k$ belong to.
We consider each case separately.

\medskip
\noindent
\textbf{Case 1:} $i_k\in J_k\setminus J'_k$ and $i'_k\in J'_k \cap J_k$. When
$|\{i'\in J'_k \setminus J_k:\tau(i') \neq c\}|\geq 1$, $i'_k\in J'_k\setminus J_k$ by
Statements (b) and (c) for $k$. Therefore, $|\{i'\in J'_k \setminus J_k:\tau(i') \neq c\}|= 0$. Furthermore,
\[J'_{k+1}\setminus J_{k+1} = (J'_k \setminus \{i'_k\}) \setminus (J_k \setminus \{i_k\})=J'_k\setminus J_k\]
and
\[J_{k+1}\setminus J'_{k+1}=(J_k \setminus \{i_k\}) \setminus (J'_k \setminus \{i'_k\})=
((J_k \setminus J'_k) \setminus \{i_k\}) \cup \{i'_k\}.\]

Since $|J'_k|=|J_k|$ and $J'_k\neq J_k$, we get
$|J'_k\setminus J_k|\geq 1$. Therefore, $J'_k\setminus J_k$
has at least one category-$c$ patient because $|\{i'\in J'_k \setminus J_k:\tau(i')\neq c\}|= 0$. Moreover, this patient has higher priority with respect to $\pi_{\tilde c}$ than any other category-$c$ patient in $J'_k \cap J_k$ because the former patient is chosen under $\tr$ and the latter is not, so $i'_k$ cannot be category $c$. Therefore, Statement (a) for $k+1$ follows from Statement (a) for $k$. Statements (b) and (c) trivially hold for $k+1$ as well because
$$\{i'\in J'_{k+1} \setminus J_{k+1}:\tau(i')\neq c\}=
\{i'\in J'_k \setminus J_k:\tau(i')\neq c\}= \emptyset.$$
Statement (d) for $k+1$ follows from Statement (d) for $k$.

\medskip
\noindent
\textbf{Case 2:} $i_k \in J_k\setminus J'_k$ and $i'_k\in J'_k \setminus J_k$. Then
\[J'_{k+1}\setminus J_{k+1} = (J'_k \setminus \{i'_k\}) \setminus (J_k \setminus \{i_k\})=(J'_k\setminus J_k)
\setminus \{i'_k\}\]
and
\[J_{k+1} \setminus J'_{k+1} = (J_k \setminus \{i_k\}) \setminus (J'_k \setminus \{i'_k\})=
(J_k \setminus J'_k) \setminus \{i_k\}.\]
Since $J_{k+1} \setminus J'_{k+1}\subseteq J_k\setminus J'_k$, Statement (a) for $k+1$ follows from Statement (a) for $k$.
Likewise, Statements (b), (c), and (d) for $k+1$ follow from
the corresponding statements for $k$ because
$J'_{k+1}\setminus J_{k+1}\subseteq J'_k \setminus J_k$,
$J'_{k+1} \cup J_{k+1} \subseteq J'_{k} \cup J_k$, and $J_{k+1}\subseteq J_{k}$.

\medskip
\noindent
\textbf{Case 3:} $i_k \in J_k\cap J'_k$ and $i'_k \in J'_k \cap J_k$. In this case, $i_k=i'_k$, then
\[J'_{k+1}\setminus J_{k+1} = (J'_k \setminus \{i'_k\}) \setminus (J_k \setminus \{i_k\})=J'_k\setminus J_k\]
and
\[J_{k+1} \setminus J'_{k+1} = (J_k \setminus \{i_k\}) \setminus (J'_k \setminus \{i'_k\})=J_k \setminus J'_k.\]
In this case, Statements (a), (b), (c), and (d) for $k+1$ follow from their respective
statements for $k$.

\medskip
\noindent
\textbf{Case 4:} $i_k \in J_k\cap J'_k$ and $i'_k \in J'_k \setminus J_k$. In this case,
\[J'_{k+1}\setminus J_{k+1} = (J'_k \setminus \{i'_k\}) \setminus (J_k \setminus \{i_k\})=((J'_k\setminus J_k)\setminus \{i'_k\}) \cup \{i_k\}\]
and
\[J_{k+1} \setminus J'_{k+1} = (J_k \setminus \{i_k\}) \setminus (J'_k \setminus \{i'_k\})=J_k \setminus J'_k.\]

Statement (a) for $k+1$ follows from Statement (a)
for $k$ trivially.

Statement (b) for $k+1$ follows from $i_k \mathrel{\pi} i$ for any $i\in J_{k+1}$ and also from Statement (b) for $k$.

To show Statement (d) for $k+1$, suppose that $\tilde c\neq u$. Then by
Statement (d) for $k$, $J'_k\setminus J_k$ does not have a category-$\tilde c$ patient. Since $i'_k\in J'_k\setminus J_k$, this implies that there are no category-$\tilde c$ patients in $J'_k$.
Therefore, $i_k$ does not have category $\tilde c$. We conclude that
$J'_{k+1}\setminus J_{k+1} = ((J'_k\setminus J_k)\setminus \{i'_k\}) \cup \{i_k\}$ does not have a category $\tilde c$ patient, which is the
Statement (d) for $k+1$.

To show Statement (c) for $k+1$, suppose that
$\tilde c\neq u$,  $i'\in J'_{k+1} \setminus J_{k+1}$,
$\tau(i') \neq c$, and $i\in J_{k+1}$. If $i'\neq i_k$, then
$i'\in J'_k\setminus J_k$ since $J'_{k+1}\setminus J_{k+1} = ((J'_k\setminus J_k)\setminus \{i'_k\}) \cup \{i_k\}$ and Statement (c) for $k+1$
follows from Statement (c) for $k$ because $i\in J_{k+1}
\subseteq J_{k}$.
Otherwise, suppose that $i'=i_k$. By Statement (d) for $k+1$, $i'$ does not
have category $\tilde c$, which implies that there are no category-$\tilde c$ patients in $J_k$; this in turn implies
there are no category-$\tilde c$ patients in $J_{k+1}$ since
$J_{k+1}\subseteq J_k$. Furthermore, by Statement (d)
for $k+1$, there are no category-$\tilde c$ patients in $J'_{k+1}\setminus J_{k+1}$. We conclude that there are no category-$\tilde c$ patients in $J'_{k+1}\cup J_{k+1}$.
Finally, $i' \mathrel{\pi_{\tilde c}} i$ for any $i\in J_{k+1}=J_k \setminus \{i'\}$ and since there are no category-$\tilde c$ patients in $J'_{k+1}\cup J_{k+1}$ we get $i' \mathrel{\pi} i$.
\medskip

Since $I(4)=J_{r_{\tilde c}}$ and $I'(4)=J'_{r_{\tilde c}}$, Statement 2 in Lemma \ref{lem:4cat} follows from Statement (a) and Statement 3 in Lemma \ref{lem:4cat} follows from Statements (b) and (c).
\end{proof}

\begin{lemma}\label{lem:5cat}
Suppose that $c\in \calc \setminus \{u\}$ and $c',c^*,\tilde c, \hat{c} \in \calc \setminus \{c\}$
are different categories. Let $\tilde I\subseteq I$ and $\tr,\trp \in \Delta(\{c,c',c^*, \tilde c, \hat c\})$ be such that
\begin{itemize}
\item $\tr$ be such that $c' \tr c \tr c^* \tr \tilde c \tr \hat{c}$,
\item $\trp$ be such that $c \trp c' \trp c^* \trp \tilde c \trp \hat{c}$,
\item $I(5)=\tilde I\setminus \vI(\tilde I)$ where $\vI=\vIR$,
\item $I'(5)=\tilde I\setminus \vpI(\tilde I)$ where $\vpI=\vpIR$, and
\item $\vI(\tilde I_c) \subsetneq \tilde I_c$.
\end{itemize}
Then the following results hold:
\begin{enumerate}
\item $|I(5) \setminus I'(5)|=|I'(5) \setminus I(5)|$ and
\item $I(5)\setminus I'(5) \subseteq \tilde I\setminus \tilde I_c$.
\end{enumerate}
\end{lemma}

\noindent{}\begin{proof}
The first statement in Lemma \ref{lem:5cat} follows as
in the proof of the first statement in Lemma \ref{lem:2cat}.

To prove the second statement, we use Lemma \ref{lem:4cat}. Let
$\hat{\tr},\hat{\tr}' \in \Delta(\{c,c',c^*,\tilde c\})$ be such that $$\hat{\tr}: \quad c' \mathrel{\hat{\tr}} c
\mathrel{\hat{\tr}} c^* \mathrel{\hat{\tr}} \tilde c \qquad \mbox{and} \qquad \hat{\tr}': \quad c \mathrel{\hat{\tr}'} c' \mathrel{\hat{\tr}'} c^* \mathrel{\hat{\tr}} \tilde c.$$
Let $I(4)=\tilde I \setminus \vh(\tilde I)$ and $I'(4)=\tilde I \setminus \vhp(\tilde I)$. Then $I(5)=I(4) \setminus \vI^{-1}(\tilde c)$ and $I'(5)=I'(4) \setminus \vpI^{-1}(\tilde c)$.

For both precedence orders $\tr$ and $\trp$ under the sequential reserve matching procedure, consider the beginning of the fifth step, at which category $\hat c$
is processed. For $\tr$, the set of available patients is $I(4)$.
For $\trp$, the set of available
patients is $I'(4)$. If $I(4)=I'(4)$, then we get $I(5)=I'(5)$, which implies
all the statements. Therefore, assume that $I(4)\neq I'(4)$. For every
precedence order, $r_{\hat c}$ patients with the highest priority
with respect to $\pi_{\hat c}$ are chosen.

We consider each patient chosen under $\tr$
and $\trp$ for category $\hat c$ one at a time in sequence with respect to the priority order $\pi_{\hat c}$. Since $\vI(\tilde I_c) \subsetneq \tilde I_c$,
$r_{\hat c}$ patients are matched with $\hat c$ under both precedence orders.
Let $i_k$ be the $k^{\mbox{\scriptsize th}}$ patient chosen under
$\tr$ for $\hat c$ and $i'_k$ be the $k^{\mbox{\scriptsize th}}$ patient chosen under $\trp$ for $\hat c$ where $k=1,\ldots,r_{\hat c}$.
Let $J_k$ be the set of patients
available when we process $\tr$ for the $k^{\mbox{\scriptsize th}}$ patient and $J'_k$ be the set of patients
available when we process $\trp$ for the $k^{\mbox{\scriptsize th}}$
patient where $k=1,\ldots,r_{\hat c}$. For $k=1$, $J_k=I(4)$ and $J'_k=I'(4)$.
By definition, $J_{k+1}=J_k \setminus \{i_k\}$ and $J'_{k+1}=J'_k \setminus \{i'_k\}$.
\medskip

We show that
\begin{enumerate}
\item[(a)] $J_k\setminus J'_k\subseteq \tilde I\setminus \tilde I_c$ and
\item[(b)] if $i'\in J'_k \setminus J_k$ and $i\in J_k \cap \tilde I_c$, then
$i' \mathrel{\pi_{\hat c}} i$
\end{enumerate}
by mathematical induction on $k$. These results trivially hold for $k=1$ by Statements 2,
3, and 4 in Lemma \ref{lem:4cat}.

Fix $k$. In the inductive step, assume that  Statements (a) and (b) hold for $k$.
Consider $k+1$. If $J_{k+1}=J'_{k+1}$, then the statements trivially hold. Assume that $J_{k+1}\neq J'_{k+1}$ which implies that
$J_{\ell} \neq J'_{\ell}$ for $\ell=1,\ldots,{k}$.
There are four cases depending on which sets $i_k$ and $i'_k$ belong to.
We consider each case separately.

\medskip
\noindent
\textbf{Case 1:} $i_k\in J_k\setminus J'_k$ and $i'_k\in J'_k \cap J_k$. Then
\[J'_{k+1}\setminus J_{k+1} = (J'_k \setminus \{i'_k\}) \setminus (J_k \setminus \{i_k\})=J'_k\setminus J_k\]
and
\[J_{k+1}\setminus J'_{k+1}=(J_k \setminus \{i_k\}) \setminus (J'_k \setminus \{i'_k\})=
\big((J_k \setminus J'_k) \setminus \{i_k\}\big) \cup \{i'_k\}.\]
If $i'_k\in \tilde I_c$, then we get a contradiction to Statement (b) for $k$. Therefore,
$i'_k\notin \tilde I_c$, which implies that Statement (a) holds for $k+1$
by Statement (a) for $k$ and the second displayed equation. Statement (b) for $k+1$ follows trivially from Statement (b) for $k$, the first displayed equation, and $J_k \supseteq J_{k+1}$.

\medskip
\noindent
\textbf{Case 2:} $i_k \in J_k\setminus J'_k$ and $i'_k\in J'_k \setminus J_k$. Then
\[J'_{k+1}\setminus J_{k+1} = (J'_k \setminus \{i'_k\}) \setminus (J_k \setminus \{i_k\})=(J'_k\setminus J_k)
\setminus \{i'_k\}\]
and
\[J_{k+1} \setminus J'_{k+1} = (J_k \setminus \{i_k\}) \setminus (J'_k \setminus \{i'_k\})=
(J_k \setminus J'_k) \setminus \{i_k\}.\]
In this case, Statements (a) and (b) for $k+1$ follow trivially from the corresponding statements for $k$.

\medskip
\noindent
\textbf{Case 3:} $i_k \in J_k\cap J'_k$ and $i'_k \in J'_k \cap J_k$. In this case, $i_k=i'_k$, then
\[J'_{k+1}\setminus J_{k+1} = (J'_k \setminus \{i'_k\}) \setminus (J_k \setminus \{i_k\})=J'_k\setminus J_k\]
and
\[J_{k+1} \setminus J'_{k+1} = (J_k \setminus \{i_k\}) \setminus (J'_k \setminus \{i'_k\})=J_k \setminus J'_k.\]
In this case, Statements (a) and (b) for $k+1$ follow trivially from the corresponding
statement for $k$.

\medskip
\noindent
\textbf{Case 4:} $i_k \in J_k\cap J'_k$ and $i'_k \in J'_k \setminus J_k$. In this case,
\[J'_{k+1}\setminus J_{k+1} = (J'_k \setminus \{i'_k\}) \setminus (J_k \setminus \{i_k\})=((J'_k\setminus J_k)\setminus \{i'_k\}) \cup \{i_k\}\]
and
\[J_{k+1} \setminus J'_{k+1} = (J_k \setminus \{i_k\}) \setminus (J'_k \setminus \{i'_k\})=J_k \setminus J'_k.\]
Then Statement (a) for $k+1$ follows from Statement (a) for $k$. Furthermore,
$i_k \mathrel{\pi_{\hat c}} i$ for any $i\in J_{k+1}$, which together with
Statement (b) for $k$ imply Statement (b) for $k+1$.

\medskip

Since $I(5)=J_{r_{\hat c}}$ and $I'(5)=J'_{r_{\hat c}}$, Statement 2 in Lemma \ref{lem:5cat} follows from Statement (a).
\end{proof} \smallskip

\noindent{}\begin{proof}[Proof of Proposition \ref{prop:comp-stat-soft-5}] Let $ |\calc| \le 5$.
Let  $\calc^*=\{ c^* \in \calc : c^* \mathrel{\tr} c'  \} $ be the set of categories processed before $c'$ under $\tr$ and before $c$ under $\trp$.
The  orders of  categories in $ \calc^*$ are the same with respect to $\tr$ and $\trp$. Thus, just before category $c'$ is processed under $\tr$ and $c$ is processed under $\trp$, the same patients are matched in both sequential reserve matching procedures. Let $\tilde{I}$ be the set of patients that are available at this point in either procedure.

Let $\th$ be the incomplete precedence order on $\calc \setminus \calc^*$
that processes categories in the same order as in $\tr$. Likewise, let
$\thp$ be the incomplete precedence order on $\calc \setminus \calc^*$
that processes categories in the same order as in $\trp$.

If $\vh(\tilde I_c)=\tilde{I}_c$ then the result is proven. Therefore, assume that $\vh(\tilde I_c) \subsetneq \tilde{I}_c$ in the rest of the proof.
 Let  $k=|\calc\setminus \calc^*|$ be the number of remaining categories.
 \begin{itemize}
     \item If $k=2$, then by Lemmas $\ref{lem:2cat}$ and $\ref{lem:2catalt}$, we obtain $\vhp(\tilde{I}_c) \subseteq \vh(\tilde I_c)$.
     \item If $k=3$, then by Lemma $\ref{lem:3cat}$, we obtain $\vhp(\tilde{I}_c) \subseteq \vh(\tilde{I}_c)$.
     \item If $k=4$, then by Lemma $\ref{lem:4cat}$, we obtain $\vhp(\tilde{I}_c) \subseteq \vh(\tilde{I}_c)$.
     \item If $k=5$, then by Lemma $\ref{lem:5cat}$, we obtain $\vhp(\tilde{I}_c) \subseteq \vh(\tilde{I}_c)$.
 \end{itemize}
These imply that
 $$\vp(I_c)\subseteq\v(I_c)$$ completing the proof.
 \end{proof}
 \bigskip

\subsection{Proofs of Results in Appendix \ref{sec:smart}}  \label{app:proofs2} \bigskip

Before providing proofs for the results regarding the smart reserve matching algorithm, we state an equivalent, descriptive (but exponential-in-time) smart reserve matching algorithm, which uses an iterative sequence of matching sets $\calm_0,\calm_1,\hdots,\calm_{|I|}$ and enables us to prove our results more efficiently. Our algorithm in Appendix \ref{sec:smart} does not require constructing these sets, and hence, it is polynomial in time.

\begin{quote}
  \noindent{}{\bf Descriptive Version of the Smart Reserve Matching Algorithm}

 Fix a parameter  $n \in \{0,1,\hdots,r_u\}$ that represents the number of unreserved units to be processed in
 the beginning of the algorithm.\footnote{For $n=0$, this algorithm gives the meritorious horizontal choice rule
 in \citet{sonmez/yenmez22}.}
 The remaining unreserved units are to be processed at the end of the algorithm.

Fix a baseline priority order $\pi$, and for the ease of description relabel patients so that
\[i_1 \mathrel{\pi} i_2 \mathrel{\pi} \hdots \mathrel{\pi} i_{\abs{I}}.\]

Proceed in two steps.

  \noindent{}\textbf{Step 1:} Iteratively construct  two sequences of patient sets $J^u_0 \subseteq J^u_1 \subseteq \hdots \subseteq J^u_{|I|}$, which determine patients to be matched with the unreserved category $u$ in this step, and $J_0 \subseteq J_1 \subseteq \hdots \subseteq J_{|I|}$, which determine 
  the patients to be matched with preferential treatment categories in $\calc\setminus \{u\}$ that they are beneficiaries of,  and a sequence of sets of matchings  $\calm_0 \supseteq \calm_1 \supseteq \hdots \supseteq \calm_{|I|}$ in $|I|$ substeps.

  Define Step 1.($k$) for any $k \in \{1,2,\hdots,|I|\}$ as follows:

  If $k=1$, let $$J^u_0=\emptyset, \quad  J_0=\emptyset,$$
  and $\calm_0$ be
   the set of all matchings that are maximal in beneficiary assignment; that is $$\calm_0=
   \underset {\nu \in \calm} {\arg \max} \big |\cup_{c \in \calc \setminus\{u\}} (\nu^{-1}(c)\cap I_c) \big |.$$

  If $k>1$, then sets of patients $J^u_{k-1}$ and $J_{k-1}$ and  set of matchings $\calm_{k-1}$ are constructed in the previous substep, Step 1.($k-1$).

  \begin{quote}
  \noindent{}\textbf{Step 1.($k$):} Process patient $i_k$. Three cases are possible:
    \begin{itemize}
      \item If $|J^u_{k-1}|<n$ and there exists a matching in $\calm_{k-1}$ that matches patient $i_k$ with the unreserved category $u$, then define
      $$J^u_k=J^u_{k-1} \cup \{i_k\},\quad J_k=J_{k-1},\quad \mbox{and} \quad \calm_k = \big\{\mu \in  \calm_{k-1} : \mu(i_k)=u\big\}.$$
      \item Otherwise: 
      \begin{itemize}
      	\item If there exists a matching in $\calm_{k-1}$ that matches patient $i_k$ with a preferential treatment category $c \in \calc \setminus \{u\}$ that she is a beneficiary of, that is $i_k \in I_c$,  then  define
      	$$J^u_k=J^u_{k-1},\quad J_k=J_{k-1} \cup \{i_k\},\quad \mbox{and}$$ $$ \calm_k = \big\{\mu \in  \calm_{k-1} : \mu(i_k)\not\in\{\emptyset,u\} \mbox{ and } i_k \in I_{\mu(i_k)}\big\}.$$
      	\item Otherwise, define
      	$$J^u_k=J^u_{k-1},\quad J_k=J_{k-1},\quad \mbox{and} \quad \calm_k = \calm_{k-1} .$$
      \end{itemize}
    \end{itemize}
  \end{quote}

  \noindent{}\textbf{Step 2:} For any  matching $\mu \in \calm_{|I|}$, construct a matching $\sigma \in \calm$ as follows:
  \begin{itemize}
    \item Assign $\mu(i)$ to every patient $i \in J_{|I|} \cup J^u_{|I|}$.     \item 
One at a time iteratively assign the remaining units to the remaining highest priority patient in $I \setminus (J^u_{|I|} \cup J_{|I|})$ who is eligible for the category of the assigned unit in the following order:
    \begin{enumerate}
      \item the remaining units of the preferential treatment categories  in an arbitrary order, and
      \item the remaining units of the unreserved category $u$.
    \end{enumerate}
  \end{itemize}
  \medskip
 \end{quote}

\noindent{}\begin{proof}[Proof of Lemma \ref{lem:smart-samepatients}] 
By the definition of the smart reserve matching algorithm induced by assigning $n$ unreserved units subsequently at the beginning, in Step 1.($k$) for every $k \in \{0,1,\hdots,|I|\}$, and for every
every matching $\mu \in \calm_k$,
\begin{itemize}
\item $\mu(i)=u$ for every $i \in J^u_k$,  and
\item $\mu(i)\not\in \{u,\emptyset\}$ and $i \in I_{\mu(i)}$ for every $i \in J_k$.
\end{itemize}
We show that for any $i \in I \setminus (J^u_{|I|}\cup J_{|I|})$ there is no matching $\mu \in \calm_{|I|}$ such that $\mu(i)\not\in\{u,\emptyset\}$ and $i \in I_{\mu(i)}$. Suppose contrary to the claim that such a patient $i$ and matching $\mu$ exist. Patient $i$ is processed in some Step 1.($k$). 
We have  $i \notin J^u_k \cup J_k \subseteq J^u_{|I|}\cup J_{|I|}$. We have $\mu \in \calm_{k-1}$ as $\calm_{k-1} \supseteq \calm_{|I|}$. Then $|J^u_{k-1}|=n$, as otherwise we can always match $i$ with $u$ even if we cannot match her with a preferential treatment category that she is a beneficiary of when she is processed under a matching that is maximal in beneficiary assignment, contradicting $i \notin  J^u_{|I|}\cup J_{|I|}$. But then as $\mu(i)\not\in \{u,\emptyset\}$ and $i\in I_{\mu(i)}$, we have $\mu \in \calm_k$ and $i \in J_k$, contradicting again $i \notin  J^u_{|I|}\cup J_{|I|}$.

Thus, in Step 2 no patient is matched with a preferential treatment categories that she is a beneficiary of.
These prove $\cup_{c \in \calc\setminus\{u\}} (\sigma^{-1}(c)\cap I_c)$ is the same set regardless of the matching $\sigma\in \cals^n$ we choose.

To prove that $\sigma^{-1}(u)$ is the same for every $\sigma\in \cals^n$, we consider two cases (for Step 2):
\begin{itemize}
\item If we have a soft reserve system:
Then all patients are eligible for the remaining $\max\{0,q-|J^u_{|I|}\cup J_{|I|}|\}$ units. Since we assign the patients in $I\setminus (J^u_{|I|}\cup J_{|I|})$ based on priority according to $\pi$ to the remaining units, we have $\sigma(I)$ is the same patient set regardless of the matching $\sigma \in \cals^n$ we choose.  Since the remaining $r_u-n$ unreserved units are assigned to the lowest priority patients that are matched in any $\sigma \in \cals^n$, if any, and $\mu^{-1}(u)$ is a fixed set regardless of which matching $\mu \in \calm_{|I|}$ we choose (as we proved above), then $\sigma^{-1}(u)$ is the same set regardless of which $\sigma \in \cals^n$ we choose.
 \item If we have a hard reserve system: Then in Step 2 no patient is matched with a unit of a category that she is not a beneficiary of. Thus, the $r_u-n$ remaining unmatched units are assigned to the highest $\pi$-priority patients in  $I\setminus (J^u_{|I|}\cup J_{|I|})$. This concludes proving that each of $\sigma(I)$ and $\sigma^{-1}(u)$ is the same patient set regardless of which $\sigma \in \cals^n$ we choose.
\end{itemize}
\end{proof}\bigskip

\noindent{}\begin{proof}[Proof of Proposition \ref{prop:smart-maximal}] For any $n\in\{0,1,\hdots,r_u\}$, we prove that every smart reserve matching in $\cals^n$ complies with eligibility requirements, is non-wasteful, respects priorities, and is maximal in beneficiary assignment. \medskip

   \noindent{}\emph{Compliance with eligibility requirements:} By construction, no patient is ever matched with a category for which she is not eligible during the procedure. \medskip

  \noindent{}\emph {Non-wastefulness:} Suppose to the contrary   of the claim that there exists some $\sigma \in \cals^n$ that is wasteful. Thus, there exists some category $c \in \calc$ and a patient $i \in I$ such that $\sigma(i)=\emptyset$, $i \mathrel{\pi_c} \emptyset$, and $|\mu^{-1}(c)|<r_c$. Then in Step 2 patient $i$ or another patient should have been matched with $c$ as we assign all remaining units to eligible patients, which is a contradiction.     \bigskip

   \noindent{}\emph{Respect for Priorities:} Let $\sigma \in \cals^n$ be a smart reserve matching.
   Suppose patients $i,j \in I$ are such that $i \mathrel{\pi} j$ and $\sigma(i)=\emptyset$. We need to show  either (i) $\sigma(j)=\emptyset$ or (ii) $i \not \in I_{\sigma(j)}$ and $j \in I_{\sigma(j)}$, which is equivalent to  $j \mathrel{\pi_{\sigma(j)}} i$. Suppose $\sigma(j)\not=\emptyset$. Suppose to the contrary that $i \in I_{\sigma(j)}$ and $j \in I_{\sigma(j)}$. Consider the smart reserve matching procedure with $n$.
     Two cases are possible:  $j \in J^u_{|I|}\cup J_{|I|}$ or not. We show that either case leads to a contradiction, showing that $\sigma$ respects priorities.
    \begin{itemize}
      \item If $j \in J^u_{|I|}\cup J_{|I|}$:
          Consider the matching $\hs $ obtained from $\sigma$   as follows: $\hs (i)=\sigma(j)$, $\hs (j)=\emptyset$, and $\hs (i')=\sigma(i')$ for every $i' \in I\setminus \{i,j\}$. Since $i,j \in I_{\sigma(j)}$, and we match $i$ instead of $j$ with $\sigma(j)$, $\hs $ is a matching that is maximal in beneficiary assignment as well. Since $i \mathrel{\pi} j$, $i$ is processed before $j$ in Step 1. Let $i$ be processed in some Step 1.($k$). Since $\sigma(i)=\emptyset$, $i\not\in J_k \cup J^u_k $.  Then $\hs  \in \calm_{k-1}$ as $\sigma \in  \calm_{k-1}$. Two cases are possible:
          \begin{itemize}
          \item if $\sigma(j)=u$: As $j$ is matched with an unreserved unit in Step 1, then an unreserved unit is still to be allocated in the procedure when $i$ is to be processed in Step 1.($k$) before $j$. We try to match $i$ with an unreserved unit first. Since $\hs \in \calm_{k-1}$ and $\hs(i)=u$  this implies $i \in J^u_k$. This contradicts $i\not\in J^u_k \cup J_{k}$.

          \item if $\sigma(j)\not=u$: Then when $i$ is to be processed in Step 1.($k$), we are trying to match her (i) if it is possible,  with unreserved category $u$ first and if not, with a preferential treatment category that she is a beneficiary of, or (ii) directly with a preferential treatment category that she is a beneficiary of without sacrificing the maximality in beneficiary assignment. However, as $\sigma(i)=\emptyset$ we failed in doing either. Since $i \in I_{\hs(i)}$ and $\hs(i)\not=u$, at least there exists a matching in $\calm_{k-1}$ that would match $i$ with a preferential treatment category that she is a beneficiary of. Hence, this contradicts $i\not\in J^u_k \cup J_{k}$.
          \end{itemize}

        \item If $j \not \in J^u_{|I|}\cup J_{|I|}$: Therefore, $j$ is matched in Step 2 of the smart reserve matching algorithm with $n$ unreserved units processed first. Since we match patients in Step 2 either with the preferential treatment categories that they are eligible but not beneficiary of or with the unreserved category $u$, and we assumed $j \in I_{\sigma(j)}$ then $\sigma(j)=u$. Since $i$ is also available when $j$ is matched, and $i \mathrel{\pi} j$, patient $i$ or another patient who has higher $\pi$-priority than $j$ should have been matched instead of $j$, which is a contradiction.
     \end{itemize} \bigskip

     \noindent{}\emph{Maximality in Beneficiary Assignment:} By construction $\cals^n \subseteq \calm_0$, which is the set of matchings that are maximal  in beneficiary assignment in the smart reserve matching procedure with $n$.
\end{proof} \bigskip

The following lemma and concepts from graph theory will be useful in our next proof. We state the lemma as follows:

\begin{lemma}[\citeauthor{mendelsohn/dulmage:58}  Theorem, 1958] \label{lem:mendelsohn/dulmage} Let $\calm^b$ be the set of matchings that match patients with only preferential treatment categories that they are beneficiaries of and otherwise leave them unmatched. If there is  a matching in $\calm^b$ that matches patients in some $J \subseteq I$  and there is another  matching $\nu \in\calm^b$ then there exists a matching in $\calm^b$  that matches all patients in $J$ and at least $|\nu^{-1}(c)|$ units of each category $c \in \calc \setminus \{u\}$.
\end{lemma}

See for example page 266 of \citet{schrijver:03} for a proof of this result.\bigskip

Let us define $I_\emptyset = \emptyset$ for notational convenience.

We define a \textbf{beneficiary alternating path from $\mu$ to $\nu$} for two matchings in $\mu,\nu \in \calm$ as a non-empty list $A=(i_1,\hdots,i_{\overline{m}})$ of patients such that
\begin{align*}
 \big[ i_1 \not\in I_{\mu(i_1)} \mbox{ or } \mu(i_1)= u\big] & \quad \& \quad \big[i_1 \in I_{\nu(i_1)} \mbox{ and } \nu(i_1)\not=u\big], \\
\big[ i_m \in I_{\mu(i_m)} \mbox{ and } \mu(i_m)=\nu(i_{m-1})\big] & \quad \& \quad  \big[i_m \in I_{\nu(i_m)} \mbox{ and } \nu(i_m)\not=u\big] \quad \mbox{for every } m \in \{2,3,\hdots,{\overline{m}}-1\}, \\
\big[i_{\overline{m}} \in I_{\mu(i_{\overline{m}})} \mbox{ and } \mu(i_{\overline{m}})=\nu(i_{\overline{m}-1})\big] &   \quad \& \quad \big[i_{\overline{m}} \not \in I_{\nu(i_{\overline{m}})} \mbox{ or } \nu(i_{\overline{m}})=u\big].
\end{align*}
A beneficiary alternating path begins with a patient $i_1$ who is not matched with a preferential treatment category that she is a beneficiary of under $\mu$ and ends with a patient $i_{\overline{m}}$ who is not matched with a preferential treatment category that she is a beneficiary of under $\nu$. Everybody else in the path is matched under both matchings with a preferential treatment category that she is a beneficiary of.
We state the following observation, which directly follows from  the finiteness of categories and patients.

\begin{observation} If $\mu$ and $\nu \in \calm$ are two matchings such that  for every $c \in \calc\setminus \{u\}$, $|\mu^{-1}(c)\cap I_c|=|\nu^{-1}(c)\cap I_c|$ and there exists some $i \in \big(\cup_{c\in \calc \setminus \{u\}} \nu^{-1}(c)\cap I_c  \big) \setminus \big( \cup_{c \in \calc \setminus \{u\}} \mu^{-1}(c)\cap I_c \big)$,   then there exists a beneficiary alternating path from $\mu$ to $\nu$ beginning with patient $i$. \label{lem:alternate}
\end{observation}

We are ready to prove our last theorem: \medskip

\noindent{}\begin{proof}[Proof of Theorem  \ref{thm:smart-cutoffs}]
By Proposition \ref{prop:smart-maximal} and Theorem \ref{thm:cutoff}, any $\sigma_0 \in \cals^0$ and $\sigma_{r_u} \in \cals^{r_u}$  are cutoff equilibrium matchings. Let $\mu \in \calm$ be any other cutoff equilibrium matching that is maximal in beneficiary assignment.

We extend the definitions of our concepts to smaller economies: given any $I^* \subseteq I$ and $r^*=(r^*_c)_{c \in \calc}$ such that $r^*_c\le r_c$ for every $c\in \calc$,
all properties and algorithms are redefined for this smaller economy $(I^*,r^*)$ by taking the restriction of the baseline priority order $\pi$ on $I^*$, and denoted using the argument $(I^*,r^*)$ at the end of the notation. For example $\calm(I^*,r^*)$ denotes the set of matchings for $(I^*,r^*)$. \bigskip

\noindent{}\textbf{Proof of $\ovf^{\sigma_{r_u}}_u \mathrel{\pieq}  \ovf^{\mu}_u $}: \medskip

We prove the following claim first.\medskip

\noindent{}{\bf Claim 1.} For any set of patients $I^* \subseteq I$ and any capacity vector $r^*\le r$, suppose matching $\nu \in \calm(I^*,r^*)$ is maximal in beneficiary assignment for $(I^*,r^*)$. Let $\sigma \in \cals^{r^*_u}(I^*,r^*)$ be a  smart reserve matching with all unreserved units processed first. Then
$$|\sigma^{-1}(u) | \ge |\nu^{-1}(u)|.$$ Moreover,  according to the baseline priority order $\pi$, for any $k\in \Big\{1,\hdots,\big|\nu^{-1}(u)\setminus\sigma^{-1}(u)\big|\Big\}$, let $j_k$ be the $k^{\mbox{\scriptsize{th}}}$ highest priority patient in $\sigma^{-1}(u)\setminus\nu^{-1}(u)$ and $j'_k$ be the  $k^{\mbox{\scriptsize{th}}}$ highest priority patient in $\nu^{-1}(u)\setminus \sigma^{-1}(u)$, then
$$j_k \mathrel{\pi} j'_k.$$ 

\noindent{}\emph{Proof.} Suppose to the contrary of the first statement $|\sigma^{-1}(u)| < |\nu^{-1}(u)|$.
Since both $\sigma$ and $\nu$ are maximal in beneficiary assignment for $(I^*,r^*)$, then the exists some patient $i \in (\cup_{c \in \calc}\nu^{-1}(c)\cap I_c)\setminus (\cup_{c \in \calc}\sigma^{-1}(c)\cap I_c)$. This patient is not committed to be matched in Step 1 of the smart reserve matching algorithm with all unreserved units first, despite the fact that there exists at least one available unreserved unit when she  was processed, which is a contradiction. Thus, $|\sigma^{-1}(u) | \ge |\nu^{-1}(u)|$.

For the rest of the proof, we use induction on the cardinality of $I^*$ and
on the magnitude of vector of  category capacity vector $r^*$:
\begin{itemize}

  \item For the base case when $I^*=\emptyset$ and $r^*_c=0$ for every $c\in \calc$, the claim holds trivially.

  \item As the inductive assumption, suppose that for all capacity vectors of categories bounded above by vector $r^*$ and all subsets of $I$ bounded above by cardinality $k^*$,  the claim holds.

  \item Consider a set of patients $I^* \subseteq I$ such that $|I^*|=k^*$ and capacity vector of categories $r^*=(r^*_c)_{c \in \calc}$. Let $\sigma \in \cals^{r_u^*}(I^*,r^*)$ be a smart reserve matching for $(I^*,r^*)$ with all unreserved units processed first  and $\nu\in \calm(I^*,r^*)$ be maximal in beneficiary assignment for $(I^*,r^*)$.  If $\sigma^{-1}(u) \supseteq \nu^{-1}(u)$ then the claim for $(I^*,r^*)$ is trivially true. Thus, suppose not. Then, there exists $j \in \nu^{-1}(u) \setminus  \sigma^{-1}(u)$. Moreover, let $j$ be the highest $\pi$-priority patient in $\nu^{-1}(u) \setminus  \sigma^{-1}(u)$. We have two cases that we consider separately:
  \begin{itemize}

    \item[Case 1.] There is no patient $i \in I^*$ such that $i \mathrel{\pi} j$ and $i \in \sigma^{-1}(u)\setminus \nu^{-1}(u)$:

    We show that this case leads to a contradiction, and hence, it cannot hold.

    When $j$ is processed in Step 1 of the smart reserve matching algorithm with all unreserved units processed first, since $\sigma(j)\not=u$, either\\
    (i) all units of the unreserved category are assigned  under $\sigma$ to patients with higher $\pi$-priority than $j$, or \\(ii) some unreserved category units are still available when $j$ is processed. \\Observe that (i) cannot hold, because it contradicts Case 1. Thus, (ii) holds.

     Since $j$ is the highest $\pi$-priority patient in $\nu^{-1}(u)\setminus \sigma^{-1}(u)$ and since we are in Case 1, for every $i \in I^*$ such that $i \mathrel{\pi} j$, we have \begin{align} \nu(i)=u \iff \sigma(i)=u. \label{eq:cutoffcomp-case1a} \end{align}

      We construct a new matching $\hn \in \calm(I^*,r^*)$ from $\nu$ and fix a patient $i \in I^*$ as follows. We check whether there exists a patient $i \in I^*$ such that \begin{align}i \mathrel{\pi} j,\quad \sigma(i)\not=u,
      \quad \mbox{and} \quad  i \not\in I_{\nu(i)}, \label{eq:cutoffcomp-case1b}\end{align}
      \begin{itemize}

        \item[(a)] If such a patient $i$ does not exist, then let $\hn =\nu$ and $i=j$.

        \item[(b)] If such a patient $i$ exists, then let her be the highest $\pi$-priority patient with the property in Equation \ref{eq:cutoffcomp-case1b}.

        We show that $i \in I_{\sigma(i)}$.  Consider the smart reserve matching algorithm. Since $\nu(j)=u$ and $i \mathrel{\pi}j$, just before $i$ is processed in Step 1, there is still at least one unreserved unit available by  Equation \ref{eq:cutoffcomp-case1a}. Since we are processing all unreserved units first and since $\sigma(i)\not=u$,  it should be the case that we had to match $i$ with a preferential treatment category that she is a beneficiary of.  Thus, $i \in I_{\sigma(i)}$.

        We create a new matching for $(I^*,r^*)$ from $\nu$, which we refer to as $\hn $, such that $\hn $ matches every patient in $I^*$ exactly as under $\nu$ except that $\hn$
         leaves patient $j$ unmatched and matches $i$ with category $u$ instead. Since $\nu$ is maximal in beneficiary assignment for $(I^*,r^*)$, so is $\hn $. 
      \end{itemize}

      So far, we have for every $i' \in I^*$ such that $i' \mathrel{\pi} i$,
      \begin{enumerate}
        \item[1.] $\sigma(i')=u \iff \nu(i')=u$ (by Equation \ref{eq:cutoffcomp-case1a} and $i \mathrel{\pieq} j$),
        \item[2.] $\sigma(i') \in I_{\sigma(i')}$ (an unreserved unit is available before $i$ is processed in Step 1 of the smart reserve matching algorithm with all unreserved units processed first; thus, every patient processed before $i$ is matched if possible, with $u$, and if not possible,  with a preferential treatment category that she is a beneficiary of under the restriction of  maximality in beneficiary assignment),  and
        \item[3.]  $\hn (i') \in I_{\hn (i')}$ (by definition of $i$ as the highest $\pi$-priority patient satisfying Equation \ref{eq:cutoffcomp-case1b}).
      \end{enumerate}
      We also have  $\hn (i)=u$ and $\sigma (i)\not=u$.

      Patient $i$ is processed in some Step 1.($k$) in the  smart reserve matching algorithm with all unreserved units processed first. As $\sigma(i)\not=u$ we have $i\notin J^u_{k}(I^*,r^*)$. On the other hand, since $\sigma \in \calm_{k-1}(I^*,r^*)$, by Statements 1, 2, and 3 above, we have $\hn  \in \calm_{k-1}(I^*,r^*)$ as well and it matches $i$ with $u$, contradicting $i \notin J^u_{k}(I^*,r^*)$. Therefore, Case 1 (ii) cannot hold either.

    \item[Case 2.] There is some  $i \in \sigma^{-1}(u)\setminus\nu^{-1}(u)$ such that $i \mathrel{\pi} j$:


    Construct a matching $\hs $ from $\sigma$ that it leaves every patient who is matched in Step 2 of the smart reserve matching algorithm with all unreserved units processed first: for every $i^* \in I^*$, $\hs (i^*)=\sigma(i^*)$ if $i^* \in I_{\sigma(i^*)}$ and $\hs (i^*)=\emptyset$ otherwise.  Clearly $\hs \in \calm(I^*,r^*)$ and is maximal in beneficiary assignment for $(I^*,r^*)$, since $\sigma$ is. By Lemma \ref{lem:mendelsohn/dulmage}, there exists a matching $\tilde{\nu}  \in \calm^b(I^*,r^*)$\footnote{As defined in the hypothesis of the lemma, $\tilde{\nu}  \in \calm^b(I^*,r^*)$ means that for every $i^*\in I^*$ and $c\in\calc$, $i^*\in \tilde{\nu} ^{-1}(c)$ implies $c\not=u$ and  $i^* \in I_c$.}
    such that under $\tilde{\nu}$ all patients in $\cup_{c \in \calc \setminus \{u\}}  \tilde{\nu}^{-1}(c)\cap I_c$ are matched with the preferential treatment categories in $\calc \setminus \{u\}$ that they are beneficiaries of,
    and for every $c \in \calc\setminus \{u\}$, $|\tilde{\nu}^{-1}(c)\cap I_c|=|\hs^{-1}(c)\cap I_c|$ (equality follows rather than $\ge$ as dictated by the lemma, because $\hs$ is maximal in beneficiary assignment for $(I^*,r^*)$). For every $i^*\in I^*$, we have $\nu(i^*)=u \implies \tilde{\nu}(i^*)=\emptyset$ as $\hn\in \calm^b(I^*,r^*)$. We modify $\tilde{\nu} $ to obtain $\hn$: For every $i^* \in \nu^{-1}(u)$, we set $\hn(i^*)=u$ and for every $i^* \in I^*\setminus \nu^{-1}(u)$, we set $\hn(i^*)=\tilde{\nu}(i^*)$.  Clearly, $\hn \in \calm(I^*,r^*)$ and is maximal in beneficiary assignment for $(I^*,r^*)$, since $\nu$ is. We will work with $\hs$ and $\hn$ instead of $\sigma$ and $\nu$ from now on. \bigskip

   Recall that $\hs(i)=u$ and $\hn(i)\not=u$. Two cases are possible: $i \in I_{\hn(i)}$ or $\hn(i)=\emptyset$.
   \begin{enumerate}
   \item If $i \in I_{\hn(i)}$: Then by Observation \ref{lem:alternate}, there exists a beneficiary alternating path $A$ from $\hs$ to $\hn$ beginning with $i$ and ending with some $i' \in I^*$ such that (i) $\hs(i')\in I_{\hs(i')}$ and $\hs(i')\not=u$, and (ii) $\hn(i')=u$ or $\hn(i')=\emptyset$.

   By the existence of the beneficiary alternating path, it is possible to match either $i$ or $i'$ with a preferential treatment category that she is a beneficiary of and match the other one with $u$ without changing the type of match of any other patient $i^* \in I^*\setminus\{i,i'\}$ has, i.e., either $i^*$ is matched with a preferential treatment category under both matchings or not. Yet, when $i$ is processed in Step 1 of the smart reserve matching algorithm with all unreserved units processed first, we chose $i$ to be matched with $u$ and $i'$ with a preferential treatment category. This means
   $$i \mathrel{\pi} i'.$$

    Let $\hat{c}\not =u$ be the category that $i$ is matched with under $\hn$. 

    If $\hs(i')\not=u$, then modify $\hn$ by assigning an unreserved unit to $i'$ instead of $j$: $\hn(j)=\emptyset$ and $\hn(i')=u$. Otherwise, do not modify $\hn$ any further.

    Consider the smaller economy $(I',r')$ such that $I'=I^* \setminus \{i,i'\}$ and for every $c\in \calc$, $r_{c}'=r^*_{c}-1$
    if $c \in \{\hat{c},u\}$ and $r'_{c}=r^*_{c}$, otherwise.

    We show that a smart reserve matching $\sigma' \in \cals^{r'_u}(I',r')$ can be obtained from the original smart reserve matching $\sigma \in \cals^{r^*_u}(I^*,r^*)$ and $\hn$.
   Consider the beneficiary alternating path $A$ we discovered above starting with patient $i$ and ending with patient $i'$ from $\hs$ to $\hn$: Suppose $A=(i,i_2,\hdots,i_{\overline{m}-1},i')$.
    Define
    \begin{align*}
      \sigma'(i^*)&=\sigma(i^*) \mbox{ for every } i^* \in I' \setminus \{i_2,\hdots,i_{\overline{m}-1}\} \mbox{ and }\\
      \sigma'(i^*)&=\hn(i^*) \mbox{ for every } i^* \in \{i_2,\hdots,i_{\overline{m}-1}\}.
    \end{align*}
    Observe that $\sigma'\in\calm(I',r')$. The existence of $\sigma'$ shows that it is feasible to match  every patient in $J^u_{|I^*|}(I^*,r^*)\setminus\{i\}$ with $u$ and it is feasible to match  every patient in $J_{|I^*|}(I^*,r^*)\setminus\{i'\}$ with a preferential treatment category that she is a beneficiary of in $(I',r')$. Thus, the smart reserve matching algorithm with all unreserved units processed first proceeds exactly in the same manner as it does for $(I^*,r^*)$ with the exception that it skips patients $i$ and $i'$ in the smaller economy $(I',r')$. Hence, $\sigma' \in \cals^{r_u'}(I',r')$.

    Let the restriction of matching $\hn$ to $(I',r')$ be $\nu'$.  Observe that $\nu'$ is a matching for $(I',r')$. Moreover, it is maximal in beneficiary assignment for $(I',r')$, since $\hn$ is maximal in beneficiary assignment for $(I^*,r^*)$.

    Now one of the two following cases holds for $\nu$:

    \begin{enumerate}
      \item If $\nu(i')\not=u$: Recall that while $\nu(j)=u$, we updated $\hn$ so that $\hn(i')=u$ and $\hn(j)=\emptyset$. Thus, $\nu'(j)=\emptyset$ as well. Since $i \mathrel{\pi} j$, this together with the inductive assumption that the claim holds for $(I',r')$ imply that the claim also holds for $(I^*,r^*)$, completing the induction.
      \item  If $\nu(i')=u$: Since $i \mathrel{\pi} i'$, this together with the inductive assumption that the claim holds for $(I',r')$ imply that the claim also holds for $(I^*,r^*)$, completing the induction. \medskip
    \end{enumerate}
    \item If $\hn(i)=\emptyset$: Recall that $\hn(j)=u$. We modify $\hn$ further that $\hn(i)=u$ and $\hn(j)=\emptyset$.
    Consider the smaller economy $(I',r')$ where $I'=I^*\setminus\{i\}$, $r'_u=r^*_u-1$, and $r'_c=r^*_c$ for every $c \in \calc\setminus\{u\}$.

     Let $\sigma'$ and $\nu'$ be the restrictions of $\sigma$ and $\hn$ to $(I',r')$, respectively. Since, $\sigma(i)=\hn(i)=u$ both $\sigma'$ and $\nu'$ are matchings for $(I',r')$. Since the capacity of category $u$ is decreased by one, $\sigma'$ is a smart reserve matching with all unreserved units processed first for $(I',r')$. To see this observe that the algorithm proceeds as it does for $(I^*,r^*)$ with the exception that it skips $i$. Matching $\nu'$ is maximal in beneficiary assignment for $(I',r')$. Therefore, by the inductive assumption, the claim holds for $(I',r')$. This together with the fact that $i \mathrel{\pi} j$ imply the claim holds for $(I^*,r^*)$, completing the induction.
     $\diamond$
    \end{enumerate}

\end{itemize}
 \end{itemize}

If $|\mu^{-1}(u)|<r_u$ then
$\ovf^{\sigma_{r_u}}_u\mathrel{\pieq}\ovf^\mu_u= \emptyset$. On the other hand, if $|\mu^{-1}(u)|=r_u$, Claim 1 implies that  $\ovf^{\sigma_{r_u}}_u =\min_{\pi}\sigma_{r_u}^{-1}(u)  \; \mathrel{\pieq} \; \ovf^\mu_u=\min_{\pi}\mu^{-1}(u)$.
\bigskip

\noindent{}\textbf{Proof of  $\ovf^{\mu}_u  \mathrel{\pieq}\ovf^{\sigma_{0}}_u$:}  \medskip

We prove the following claim first.\medskip

\noindent{}\textbf{Claim 2.}
For any set of patients $I^*\subseteq I$ and any capacity vector $r^*\le r$, suppose $\nu \in \calm(I^*,r^*)$ is a matching  that is maximal in beneficiary assignment for $(I^*,r^*)$. Let $\sigma \in \cals^{0}(I^*,r^*)$ be a smart reserve matching with all unreserved units processed last, $$J= \cup_{c\in \calc\setminus \{u\}}(\sigma^{-1}(c)\cap I_c), \quad \mbox{and} \quad
J'= \cup_{c\in \calc\setminus \{u\}}(\nu^{-1}(c)\cap I_c).$$
According to the baseline priority order $\pi$, for any $k \in \{1,\hdots, |J' \setminus J|\}$, let $j_k$ be the $k^{\mbox{\scriptsize{th}}}$ highest priority patient in $J \setminus J'$ and $j'_k$ be the $k^{\mbox{\scriptsize{th}}}$ highest priority patient in $J' \setminus J$, then $$j_k \mathrel{\pi} j'_k.$$\medskip

\noindent{}\emph{Proof.}
 We use induction on the cardinality of $I^*$ and
on the magnitude of vector of capacities of  categories $r^*$:\begin{itemize}
  \item For the base case when $I^*=\emptyset$ and
$r^*_c=0$ for every $c \in \calc$, the claim holds trivially.
\item In the inductive step, suppose for every capacity of categories bounded above by vector $r^*$ and subsets of patients in $I$ bounded above by cardinality $k^*$ the claim holds.

\item Consider set of patients $I^* \subseteq I$ such that $|I^*|=k^*$ and capacity vector for categories $r^*=(r^*_c)_{c \in \calc}$. If $J\setminus J'=\emptyset$ then the claim holds trivially. Suppose $J \setminus J' \not=\emptyset$. Let $i \in J \setminus J'$ be the highest priority patient in $J  \setminus J'$ according to $\pi$.

 We have $|J|=|J'|$ by maximality of $\sigma$ and $\nu$ in beneficiary assignment for $I^*$. Thus, $|J\setminus J'|=|J'\setminus J|$, which implies  $J' \setminus J \not =\emptyset$.

    By Lemma, \ref{lem:mendelsohn/dulmage} there exists a matching $\hn  \in \calm^b(I^*,r^*)$ that matches  patients only with preferential treatment categories that they are beneficiaries of such that $\hn $ matches patients in $J'$ and $|\sigma^{-1}(c)\cap I_c|$ units reserved for every preferential treatment category $c \in \calc \setminus \{u\}$.\footnote{Although both $\sigma$ and $\nu$ may be matching some patients with categories that they are not beneficiaries of or with the unreserved category, we can simply leave those patients unmatched in $\sigma$ and $\nu$ and apply Lemma \ref{lem:mendelsohn/dulmage} to see such a matching $\hn$ exists.} Since both $\nu$ and $\sigma$ are maximal in beneficiary assignment for $(I^*,r^*)$, then only patients in $J'$ should be matched under $\hn $ and no other patients (as otherwise $\nu$ would not be maximal in beneficiary assignment for $(I^*,r^*)$).

    Since $\hn(i)=\emptyset$, $i\in I_{\sigma(i)}$, and $\sigma(i)\not=u$, by Observation \ref{lem:alternate}, there exists a beneficiary alternating path $A$ starting with $i$ from $\hn$ to $\sigma$ and ending with a patient $i' \in I_{\hn(i')}$ (and $\hn(i')\not=u$ by its construction), and yet $i' \notin I_{\sigma(i)}$ or $\sigma(i')=u$.

    Existence of the beneficiary alternating path shows that it is possible to match $i$ or $i'$ (but not both) with preferential treatment categories that they are beneficiaries of without affecting anybody else's status as  committed or uncommitted in Step 1 of the smart reserve matching algorithm with all unreserved units processed last.  Since $\sigma$ matches $i$ with a preferential treatment category that she is a beneficiary of at the cost of patient $i'$, we have $$i \mathrel{\pi} i'.$$

    Next consider the smaller economy $(I',r')$ in which we  remove (i) $i$ and $i'$ and set $I'=I^*\setminus\{i,i'\}$, (ii) one of the units associated with preferential treatment category $\hat{c}=\hn(i')$  and set   $r'_{\hat{c}}=r^*_{\hat{c}}-1$, and (iii) keep the capacity of every other category $c \in \calc\setminus \{\hat{c}\}$ the same and set $r'_{c}=r^*_{c}$.

    Let $\nu'$ be the restriction of $\hn $ to $(I',r')$. As $\hn(i)=\emptyset$, $\hn (i')=\hat{c}\not=u$ such that $i \in I_{\hat{c}}$ and the capacity of $\hat{c}$ is reduced by 1 in the smaller economy, $\nu' \in \calm(I',r')$, and furthermore, it is maximal in beneficiary assignment for $(I',r')$.

     We form a matching $\sigma' \in \calm(I',r')$ by modifying $\sigma$ and $\hn$ using the beneficiary alternating path $A$ we found before. Recall that $A$ is the beneficiary alternating path from $\hn$ to $\sigma$ beginning with $i$ and ending with $i'$. Suppose $A=(i,i_2,\hdots,i_{\overline{m}-1},i')$.
    Define
    \begin{align*}
      \sigma'(i^*)&=\sigma(i^*) \mbox{ for every } i^* \in I' \setminus \{i_2,\hdots,i_{\overline{m}-1}\} \mbox{ and }\\
      \sigma'(i^*)&=\hn(i^*) \mbox{ for every } i^*\in \{i_2,\hdots,i_{\overline{m}-1}\}.
    \end{align*}
    Observe that $\sigma'\in\calm(I',r')$. The existence of $\sigma'$ shows that it is possible to match every patient in $J_{|I^*|}(I^*,r^*)\setminus \{i\}$ with a preferential treatment category that she is a beneficiary of in $(I', r')$. Thus, the smart reserve matching algorithm with all unreserved units processed last proceeds as it does for $(I^*,r^*)$ with the exception that it skips patients $i$ and $i'$. Therefore, $\sigma' \in \cals^0(I',r')$.

     By the inductive assumption, the claim holds for  $\sigma'$ and $\nu'$ for $(I',r')$. This completes the induction, as we already showed  $i \mathrel{\pi} i'$. $\diamond$
\end{itemize}

Thus, we showed that at the end of  Step 1 of the smart matching algorithm with all unreserved units processed last, weakly lower priority patients have remained uncommitted in $J^*=I \setminus J_{|I|} = I \setminus \cup_{c \in \calc\setminus\{u\}} (\sigma_0^{-1}(c)\cap I_c)$ than in $\hat{J}=I \setminus \cup_{c \in \calc\setminus\{u\}} (\mu^{-1}(c)\cap I_c)$.

Two cases are possible:
\begin{itemize}
\item If we have a soft reserves system: As both $\sigma_0$ and $\mu$ are maximal in beneficiary assignment, an equal number of units are assigned to the highest $\pi$-priority patients in $\hat{J}$ (by Step 2 of the smart reserve matching algorithm) and $J^*$ (as by Theorem \ref{thm:cutoff}, $\mu$ respects priorities and is non-wasteful). Under $\sigma_0$, the unreserved units are assigned last in order in Step 2 of the algorithm. On the other hand, the remainder of $\mu$, i.e., the units assigned to the non-beneficiaries of preferential treatment categories  and beneficiaries of $u$, can be constructed by assigning the rest of the units sequentially to the highest priority patients in $J^*$ one by one when unreserved units are not necessarily processed last.

Therefore, if $|\mu^{-1}(u)|<r_u$ then $|\sigma_0^{-1}(u)|<r_u$, in turn implying $\ovf^{\mu}_u= \ovf^{\sigma_{0}}_u=\emptyset$. If $|\mu^{-1}(u)|=r_u$ then $|\sigma_0^{-1}(u)|\le r_u$ and  $ \ovf^\mu_u = \min_{\pi}\mu^{-1}(u)  \mathrel{\pieq} \ovf^{\sigma_{0}}_u\in\{\emptyset,\min_{\pi}\sigma_0^{-1}(u)\}$.
\item If we have a hard reserves system: The proof is identical as the above case with the exception that now only unreserved units are assigned as both $\sigma_0$ and $\mu$ comply with eligibility requirements. \end{itemize}
\end{proof}

\end{document}